\documentclass[11pt]{article}

\usepackage{amsfonts}
\usepackage{amsmath}
\usepackage{amssymb}
\usepackage{amsthm}
\usepackage[mathscr]{eucal}
\usepackage{bbold}
\usepackage{braket}
\usepackage{color}
\usepackage{comment}
\usepackage{dsfont}
\usepackage{enumitem}
\usepackage{framed}
\usepackage{jheppub}
\usepackage{mathtools}
\usepackage{physics}
\usepackage[normalem]{ulem}
\usepackage{tensor}
\usepackage{thmtools}
\usepackage{thm-restate}
\usepackage{ulem}
\usepackage{tikz}
\usetikzlibrary{math} %needed tikz library for setting variables

\definecolor{acolor}{rgb}{0.1,.6,0}

\definecolor{rcolor}{rgb}{0.9,0.1,0.1}

\definecolor{bcolor}{rgb}{0.1,0,1}

\newtheorem{lemma}{Lemma}

\newtheorem{assumption}{Assumption}
\newtheorem{conjecture}{Conjecture}

\def\Tr {\operatorname{Tr}}

\title{A trace inequality for Euclidean gravitational path integrals
(and a new positive action conjecture)}

\author[a]{Eugenia Colafranceschi,} \emailAdd{ecolafranceschi@ucsb.edu }

\author[a]{Donald Marolf} \emailAdd{marolf@ucsb.edu}

\author[a,b]{and Zhencheng Wang} \emailAdd{zcwang1@illinois.edu}

\affiliation[a]{Department of Physics, University of California, Santa Barbara, CA 93106, USA}
\affiliation[b]{Department of Physics, University of Illinois, Urbana-Champaign, 1110 W. Green St., Urbana, IL 61801, USA}

\abstract{
The AdS/CFT correspondence states that certain conformal field theories are equivalent to string theories in a higher-dimensional anti-de Sitter space. One aspect of the correspondence is an equivalence of density matrices or, if one ignores normalizations, of positive operators. On the CFT side of the correspondence, any two positive operators $A,B$ will satisfy the trace inequality $\Tr(AB) \le \Tr(A) \Tr(B)$. This relation holds on any Hilbert space ${\cal H}$ and is deeply associated with the fact that the algebra $B({\cal H})$  of bounded operators on ${\cal H}$ is a type I von Neumann factor.   Holographic bulk theories must thus satisfy a corresponding condition, which we investigate below.  In particular, we argue that the Euclidean gravitational path integral respects this inequality at all orders in the semi-classical expansion and with arbitrary higher-derivative corrections.   The argument relies on a conjectured property of the classical gravitational action, which in particular implies a positive action conjecture for quantum gravity wavefunctions.  We prove this conjecture for Jackiw-Teitelboim gravity and we also motivate it for more general theories.   }
\begin{document}

% Title Page
% \input{editionlegend.tex}
\maketitle

\section{Introduction}
\label{sec:intro}

The Anti-de Sitter/Conformal Field theory correspondence  (AdS/CFT) \cite{Maldacena:1997re} predicts exact equivalence between appropriate conformal field theories and their dual bulk string theories. Using the bulk to reproduce detailed properties of specific CFTs typically requires using intricate properties of the stringy description.  However, it is often the case that fundamental properties of CFTs can already be seen in the approximation where the bulk theory is described by semiclassical gravity, perhaps coupled to appropriate matter fields.  Important examples of such properties include CFT microcausality, strong subadditivity of entropy, and the fact that larger regions of the CFT define larger algebras of observables.  In particular, these features are associated with results for asymptotically locally anti-de Sitter (AlAdS) bulk spacetimes satisfying the null energy condition.  The corresponding bulk results are,  first, that any causal bulk  curve between boundary points is deformable to a causal curve lying entirely within the boundary \cite{Gao:2000ga}, second that strong subadditivity holds for HRT surfaces \cite{Wall:2012uf}, and third that entanglement wedges nest appropriately \cite{Wall:2012uf,Czech:2012bh,Akers:2016ugt}.  Quantum effects in the bulk typically preserve such properties so long as they satisfy the quantum focussing conjecture \cite{Bousso:2015mna}.

The goal of the present work is to study the dual bulk implementation of the CFT inequality
\begin{equation}
\label{eq:trineq}
\Tr_{\scriptscriptstyle {\cal D}} (BC) \le \Tr_{\scriptscriptstyle {\cal D}}(B) \Tr_{\scriptscriptstyle {\cal D}}(C),
\end{equation}
which relates traces of positive operators $B,C$ on any Hilbert space ${\cal H}$. For positive $B=b^\dagger b$, $C=c^\dagger c$, some readers may prefer to write this in the form
\begin{equation}
\label{eq:trineq1}
\Tr_{\scriptscriptstyle {\cal D}} (bc^\dagger c b^\dagger) \le \Tr_{\scriptscriptstyle {\cal D}}(b^\dagger b) \Tr_{\scriptscriptstyle {\cal D}}(c^\dagger c),
\end{equation}
so that the argument of the trace on the left-hand-side is also a positive operator.
Recall that positive operators are self-adjoint by definition \cite{Blackadar:2006zz}, and that `positivity' requires the eigenvalues to be non-negative.  In \eqref{eq:trineq} and \eqref{eq:trineq1}, we use the subscript ${\cal D}$ to denote the non-gravitational CFT dual of a bulk theory, and we write $\Tr_{\scriptscriptstyle {\cal D}}$ to emphasize that the trace is the standard trace on the ${\cal D}$ side of the duality.  In particular,
$\Tr_{\scriptscriptstyle {\cal D}}$ denotes the familiar operation computed by introducing any orthonormal basis $|i\rangle$ on the Hilbert space for ${\cal D}$ and  performing the sum
\begin{equation}
\label{eq:trdef}
\Tr_{\scriptscriptstyle {\cal D}}({\cal O}) := \sum_i \ \langle i |{\cal O} |i\rangle.
\end{equation}
For simplicity of presentation we confine ourselves to the AdS/CFT context below, but similar discussions clearly apply to other gauge/gravity dualities as well, such as those described in e.g. \cite{Banks:1996vh,Itzhaki:1998dd}.
To avoid infrared divergences, we assume ${\cal D}$ to be defined on a spatially compact spacetime.  Since we consider path integrals dual to some $\Tr_{\scriptscriptstyle {\cal D}} B$, we may then take all of our Euclidean boundaries to be compact.

The inequality \eqref{eq:trineq} is easily proven using standard Hilbert space operations in ${\cal D}$.  One first notes that the inequality is trivial when $\Tr_{\scriptscriptstyle {\cal D}}(B)=+\infty$, so this leaves only the case of finite $\Tr_{\scriptscriptstyle {\cal D}}(B)$.  One then observes that, when $\Tr_{\scriptscriptstyle {\cal D}}(B)$ is finite, the positivity of  $B$ requires the operator $B$ to have a largest eigenvalue $B_{max}$.  We then simply choose the $\{|i\rangle\}$ in \eqref{eq:trdef} to be eigenstates of $C$ with eigenvalues $C_i \ge0$ and write
\begin{equation}
\label{eq:Trinder}
\Tr_{\scriptscriptstyle {\cal D}}(BC) = \sum_i C_i\, \langle i | B | i \rangle \le \sum_i B_{max} C_i  = B_{max} \Tr(C) \le \Tr_{\scriptscriptstyle {\cal D}}(B) \Tr_{\scriptscriptstyle {\cal D}}(C).
\end{equation}
Indeed, this argument also shows that the bound \eqref{eq:trineq} is quite weak, and that it is saturated only when $B,C$ are both proportional to a common projection of rank one.  For $B=C$, this latter observation is equivalent to the familiar statement that the purity of a density matrix is $1$ only when the density matrix is pure, and thus when it is proportional to a projection of rank one.

While the bound \eqref{eq:trineq} may be weak,  stronger bounds typically involve further details of the spectrum of $B,C$ and are thus more difficult to study. One example is the bound $B_{max} \Tr(C)$ also derived in \eqref{eq:Trinder}.  Another is the even stronger von Neumann trace inequality $\Tr(BC) \le \sum_i B_i C_i$, where we have now introduced the full set of eigenvalues $B_i$ of $B$, and both $C_i$ and $B_i$ have been ordered so that $C_i \ge C_j$ and $B_i \ge B_j$ when $i \ge j$.  These more intricate bounds on the CFT trace are correspondingly more awkward to study on the gravitational side of the AdS/CFT duality.

However, despite its weakness, the bound \eqref{eq:trineq} can be used to derive fundamental consequences.  One example is the fact that the algebra $B({\cal H})$ of bounded operators on any Hilbert space is a type I von Neumann factor.  This can be shown by first noting that the commutant of $B({\cal H})$ is trivial, so that $B({\cal H})$ must be a factor of some type.  One then considers any projection $P$ and sets $C = B =P$.  Since $P^2 = P$, the bound \eqref{eq:trineq} requires $\Tr(P) \ge 1$ for any $P$.  In contrast, when factors of some type other than I are present in a von Neumann algebra, any faithful normal semi-finite trace on the algebra will always assign arbitrarily small traces to some family of projections having arbitrarily small trace \cite{KR:1997}.  This result is a key motivation for our study.

Our goal here is to show how \eqref{eq:trineq} arises from the bulk point of view.  In doing so we will work at the level of the semiclassical approximation to the Euclidean path integral for a low-energy bulk effective theory. The semiclassical bulk description will necessarily involve gravity, but our analysis will not depend on the details of any UV completion.

Now, in fact, gravitational path integrals that include sums over topology are generally {\it not} dual to single CFTs as they fail to factorize over disconnected boundaries (see e.g. the classic discussion of \cite{Maldacena:2004rf}). However, if a non-factorizing bulk path integral makes sense, we expect it to behave like those discussed in \cite{Marolf:2020xie,Blommaert:2022ucs} where the path integral decomposes into a sum over so-called baby universe $\alpha$-sectors in which factorization holds; see also \cite{Coleman:1988cy,Giddings:1988cx} for earlier discussions of this idea.    We then expect \eqref{eq:trineq} to be satisfied separately in each $\alpha$-sector.

Furthermore, if an inequality of the form \eqref{eq:trineq1} holds in each member of an ensemble then, so long as the ensemble has non-negative probabilities to realize each of its members, a similar inequality will hold for ensemble averages.  We might write this averaged inequality in the form
\begin{equation}
\label{eq:trineq2}
\langle \Tr_{\scriptscriptstyle {\cal D}} (bc^\dagger c b^\dagger) \rangle \le  \langle \Tr_{\scriptscriptstyle {\cal D}}(b^\dagger b) \Tr_{\scriptscriptstyle {\cal D}}(c^\dagger c) \rangle.
\end{equation}
Here it will of course be important that the right-hand side is a single ensemble correlation function and {\it not} a product of ensemble averages.

It is therefore interesting to understand if a given bulk theory satisfies a corresponding inequality, which we might write in the form
\begin{equation}
\label{eq:trineq3}
\zeta\left( \tilde M_{bc^\dagger c b^\dagger} \right) \le  \zeta\left( \tilde M_{b^\dagger b} \sqcup \tilde M_{c^\dagger c} \right).
\end{equation}
Here $\zeta(\tilde M)$ denotes the Euclidean Asymototically locally Anti-de Sitter gravitational path integral defined by the boundary conditions $\tilde M$, which we think of as a boundary manifold equipped with various fields.  The symbol $\sqcup$ denotes disjoint union and $\tilde M_{bc^\dagger c b^\dagger}$ is a smooth closed manifold that can be broken into four pieces $M_b, M_{b^\dagger}, M_c, M_{c^\dagger}$.  Since  the operators $b^\dagger b$ and $c^\dagger c$ on the dual ${\cal D}$ are positive, we also require that gluing together $M_b$ and  $M_{b^\dagger}$ gives a new smooth closed manifold  $M_{bb^\dagger}$ that is invariant under a reflection-symmetry that exchanges the $b$ and $b^\dagger$ pieces (and which complex-conjugates any complex boundary conditions) and similarly for $M_{cc^\dagger}$.  This notation will be explained in more detail (and with appropriate figures) in the sections below.  Note that, in general, the right-hand side of \eqref{eq:trineq3} may involve spacetime wormholes, and we should expect there to be cases where such wormholes are important in enforcing the inequality. In \eqref{eq:trineq3}, we are allowed to consider the case where $\tilde M_{bc^\dagger c b^\dagger}$ is not connected, though when $\tilde M_{bc^\dagger c b^\dagger}$ is a disjoint union of its $bb^\dagger$ and $cc^\dagger$ parts, those parts are precisely
$ \tilde M_{b^\dagger b}$, $\tilde M_{c^\dagger c}$ and \eqref{eq:trineq3} becomes a trivial equality.

There is, however, a normalization issue that remains to be addressed.  The reader will immediately note that equation \eqref{eq:trineq} fails to be invariant under rescaling the trace by a constant factor $\lambda$.  This failure arises because the left-hand-side scales with $\lambda$ while the right-hand-side scales with $\lambda^2$.  This is not a problem since the trace of an operator is defined by summing diagonal matrix elements over an orthonormal basis and so comes to us with a preferred normalization.

In contrast, the conjecture \eqref{eq:trineq3} for the bulk path integral $\zeta$ remains invariant if we change the normalization of $\zeta$ since there is only a single $\zeta$ on each side. Nevertheless, if we would like our path integral to factorize  in the sense that for disjoint unions $\tilde M_1 \sqcup \tilde M_2$ we have
\begin{equation}
\label{eq:fact}
\zeta(\tilde M_1 \sqcup \tilde M_2) = \zeta(\tilde M_1)\zeta(\tilde M_2),
\end{equation}
then the relation \eqref{eq:fact}  will in fact require a preferred normalization for $\zeta$.  Indeed, even if our path integral is to satisfy \eqref{eq:fact} only to some approximation this condition will still greatly restrict the possible choice of normalization for $\zeta$.  

Since the emptyset $\emptyset$ satisfies $\emptyset = \emptyset \sqcup \emptyset$, setting $\tilde M_1 = \tilde M_2 = \emptyset$ in \eqref{eq:fact} shows that the required normalization gives $\zeta(\emptyset)=1$.  Thus the path integral over all compact Euclidean spacetimes (with no boundaries) should give the value $1$.  This is, of course, equivalent to first allowing an arbitrary normalization and then dividing the result by the norm of the Hartle-Hawking no boundary state \cite{Hartle:1983ai}.  It is thus also equivalent to simply defining our path integral to sum {\it only} over Euclidean spacetimes in which every bulk point is connected by some path to a point on the (asymptotically locally AdS) boundary at which boundary conditions are specified.  This coincides with the traditional treatment of the gravitational path integral in AdS/CFT \cite{Witten:1998qj}.  We adopt this normalization in all discussions below.

Our discussion begins with an analysis of simple cases and simple bulk theories in section \ref{sec:simple}. We first show that, in the context of black hole thermodynamics,
standard results for either Jackiw-Teitelboim (JT) \cite{Jackiw:1984je,Teitelboim:1983ux} or Einstein-Hilbert gravity imply the bulk version of the inequality \eqref{eq:trineq} to hold at all orders in the semiclassical expansion and at all orders in any perturbative higher-derivative corrections.  By referring to the black hole thermodynamics context above, we mean that the operators $B,C$ in \eqref{eq:trineq3} are both functions of the Hamiltonian $H$ and that relevant path integrals are dominated by Euclidean black hole saddles.  We focus on the simple case of pure JT gravity, where there are no other operators to consider and where all bulk saddles contain black holes.   However, the arguments in \ref{subsec:pureJT} also apply to black hole thermodynamics more generally.  We then also show that, due to the simplicity of JT gravity, for any UV completion where the path integral can be studied in the manner described by Saad, Shenker, and Stanford \cite{Saad:2019lba} and for interesting semiclassical limits, \eqref{eq:trineq3} holds even when the theory is coupled to matter (so long as the matter coupling is dilaton-free and the matter satisfies a positive action condition).

In Section \ref{sec:gen2}, we then proceed to discuss \eqref{eq:trineq3} for operators $B,C$ in more general theories and more general phases (perhaps not dominated by black holes).  Since the inequality \eqref{eq:trineq} holds for any quantum theory, it will be enlightening to
look again at the ${\cal D}$ side of the duality to see how the standard non-gravitational Euclidean path integral for ${\cal D}$ can be used to provide an alternate derivation of \eqref{eq:trineq3} at leading order in the semiclassical approximation (without yet invoking any possible gravitating bulk dual).  This is done in section \ref{sec:nongrav},  where we assume only that each member of the relevant class of Hamiltonians for the theory is bounded below and that the theory is 2nd order in derivatives.  Higher derivative corrections can then be incorporated perturbatively.  We do not study quantum corrections in this context since we will treat such corrections by a different argument in our discussion of gravitating bulk duals.

The above discussion sets the stage for us to address the general derivation of \eqref{eq:trineq3} from the gravitational side of the duality.  We open this discussion in section \ref{sec:bulk} by showing that the basic outline of the non-gravitational argument of section \ref{sec:nongrav} can be easily adapted to the gravitational context.  However, a crucial ingredient in the non-gravitational argument turns out to be the fact that the non-gravitational Euclidean action is bounded below.  This property is of course well-known to fail off-shell in gravitational theories; see e.g. \cite{Gibbons:1978ac}.  We deal with this issue in stages by phrasing the argument of section \ref{sec:bulk} in terms of a series of assumptions about the gravitational path integral which will turn out to be plausible (and, in some cases, provably true) despite the fact that the gravitational action is unbounded below.  The main discussion focuses on two-derivative theories of gravity (like Einstein-Hilbert or JT), though arbitrary higher derivative corrections are allowed so long as they are treated perturbatively.  When our assumptions are satisfied, the argument establishes \eqref{eq:trineq3} at all orders in the semiclassical expansion.

We then separate out discussion of the status of those assumptions (and the associated issues surrounding the conformal factor problem of Euclidean gravity), placing this material in section \ref{sec:asstatus}.  These assumptions imply a new positive action conjecture that generalizes the original conjecture of Hawking \cite{Gibbons:1978ac} in several ways.  We prove this conjecture to hold in JT gravity minimally-coupled to positive-action matter, and we also motivate the conjecture more generally in section \ref{sec:posI}.  Finally, we close in section \ref{sec:disc} with a summary and brief discussion of future directions.

\section{Simple cases and simple theories}
\label{sec:simple}

Asymptotically Anti-de Sitter Jackiw-Teitelboim gravity is a simple $2d$ toy model of gravitational systems in which many explicit computations are possible.  Section \ref{subsec:pureJT} considers the theory of ``pure'' JT gravity which contains only a metric $g$ and a dilaton $\phi$, with no additional matter fields. The addition of matter fields will be discussed in section \ref{subsec:SSS} using ideas from \cite{Saad:2019lba}.
 We use conventions in which the pure JT action on a disk takes the form
\begin{eqnarray}
\label{eq:JTEI1}
I &=& - \phi_0 \left[ \int_{{\cal M}_{ }} \sqrt{g} R + 2 \int_{\partial_{as} {\cal M}_{ }} \sqrt{h} K + 2 \int_{\partial_{f} {\cal M}_{ }} \sqrt{h} K \right] \cr
&-& \left[ \int_{{\cal M}} \sqrt{g} \phi (R+2)   + 2 \int_{\partial {\cal M}_{ }} \sqrt{h} \phi (K-1)\right] .\ \ \ \ \ \
\end{eqnarray}
Here $\phi_0$ is a constant, $h$ is the induced metric on a boundary, and $K$ is the extrinsic curvature (a scalar, since the boundary is one-dimensional) defined by the outward-pointing normal. The detailed boundary conditions to be used will be described in appendix \ref{subsec:JTBC}.

\subsection{The Trace Inequality in gravitational thermodynamics: Jackiw-Teitelboim gravity and beyond}
\label{subsec:pureJT}

Pure JT gravity has no local degrees of freedom, and in fact there is very little to compute.    In particular,  our 2-dimensional bulk must have a 1-dimensional boundary, so the only compact connected boundary is a circle.  The JT path integral is then specified by the constant $\phi_0$ in \eqref{eq:JTEI1}, a function $\phi_b$ on this circle having dimensions of length and prescribing boundary conditions for the dilaton, and the length $\beta$ of the circle (as defined using a rescaled unphysical metric). However, one may change the conformal frame at infinity without changing the path integral and, by doing so, one can reduce the general computation to the case where $\phi_b$ is any given positive constant $\bar \phi_b$ \cite{Maldacena:2016upp}.  This result is reviewed in appendix \ref{subsec:JTpos}.  As a result, in the rest of this section we simply choose some fixed value of this constant  $\bar \phi_b$ and consider all circles to be labelled only by their length $\beta$ in the corresponding conformal frame.

If one were to treat JT gravity non-perturbatively at a level where it is equivalent to a theory of a single matrix (see e.g. \cite{Blommaert:2022ucs}), then \eqref{eq:trineq3} would follow by using this equivalence to transcribe into bulk language the quantum mechanical derivation of  \eqref{eq:trineq} given in section \ref{sec:intro}.  Here we instead wish to focus on semiclassical treatments of JT gravity.  The idea is to gain insight into calculations we can also hope to control in higher dimensional gravitational theories.

In higher dimensions,  the semiclassical limit can be characterized by taking $G \rightarrow 0$.  However, in JT gravity {\it two} of the above-mentioned parameters, $\phi_0$ and $\bar \phi_b$, each take on aspects of the role played by $G$ in higher dimensions.  As a result, JT gravity admits various notions of semiclassical limit.  One of these is given by taking  $\phi_0$ large with $\bar \phi_b$ fixed, while another is the limit of large $\bar \phi_b$ with fixed $\phi_0$.  Establishing \eqref{eq:trineq3} in both cases then clearly also establishes the desired result in any limit where both $\phi_0$ and $\bar \phi_b$ become large.

As one can see from \eqref{eq:JTEI1}, the entire affect of $\phi_0$ is to weight spacetimes in the path integral by $e^{4\pi \phi_0 \chi}$, where $\chi$ is the Euler character of the spacetime.  As a result, since we use the normalization described in the introduction in which disconnected compact universes do not contribute,
the limit $\phi_0 \rightarrow \infty$ with all other parameters held fixed is dominated by disk contributions.   Furthermore, there is a factor of $e^{4\pi \phi_0}$ for each disk.  The number of disks is determined by the number of circular boundaries for the path integral, which is necessarily {\it larger}\footnote{Except in the trivial case where $\tilde M_{bc^\dagger cb^\dagger}$ is already disconnected with $\tilde M_{bc^\dagger cb^\dagger} = \tilde M_{b^\dagger b} \sqcup \tilde M_{c^\dagger c}$.  In this case the two sides of \eqref{eq:trineq3} are manifestly identical so that the inequality is a tautology.}
on the right-hand-side of \eqref{eq:trineq3} (where $\tilde M_{bc^\dagger cb^\dagger}$ has been split into $\tilde M_{b^\dagger b}$ and $\tilde M_{c^\dagger c}$) than on the left (where $M_{bc^\dagger cb}$ remains intact). The right-hand-side is thus clearly larger than the left-hand-side in the limit where $\phi_0$ is taken large with all else fixed.  This establishes the desired inequality \eqref{eq:trineq3} in this context.

However, as mentioned above, we can instead choose to keep $\phi_0$ finite and to study the limit $\bar \phi_b \rightarrow \infty$ with all else fixed (including the inverse temperature $\beta$, which we henceforth require to be finite).  Let us use $Z(\beta)$ to denote the path integral defined by a circular boundary of length $\beta$.  In the dual quantum mechanical system one would write

\begin{equation}
\label{eq:ZbJT}
Z(\beta) = \Tr_{\scriptscriptstyle {\cal D}}(e^{-\beta H}).
\end{equation}
Since the only objects we can compute are linear combinations of \eqref{eq:ZbJT} with different values of $\beta$,  the only operators in ${\cal D}$ that we can study are functions of $H$, where $H$ is the Hamiltonian of ${\cal D}$.  The change of conformal frame mentioned above that removes the dependence on general functions $\phi_b$ is sufficiently local that no further operators would have been found for more general (position-dependent) choices $\phi_b$. For later use, we note that at leading semiclassical order (with the above normalization of the action) one finds \cite{Maldacena:2016upp}
\begin{equation}
\label{eq:scJTZ}
Z(\beta) \approx e^{4\pi \phi_0} e^{4\pi^2 \bar \phi_b/\beta}.
\end{equation}

We will first discuss the trace inequality \eqref{eq:trineq3} for the simple case where $B = e^{-\beta_1 H}$ and $C = e^{-\beta_2 H}.$   In doing so, it will be useful to recall that a partition function $Z(\beta)$ allows one to compute an associated entropy $S(\beta)$ using
\begin{equation}
\label{eq:SfromZ}
S(\beta) := -\beta^2 \frac{d}{d\beta} \left( \beta^{-1}  \ln Z(\beta) \right) \approx 4\pi(\phi_0 + 2 \pi \bar \phi_b/\beta).
\end{equation}

It turns out that the condition $S \ge 0$ is sufficient to derive the trace inequality \eqref{eq:trineq3} in the current context.  To see this note that, for $B, C$ as above, our  \eqref{eq:trineq3} is equivalent to
\begin{equation}
\label{eq:lnZsubadd}
\ln Z(\beta_1 + \beta_2) \le \ln Z(\beta_1) + \ln Z(\beta_2).
\end{equation}
In other words, \eqref{eq:trineq3} is equivalent to the requirement that $\ln Z(\beta)$ be a superadditive function of $\beta$.  However, since $\beta >0$, non-negativity of \eqref{eq:SfromZ} is equivalent to stating that $\beta^{-1}  \ln Z(\beta)$ decreases monotonically for $\beta \in (0, \infty)$.
We may thus derive \eqref{eq:lnZsubadd} from such non-negativity as follows:
\begin{eqnarray}
\ln Z(\beta_1 + \beta_2)&=& \beta_1 \frac{ \ln Z(\beta_1 + \beta_2)}{\beta_1 + \beta_2 } + \beta_2 \frac{\ln Z(\beta_1 + \beta_2)}{\beta_1 + \beta_2 }\cr &\le& \beta_1 \frac{ \ln Z(\beta_1)}{\beta_1} + \beta_2 \frac{\ln Z(\beta_2)}{\beta_2 } = \ln Z(\beta_1) + \ln Z(\beta_2).
\end{eqnarray}
Furthermore, for $S >0$ we see that $\eqref{eq:trineq3}$ becomes a strict inequality.

Since \eqref{eq:SfromZ} is in fact positive, it follows that \eqref{eq:trineq3} is satisfied at this order for $B = e^{-\beta_1 H}$, $C = e^{-\beta_2 H}.$  Indeed, we see that the inequality cannot be saturated for any $\beta_1, \beta_2$.  As a result, when treated perturbatively, higher order corrections cannot lead to violations of \eqref{eq:trineq}.

Now, as described in \cite{Engelhardt:2020qpv}, negative entropies {\it do} arise in non-perturbative regimes if one takes the path integral for the no boundary baby universe state to compute the entropy \eqref{eq:SfromZ}.  But the entropies  in individual super-selection sectors (which are dual to entropies of individual CFTs) should be positive even at the non-perturbative level; see again the discussion of superselection sectors, ensembles, and factorization in section \ref{sec:intro}.  Furthermore, as described there, we would still expect the trace inequality to hold in the form \eqref{eq:trineq3}, which requires us to include contributions from spacetime wormholes on the right-hand-side.  Including simple such wormholes did indeed ameliorate the negative entropy issues discussed in \cite{Engelhardt:2020qpv}.  Consistency with the dual matrix ensemble of \cite{Saad:2019lba} then requires that the remaining issue to be resolved by the inclusion of higher topologies and the appropriate non-perturbative completion, though this remains to be explicitly analyzed.

The simple argument given above for the case $B= e^{-\beta_1 H}$, $C= e^{-\beta_2 H}$ can be extended to general functions of $H$ constructed as linear combinations of the $e^{-\beta H}$.  A straightforward way to do so is to realize that, in any dual quantum-mechanics theory, we may first analytically continue $e^{-\beta H}$ in $\beta$ to construct the operators $e^{itH}$, whence for each real $E$ one may define the operators
\begin{equation}
\label{eq:deltadef}
\delta(H-E) := \frac{1}{2\pi} \int_{\Gamma} dt\, e^{itH}e^{-itE}.
\end{equation}
In \eqref{eq:deltadef}, since we wish $\delta (H-E)$ to be the inverse Laplace transform of $e^{-\beta H}$, we should take the contour of integration $\Gamma$ to be {\it above} any singularities that may arise.  This is equivalent to choosing the contour for $\beta = -it$ to be to the {\it right} of any singularities.
Linearity then implies the traces of the operators \eqref{eq:deltadef} to be given by the Fourier transform of $Z(-it) := \Tr_{\scriptscriptstyle {\cal D}} e^{itH}$, which we take to be given by the continuation of \eqref{eq:scJTZ}; i.e.
\begin{equation}
\label{eq:Zit}
Z(-it) \approx
Z(\beta) \approx e^{4\pi \phi_0} e^{i 4\pi^2 \bar \phi_b/t}.
\end{equation}
Combining these results yields
\begin{equation}
\label{eq:deltatr}
\Tr_{\scriptscriptstyle {\cal D}} \ \delta(H-E) := \frac{1}{2\pi} \int dt\, Z(-it)e^{-itE} \approx \frac{e^{4\pi \phi_0}}{2\pi} \int dt\,   e^{i 4\pi^2  {\bar \phi}_b/t}e^{-itE},
\end{equation}
where as in \eqref{eq:deltadef} the contour is taken to lie above the singularity at $t=0$, though we may otherwise choose it to run along the real $t$-axis.

For fixed real $E >0$, in the limit of large $\bar \phi_b$, we may then evaluate the remaining integral using the leading-order stationary phase approximation.  The exponent is stationary at $t = \pm i \sqrt{4\pi^2 \bar {\phantom{|}\phi_b}/E}$, where the integrand on the far right of \eqref{eq:deltatr} takes the values $e^{\pm 4\pi \sqrt{\bar {\phantom{|}\phi_b} E} }$.  Since our contour lies above the singularity at $t=0$, it is then clear that the contour can be deformed to run through the saddle at
$t = i \sqrt{4\pi^2 \bar {\phantom{|}\phi_b}/E}$, which would in any case give the larger saddle-point contribution. A more detailed analysis also shows that the steepest ascent curve from this saddle lies along the positive $t$-axis and thus intersects the contour of integration as desired\footnote{Interestingly, in this case the steepest descent curve is just the circle  $tt^* =\frac{4\pi^2 \bar{\phantom{|}\phi_b}}{E}$, which can be thought of as running from the upper saddle at $t = i \sqrt{4\pi^2 \bar {\phantom{|}\phi_b}/E}$ to the lower saddle at $t = - i \sqrt{4\pi^2 \bar {\phantom{|}\phi_b}/E}$.   But the part of the axis inside the circle can nevertheless be deformed to pass through the above saddle. See e.g. \cite{Witten:2010cx} for a more complete discussion of the general theory of steepest descent and ascent curves.}.  As a result, the leading semiclassical approximation gives
\begin{equation}
\label{eq:deltatr2}
\Tr_{\scriptscriptstyle {\cal D}} \ \delta(H-E) \approx {e^{4\pi \phi_0}} e^{4\pi \sqrt{\bar {\phantom{|}\phi_b} E} } \theta(E)= e^{S(E)}\theta(E),
\end{equation}
where in the first step we have dropped a factor of $1/2\pi$ since it is subleading at leading semiclassical order.  In \eqref{eq:deltatr2} we have used the symbol $\theta(E)$ to  denote the usual Heaviside step function and $S(E)$ is defined as
\begin{equation}
\label{eq:SE}
S(E) : = S(\beta)\Big|_{\beta = 2\pi \sqrt{ \! \bar {\phantom{\frac{1}{1}}\!\!\phi_b}/E}}.
\end{equation}
As usual the definition \eqref{eq:SE} is made because, at leading semiclassical order, the expectation value of $E$ in the ensemble defined by $e^{-\beta H}$ is given by
\begin{equation}
\label{eq:Ebeta}
E = \beta^{-1}(S-\ln Z) \approx 4 \pi^2 \bar {\phantom{|}\phi_b}/\beta^2,
\end{equation}
and because solving \eqref{eq:Ebeta} for $\beta$ yields the relation $\beta = 2\pi \sqrt{\bar {\phantom{|}\phi_b}/E}$ used in \eqref{eq:SE}.

Given any function $f$ on the real line, we can now define an operator $f(H)$ via
\begin{equation}
\label{eq:f}
f(H) :=  \int dE \, f(E) \delta(H-E).
\end{equation}
Let us do so for two functions $f_1, f_2$, and let $f_1f_2$ denote the product of these functions.
As a consequence of \eqref{eq:deltadef} one finds
\begin{equation}
\label{eq:deltamult}
\delta(H-E_1) \delta(H-E_2)= \delta(H-E_1) \delta(E_1-E_2),
\end{equation}
which further implies that
we have
\begin{equation}
f_1(H)f_2(H) = (f_1f_2)(H),
\end{equation}
where $(fg)(H)$ is again defined as in \eqref{eq:f} but using the function $(fg)(E):= f(E)g(E)$ in the integral over $E$.

Linearity and \eqref{eq:deltatr} then require
\begin{eqnarray}
\Tr_{\scriptscriptstyle {\cal D}} \ f_1(H) &=& \int_{E>0} dE\, f_1(E) e^{S(E)},  \ \ \
\Tr_{\scriptscriptstyle {\cal D}} \ f_2(H) = \int_{E>0} dE\, f_2(E) e^{S(E)},  \ \ \ {\rm and}\cr
\Tr_{\scriptscriptstyle {\cal D}} \ (f_1f_2)(H) &=& \int_{E>0} dE\, \left(f_1(E)f_2(E)\right) e^{S(E)}.
\end{eqnarray}
Furthermore, for fixed $f_1,f_2$, in the limit of large $\bar \phi_b$ these integrals can be performed in the saddle point approximation.  Since each integral is real, it must be dominated by the largest saddle on the positive real axis.  Denoting the relevant saddle-point values of $E$ as $E_1, E_2, E_{12}$, we then have
\begin{eqnarray}
\Tr_{\scriptscriptstyle {\cal D}} \ [f_1(H)] &\approx& f_1(E_1) e^{S(E_1)},  \ \ \
\Tr_{\scriptscriptstyle {\cal D}} \ [f_2(H)] \approx f_2(E_2) e^{S(E_2)}, \cr
\Tr_{\scriptscriptstyle {\cal D}} \ [(f_1f_2)(H)] &\approx&  \left(f_1(E_{12})f_2(E_{12})\right) e^{S(E_{12})}.
\end{eqnarray}
But since $E_1$ dominates the first integral, we have $f_1(E_{1}) e^{S(E_{1})} \ge f_1(E_{12}) e^{S(E_{12})}$, and similarly $f_2(E_{2}) e^{S(E_{2})} \ge f_2(E_{12}) e^{S(E_{12})}$.  Thus we find
\begin{eqnarray}
\Tr_{\scriptscriptstyle {\cal D}} \ [(f_1f_2)(H)] &\le&  \left(f_1(E_{1})f_2(E_{2})\right) e^{S(E_{1}) +S(E_2) - S(E_{12})} \cr &\approx & e^{-S(E_{12})} \left( \Tr_{\scriptscriptstyle {\cal D}}\  [f_1(H)] \right)\left( \Tr_{\scriptscriptstyle {\cal D}}\  [f_2(H)] \right). \ \
\end{eqnarray}
Finally, we note that in our semiclassical limit the quantity $S(E_{12})$ will be large and positive for any fixed $E_{12} >0$.  Thus we have  $e^{-S(E_{12})} \ll 1$.  In particular, this factor will be much more important than any subleading terms in our approximations.

This then establishes the trace inequality \eqref{eq:trineq3} for arbitrary $f_1, f_2$ at leading order in the limit of large $\bar \phi_b$.  In fact, we have shown the inequality to hold strictly in this limit, in the sense that it cannot be saturated.    As a result,  quantum corrections cannot violate the trace inequality \eqref{eq:trineq3} when they are treated perturbatively.  The same is true for any perturbative higher-derivative corrections one may wish to add.

While the above discussion was phrased in terms of JT gravity, the only properties we actually used were that $B,C$ were chosen to be functions of $H$ and that $S(E) >0$ for all $E$.  As a result, the same arguments also apply verbatim to such $B,C$ when ${\cal D}$ is dual to a higher-dimensional gravity theory so long as each path integral is dominated by a black hole saddle (so that $S=A/4G >0$). The one subtlety is that, due to the Hawking-Page transition, if one wishes to see the fact that $S(E) >0$ at small $E$ one will need to appropriately analytically continue to low energies the large-energy saddles that dominate the high-temperature phase; see e.g. the discussion of microcanonical entropy from the gravitational path integral in \cite{Marolf:2018ldl}.

The above analysis considered choices of operators $B,C$ that each define connected parts of the  AlAdS boundary.  For example, the operators  $B=e^{-\beta_1 H}$ and $C=e^{-\beta_2 H}$ are each associated with a single line-segment on the boundary.  As described in the introduction, when e.g. $B$ instead contains several disconnected components, it may be important to include spacetime wormholes in the analysis.  Since such cases quickly become cumbersome, we will not attempt to treat them via explicit calculations of the form described above.  However, such cases are readily included in the analysis of section \ref{subsec:SSS} below.

\subsection{Adding matter using the Saad-Shenker-Stanford pardigm}
\label{subsec:SSS}

Jackiw-Teitelboim gravity turns out to be a simple enough  theory that we can also establish \eqref{eq:trineq3} for the case where it has a dilaton-free coupling to positive-action matter.  Here we require only that the theory admit a UV-completion in which the JT path integral can be treated in the manner described by Saad, Shenker, and Stanford in \cite{Saad:2019lba} and in which a semiclassical treatment remains valid.    By a `dilaton-free coupling,' we simply mean that the matter action depends {\it only} on the JT metric and does {\it not} depend on the dilaton.  Furthermore, the specific positive-action matter requirement  is that the classical matter action should be bounded below by zero on all asymptotically AdS$_2$ Euclidean spacetimes with arbitrary topology and arbitrary number of boundaries.

The present discussion will require certain details regarding the formulation and properties of JT gravity. In order not to distract from the main thrust of this work, we relegate the more technical analyses to appendix \ref{app:JT}.  However, we recall here that the action for pure JT gravity on an asymptotically AdS$_2$ spacetime takes the form
\begin{eqnarray}
\label{eq:JTaction}
I &=& - \phi_0 \left[ \int_{{\cal M}} \sqrt{g} R + 2 \int_{\partial {\cal M}} \sqrt{h} K\right] \cr
&-&  \int_{{\cal M}} \sqrt{g} \phi (R+2)   + 2 \int_{\partial {\cal M}} \sqrt{h} \phi (K-1).\ \ \ \ \ \
\end{eqnarray}
We refer the reader to appendix \ref{app:JT} for a discussion of the boundary conditions under which \eqref{eq:JTaction} can be used, though in this section we will refer to the associated conditions as the requirement that the AdS$_2$ boundary be ``smooth.''

As in section \ref{subsec:pureJT}, there are various possible notions of a semiclassical limit for this theory.  And, again as in section \ref{subsec:pureJT}, the effect of $\phi_0$ is to weight topologies by a factor of $e^{4\pi \phi_0 \chi}$ so that taking $\phi_0$ large with all else fixed immediately yields \eqref{eq:trineq3}.  We will thus follow section \ref{subsec:pureJT} in showing that \eqref{eq:trineq3} {\it also} holds when we take $\bar \phi_b$ large while holding all other parameters fixed (including both $\phi_0$ and the inverse temperature $\beta$).

If one examines the action \eqref{eq:JTaction}, one sees that it is strictly linear in $\phi$.  This remains true in the presence of our  dilaton-free matter couplings.  Following \cite{Saad:2019lba}, it is then natural to define the ``Euclidean" JT path integral by integrating $\phi$ over strictly {\it imaginary} values, so that the integrals over $\phi$ give delta-functions of $R+2$.  The path integral then reduces to an integral over (real) $R=-2$ constant curvature Euclidean spacetimes and over any matter fields.

The bulk term in \eqref{eq:JTaction} vanishes for such spacetimes, so the JT action becomes a sum of boundary terms -- one for each $S^1$ connected component of the boundary -- each of which that can be written in terms of a Schwarzian action \cite{Maldacena:2016upp}.  We will assume the remaining integrals to have a good semiclassical limit in the sense that, when $\bar \phi_b \rightarrow \infty$ with all else fixed, the result of the integrals is well approximated by $e^{-I_0}$ where $I_0$ is the minimum-action configuration of metrics and matter fields that satisfy the boundary conditions.

Due to the simplicity of JT gravity, with this assumption one can again quickly see that \eqref{eq:trineq3} holds as $\bar \phi_b \rightarrow \infty$.  The point is that, as shown in appendix \ref{subsec:JTpos}, in this context the action  is bounded below by $- 4\pi \phi_0 \chi -\sum_j 4\pi^2 \bar \phi_b/\beta_j$   where $\chi$ is the spacetime's Euler character and $\beta_j$ is the preiod of the $j$th circular boundary. Since $\chi \le n$ for any $2d$ manifold with $n$ circular boundaries, we thus find a topology-independent lower bound
$-\sum_j \left(4\pi \phi_0 +4\pi^2 \bar \phi_b/\beta_j \right)$.   It should be no surprise that this is just the action of the Euclidean black hole with inverse temperature $\beta$.

Furthermore, since the matter action is non-negative, the full coupled matter-plus-gravity action is also bounded below by $-\sum_j \left(4\pi \phi_0 +4\pi^2 \bar \phi_b/\beta_j \right)$.  It turns out that this is also a good estimate of the actual minimum of the action at large $\bar \phi_b$.  To see this, let us use $g_{min}$ to denote the Poincar\'e disk metric (representing Euclidean JT black holes with periods $\beta_j$) that saturates this bound.  We then choose any  matter field configuration that satisfies the required boundary conditions when taken together with $g_{min}$.   Since $\bar \phi_b$ is just an overall coefficient in front of the Schwarzian action \eqref{eq:Schwform1}, our $g_{min}$ cannot depend on $\bar \phi_b$.
Our full field configuration thus has action $I= -\sum_j \left(4\pi \phi_0 +4\pi^2 \bar \phi_b/\beta_j \right) + I_{matter,0}$ where the last term is manifestly independent of $\bar \phi_b$.  But the true minimum $I_0$ of the full action must be less than or equal to this result, showing that $I_0$ satisfies
\begin{equation}
-\sum_j \left(4\pi \phi_0 +4\pi^2 \bar \phi_b/\beta_j \right)   \le I_0 \le -\sum_j \left(4\pi \phi_0 +4\pi^2 \bar \phi_b/\beta_j \right) + I_{matter,0}.
\end{equation}
In particular, in the limit of large $\bar \phi_b$ we see that $I_0$ will scale linearly in $\bar \phi_b$ with coefficient $-\sum_j 4\pi^2 \bar \phi_b/\beta_j$.  The inequality \eqref{eq:trineq3} then follows immediately by noting that if $\beta_B, \beta_C, \beta_{BC}$ are the lengths of the relevant boundaries then we must have $\beta_{BC} = \beta_B + \beta_C$, and thus also
\begin{equation}
\frac{1}{\beta_{BC}} = \frac{1}{\beta_{B} + \beta_C} < \frac{1}{\beta_B} + \frac{1}{\beta_C}.
\end{equation}
Again, since this also forbids saturation of the trace inequality at this order, \eqref{eq:trineq3} must continue to hold in the presence of both higher-order semiclassical corrections and perturbative higher-derivative corrections.

\section{The trace inequality in general semiclassical gravity theories}
\label{sec:gen2}

The remainder of this work is devoted to arguing that the bulk analog \eqref{eq:trineq3} of the trace inequality \eqref{eq:trineq} should hold in general semiclassical theories of gravity and for general operators $B,C$.  After the discussion of sections \ref{sec:intro} and \ref{sec:simple}, this should not be a surprise.  When the path integrals are dominated by black holes, it is natural to expect the behavior seen in section \ref{sec:simple} where very general computations are semiclassically controlled by black hole thermodynamics whence (as described in section \ref{subsec:pureJT}) the trace inequality follows from positivity of the microcanonical entropy $S(E)$.  Furthermore, when the gravitational path integrals are {\it not} dominated by black holes, it is natural for the bulk to behave like a standard quantum system  so that the argument in \eqref{eq:Trinder} should apply.  Putting these together should be expected to yield an argument for general theories of gravity.

What makes this discussion subtle is our lack of understanding of the Euclidean gravitational path integral, as well as the associated conformal factor problem that makes the Euclidean action unbounded below (see e.g. \cite{Gibbons:1978ac}).  We therefore address these issues in stages below.  We first return to the non-gravitational setting  in section \ref{sec:nongrav} and find a path integral derivation of our trace inequality in the semiclassical limit.  We then show in section \ref{subsec:main} that the broad outline of this non-gravitational path-integral argument can be transcribed to the gravitational case, so long as one makes a number of assumptions concerning both the gravitational action and the treatment of the conformal factor problem.  We will take care, however, to formulate such assumptions in such a manner that they remain plausible despite the above-mentioned fact that the gravitational action is {\it not} bounded below.  This plausibility argument is then made in section \ref{sec:asstatus}, which in particular shows these assumptions to imply a new positive-action conjecture that extends the original positive-action conjecture of Hawking \cite{Gibbons:1976ue,Gibbons:1978ac} in several ways.  The conjecture can then be verified for JT gravity with minimal (or, more generally, dilaton-free) couplings to positive-action matter. Furthermore, in simple contexts, for more general theories it can be related to positivity of the Hamiltonian with general boundary conditions.

\subsection{Trace Inequality from the Semiclassical Euclidean Path Integral:  The non-gravitational case}
\label{sec:nongrav}

We now return to the non-gravitational context to describe a Euclidean path integral derivation of \eqref{eq:trineq} in the semiclassical limit.  We restrict ourselves to the case where both $B$ and $C$ are defined by real sources. We will also assume that, with any fixed set of allowed real-valued sources, the corresponding Euclidean action is both real and bounded below.

We will also assume that each such path integral is dominated by a saddle (or by a set of saddles) in the semiclassical limit, and in particular that the action of any configuration is always greater than or equal to the action of the dominant saddle. The latter will be true under assumptions that prevent the action from being minimized on the boundaries of the space of allowed field configurations.  Such assumptions are reasonable since regions near such boundaries typically have infinite measure, so that minimizing the action on such boundaries typically causes the path integral to diverge.  We leave the exceptional cases open for future study.

Our restriction to real sources means that our path integrals are manifestly real.  Such integrals can only be dominated in the semiclassical limit by real saddles corresponding to global minima of the action over the contour of integration.  In particular, all saddles (as well as more general configurations) discussed below will lie on the contour of integration that defines the path integral.  This means that no issues of contour deformations can be relevant to our discussion.

We will also consider only cases where the right side of \eqref{eq:trineq} is dominated by a {\it single} saddle.
Contexts with more than one equally-dominant saddle typically describe phase transitions; see e.g. the classic discussion of Hawking and Page \cite{HawkingPage}.  Close to such a phase transition one typically finds that formally non-perturbative effects associated with additional saddles and/or mixing between saddles are more important than perturbative corrections; see e.g. recent discussions in \cite{Vidmar:2017pak,Murthy} for condensed matter analogues and in \cite{Penington:2019kki,Marolf:2020vsi,Dong:2020iod,Akers:2016ugt}.  We thus save further consideration of this case for future study.

It will be enough for our purposes to work at leading order in the semi-classical expansion, so that the path integral is approximated by $e^{-I}$, where $I$ is the Euclidean action of the dominant saddle. This is to be a model for the leading-order analysis of (gravitating) bulk duals in section \ref{subsec:main}, though in that section we will use a rather different method to include quantum corrections.

We have already mentioned that we are interested in quantum field theories with sources, say in $d$ Euclidean spacetime dimensions.  In quantum field theories, the UV structure of the Hilbert space can be sensitive to the choice of sources, and in fact to various time-derivatives of such sources when $d$ is large.
 We will therefore further restrict discussion to the case where the Hilbert space of interest can be thought of as being defined by a set of time-translation-invariant boundary conditions that define an associated ``cylindrical'' Euclidean manifold ${\cal C}_\infty = {\cal B} \times {\mathbb R}$ with translation-invariant sources, where ${\cal B}$ is an appropriate $(d-1)$-dimensional manifold and $\times$ denotes the Cartesian product of metrics as well as of the underlying manifolds.

 The equivalent definition in Lorentz signature would thus restrict us to considering Hilbert spaces defined by static metrics\footnote{\label{foot:Trev} While it is also of interest to discuss time-dependent Lorentz-signature QFTs, upon analytic continuation to Euclidean signature such time-dependence is generally associated with complex-valued Euclidean sources.  Complex sources raise further issues for the saddle-point approximation associated with the possible existence of saddles at points in the complex plane away from the original contour of integration.  Such issues are beyond the scope of the current work, so we save that setting for future investigation.  The same comment applies to stationary but non-static background metrics, to vector-valued sources with non-trivial time-components, and to other boundary conditions not naturally described as being of the form $B\times {\cal R}$ due to breaking time-reversal symmetry. }.
 In particular, note that the ${\mathbb Z}_2$ relection symmetry of the ${\mathbb R}$ factor implies that ${\cal C}_\infty$ also admits a ${\mathbb Z}_2$ ``time-reversal'' symmetry. We refer to ${\cal C}_\infty$ as the infinite cylinder. It will be useful to define corresponding finite cylinders ${\cal C}_\epsilon = {\cal B} \times [0,\epsilon]$,  and to define ${\cal B}_0$ to be the boundary of ${\cal C}_\epsilon$ at the zero of the interval.  Due to the time reversal symmetry of ${\cal C}_\infty$, we will need this definition only for positive values of $\epsilon$.

 We emphasize that this is a restriction on the background fields that define the Hilbert space and not on the background fields used to construct any particular state.  Of course, the two must be compatible, so in practice we will consider only states that are prepared by manifolds-with-boundary where a neighborhood of each boundary contains a {\it rim} diffeomorphic to a finite cylinder ${\cal C}_\epsilon$ (or, more properly, to the part of this cylinder associated with the half-open interval $[0, \epsilon)$).

In most of this section we will also assume that the Euclidean action $I$ for our theory is the integral of a local Lagrangian $L$, with $L$ built from fields and their {\it first} derivatives only. In particular,  assume that $I = \int_M L$ without additional boundary terms at $\partial M$. Of course, many potential such boundary terms can be absorbed into $L$ by the addition of a total divergence (so long as it is again built from fields and their first derivatives). The above condition implies that the equations of motion are of no more than 2nd order, but at the end of this section we will see how to include perturbative higher derivative corrections.

We begin by considering positive operators $B$ and $C$.  Such operators may always be written $B = b^\dagger b$, $C =c^\dagger c$ for appropriate $b,c$.  We assume that $b,c$ are computed by some (perhaps complex) linear combination of Euclidean path integrals with real sources.  For simplicity, we begin with the case where each operator $b,c$ is computed by a single Euclidean path integral, saving non-trivial linear combinations for later.  However, in contrast to section \ref{subsec:pureJT}, we now include the case where the boundary associated with $\Tr(BC)$ may be disconnected.

\begin{figure}[h!]
	\centering
\includegraphics[width=.7\linewidth]{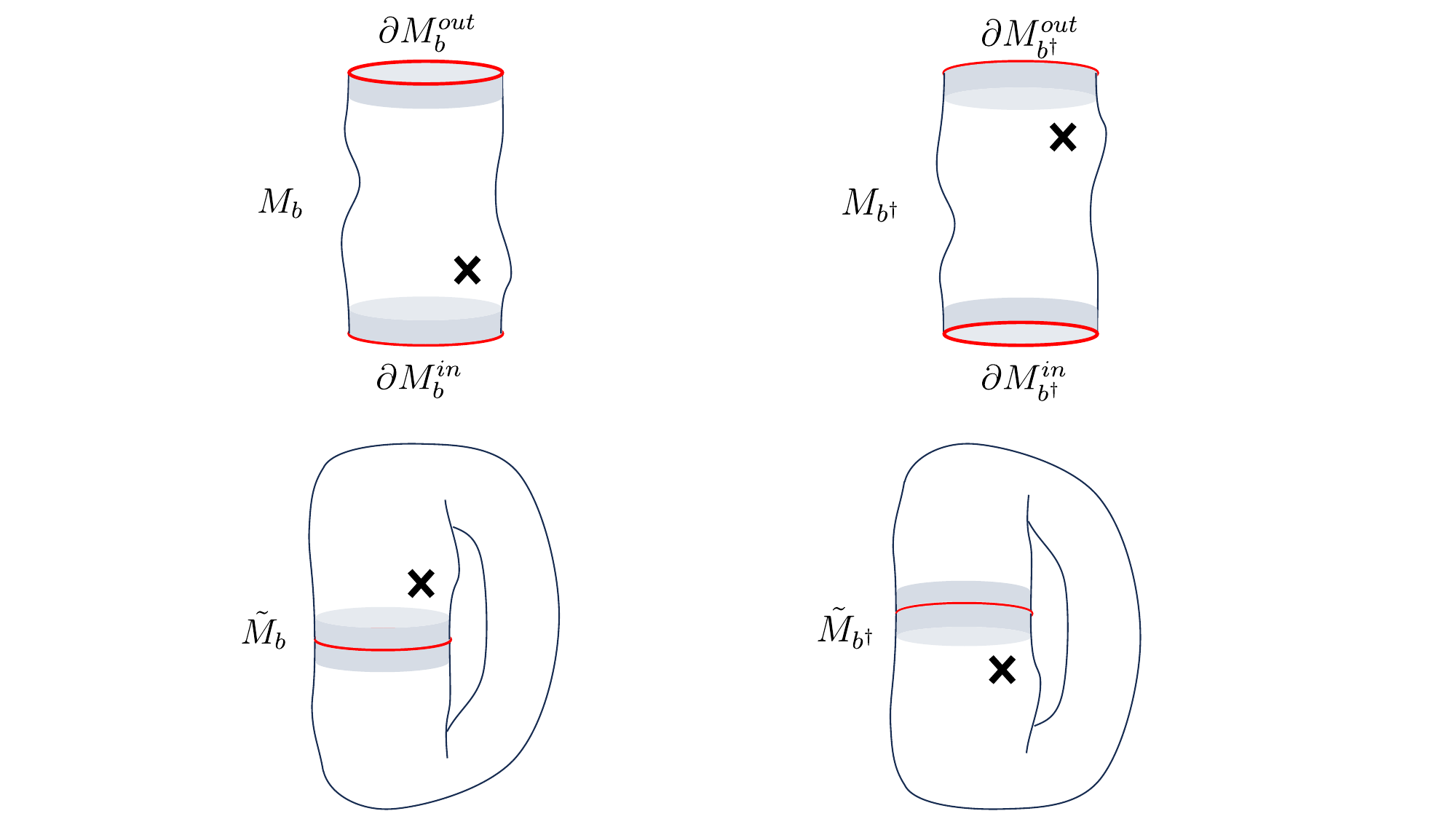}
\caption{{\bf Upper left panel:} A source-manifold-with-boundary $M_b$ is shown with its input $\partial M_b^{in}$ and output $\partial M_b^{out}$. Each boundary (red) has a $C_\epsilon$ rim (grey).   The $\times$ symbol denotes a localized feature of a source. {\bf Upper right panel:} $M_{b^\dagger}$ is constructed from $M_{b}$ by interchanging the labels $\partial M_b^{in}, \partial M_b^{out}$ but keeping all sources unchanged. If one always draws $\partial M_b^{in}$ at the bottom and  $\partial M_b^{out}$ at the top, then $M_{b^\dagger}$ is obtained from $M_b$ by acting with a reflection in the vertical direction.    {\bf Lower panels:} The closed source manifolds $\tilde M_b$ and $\tilde M_{b^\dagger}$ (which are diffeomorphic for real sources). }\label{fig:Mb}
\end{figure}

We use the notation $M_b,M_c,M_B,M_C$ to denote the manifolds over which the path integrals for $b,c,B,C$ are performed, together with the appropriate set of sources.  To remind the reader of this, we will sometimes refer to
$M_b,M_c,M_B,M_C$ as {\it source manifolds}.  In particular, we take such source manifolds to specify the full set of background structures (e.g., spin-structures, etc.) which are required to define the theory\footnote{The restrictions imposed above, and in particular the implicit assumption that the theory be invariant under time-reversal, imply that our theory cannot depend on a choice of orientation.  See related comments in footnote \ref{foot:Trev}. }. A pictorial representation of such a source manifold is provided on the upper left panel of figure \ref{fig:Mb}.

%Since the space of operators is linear, it is in fact useful to further extend this notation to allow $M_B,M_C$ to represent formal linear combinations of such manifolds with complex coefficients.   In a slight abuse of terminology, we may think of the associated coefficients as part of the specification of sources on $M_B$ and $M_C$.   Such coefficients should thus be understood to be included whenever `sources on manifolds' are discussed below.  Because these coefficients can be complex, from this point forward we will refer to the sources as being complex.  However, it is important to understand that the coefficients are in fact the only complex sources we allow, and that all other sources are still required to be real.

Since $b$ is an operator on a given Hilbert space, we may take the boundary $\partial M_b$ to be the disjoint union of two parts
$\partial M_b^{in}$, $\partial M_b^{out}$ describing the input and output of $b$, and where the sources near both  $\partial M_b^{in}$ and $\partial M_b^{out}$ are those associated with the given Hilbert space.   In particular, we assume that $M_b$ may be chosen to be some $C_\epsilon$ in some neighborhood of each of
$\partial M_b^{in}$ and $\partial M_b^{out}$ so that the boundary $\partial M_b^{in}$  (or $\partial M_b^{out}$) agrees with ${\cal B}_0$. As mentioned above, we refer to this as requiring $M_b$ to have {\it rims}, and we make analogous requirements for the source-manifolds with boundary associated with any operator discussed below.

Since the theory is non-gravitational, one should regard points on $M_b$ and $C_\epsilon$ as being labelled.  The agreement of
$\partial M_b^{in}$, $\partial M_b^{out}$ with ${\cal B}_0$ thus defines a particular diffeomorphism $\phi_b: \partial M_b^{in} \rightarrow \partial M_b^{out}$.  We then define a closed source manifold $\tilde M_b$ (without boundary, so that $\partial \tilde M_b = \emptyset$) by using $\phi_b$ to identify $\partial M_b^{in}$ with $\partial M_b^{out}$.  The trace of $b$ ($\Tr(b)$) is then computed by the path integral over the resulting $\tilde M_b$; see figure \ref{fig:Mb}.

It is useful to take the definition of $M_b$ to include the partition of $\partial M_b$ into $\partial M_b^{in}$ and $\partial M_b^{out}$.   We may then describe $b^\dagger$ as being computed by the path integral over $M_{b^\dagger}$, where (since we restrict to real sources) $M_{b^\dagger}$ is constructed from  $M_b$ by interchanging the labels $\partial M_b^{in}$, $\partial M_b^{out}$ but keeping all sources unchanged; see again figure \ref{fig:Mb}.

Corresponding assumptions and definitions will also be made for any other operator $c$ and the associated $M_{c}$, $\partial M_{c}$, and $\tilde M_{c}$.  In particular, since $b$ and $c$ both act on the same Hilbert space ${\cal H}$, the inputs of $b$ must be identical to those of $c$, and similarly for the outputs.  As a result, the labelling of points on ${\cal B}_0$ also defines source-preserving diffeomorphisms $\phi_{bc}:  \partial M_b^{in} \rightarrow \partial M_c^{out}$ and $\phi_{cb}:  \partial M_c^{in} \rightarrow \partial M_b^{out}$.  We may then use $\phi_{bc}$ (or $\phi_{cb}$) to define the source manifold $M_{bc}$ (or $M_{cb}$) by identifying the input of $M_b$ with the output $M_c$ (or vice versa).  The path integral over $M_{bc}$ then clearly computes the operator $bc$.  Using both $\phi_{bc}$ and $\phi_{cb}$ to make identifications allows us to further construct the closed source manifold $\tilde M_{bc}$, over which the path integral computes $\Tr(bc)$.  Note that swapping $b$ and $c$ would define the source manifold $M_{cb}$ associated with the operator $cb$, but that $\tilde M_{cb} = \tilde M_{bc}$ so that $\Tr(bc) = \Tr(cb)$ as expected; see figure \ref{fig:multandtrace} below.
\begin{figure}[h!]
	\centering
\includegraphics[width=0.9\linewidth]{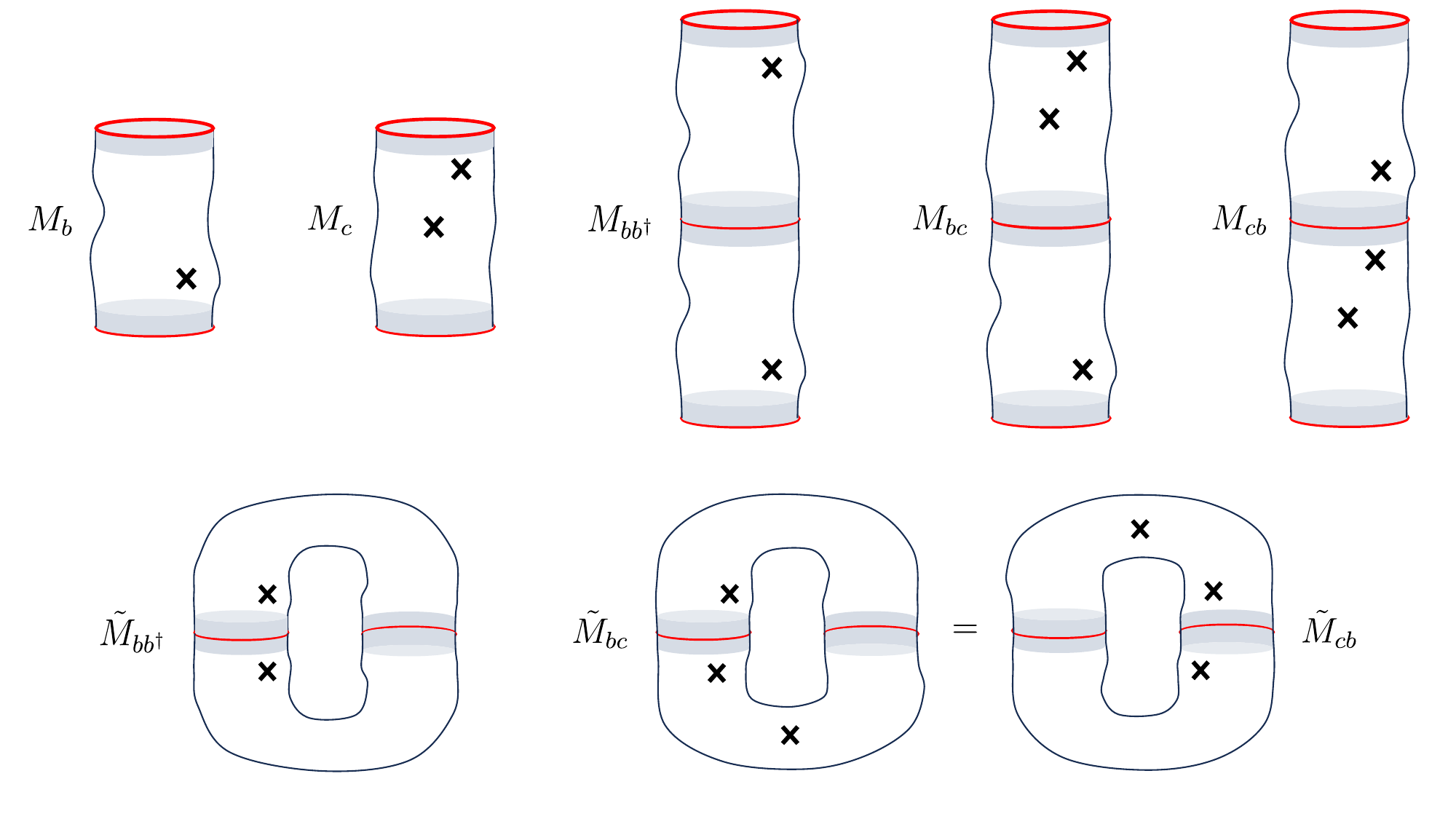}
\caption{The source manifolds with boundary $M_b$, $M_c$ can be used to construct $M_{bc}$ , $M_{cb}$, and $M_{bb^\dagger}$ as well as the closed source manifolds $\tilde M_{bc} = \tilde M_{cb}$  as shown.  In all cases the output boundaries are drawn at the top and the input boundaries are at the bottom.  Note that $M_{bb^\dagger}$ and $\tilde M_{bb^\dagger}$ are both symmetric under reflections of the vertical direction.}
\label{fig:multandtrace}
\end{figure}

In order to derive \eqref{eq:trineq3},  we will thus need to compare the Euclidean path integrals over $\tilde M_B$, $\tilde M_C$, and $\tilde M_{BC}$. Recall that we require $B=b^\dagger b$, $C=c^\dagger c$ where $b,c$ can again be written as Euclidean path integrals, say over $M_b, M_c$.  In direct parallel with the above construction of $M_{bc}$ from $M_b, M_c$, we may also choose $M_B$ to take the form $M_{b^\dagger b}$.  The trace $\Tr(B)$ is then computed via the path integral over the corresponding closed source manifold $\tilde M_{b^\dagger b}$.   Since the sources on $M_b$ are real, they must agree with those on $M_{b^\dagger}$ up to an appropriate diffeomorphism. Thus $M_B$ admits a ${\mathbb Z}_2$ symmetry that exchanges the $b$ and $b^\dagger$ regions of $M_B = M_{b^\dagger b}$; see again figure \ref{fig:multandtrace}.

\begin{figure}[t]
	\centering
\includegraphics[width=\linewidth]{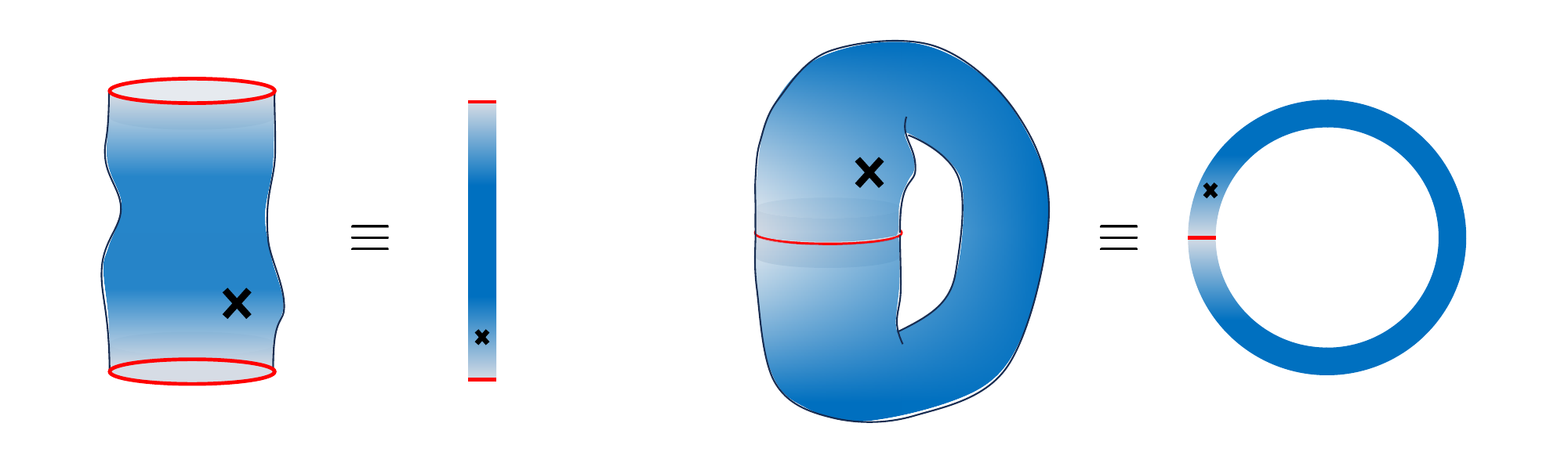}
\caption{This figure illustrates the scheme that we adopt below to depict source manifolds along with particular configurations on such manifolds that arise in the associated path integrals. It also shows the connection to the old scheme.   {\bf Left panel:} A configuration for the path integral performed over a source-manifold $M_b$ will be described by the coloration assigned to that manifold.  The left hand side of the equivalence uses the old scheme with a two-dimensional depiction of the source manifold, but now with the coloration added.  The right hand side uses the new scheme in which the source manifold is drawn as one-dimensional and the $\times$ is the only indication of structure associated with the sources.   {\bf Right panel:} A configuration for the path integral performed over the closed source-manifold $\tilde M_b$ is shown using both schemes.}
\label{fig:changeRepresentation}
\end{figure}

Before proceeding, we pause to comment on our depiction of the source manifolds $M_b$, etc. in the accompanying figures.  Below, we will wish to show features of individual configurations of fields that appear in the path integral (in addition so the source features shown thus far).  Such information makes the figures correspondingly more complicated, so that it is useful to simplify our illustrations in other ways, even at the expense of making them more abstract.  See figure \ref{fig:changeRepresentation} below for the dictionary relating figures thus far to those that will appear in the remainder of this work.

At leading semi-classical order, comparing path integrals over $\tilde M_B$, $\tilde M_C$, and $\tilde M_{BC}$ is equivalent to comparing the dominant saddles $\sigma_B$, $\sigma_C$, and $\sigma_{BC}$ on these source manifolds.   We begin with an observation, which we codify as a lemma to facilitate future reference:

\begin{lemma}
\label{lemma:sym}
Consider an operator $D = d^\dagger d$, where $d$ is computed by a Euclidean path integral over a source manifold $M_d$.  The source manifold $\tilde M_{D} = \tilde M_{d^\dagger d}$ then clearly enjoys a ${\mathbb Z}_2$ reflection symmetry as discussed above. This symmetry is in fact preserved by any saddle $\sigma_D$ that dominates the path integral over $\tilde M_D$.  In cases where the minimum value of the action is shared by several saddles,  the symmetry is preserved by at least one such $\sigma_D$.
\end{lemma}

To prove Lemma \ref{lemma:sym}, we begin by considering an arbitrary saddle $\sigma^0_D$ for $\Tr D$.  Let $k^0_d$ be the part of this saddle on $M_d$, and let $k^0_{d^\dagger}$ be the part on $M_{d^\dagger}$.  Furthermore, let $\Phi_d: M_d \rightarrow M_{d^\dagger}$ be the map that defines the ${\mathbb Z}_2$ symmetry of $\tilde M_D$.  Here we use the symbol $k$ (with subscripts) for configurations that are not  given to us as saddles of the original path integrals.

\begin{figure}[h!]
	\centering
\includegraphics[width=\linewidth]{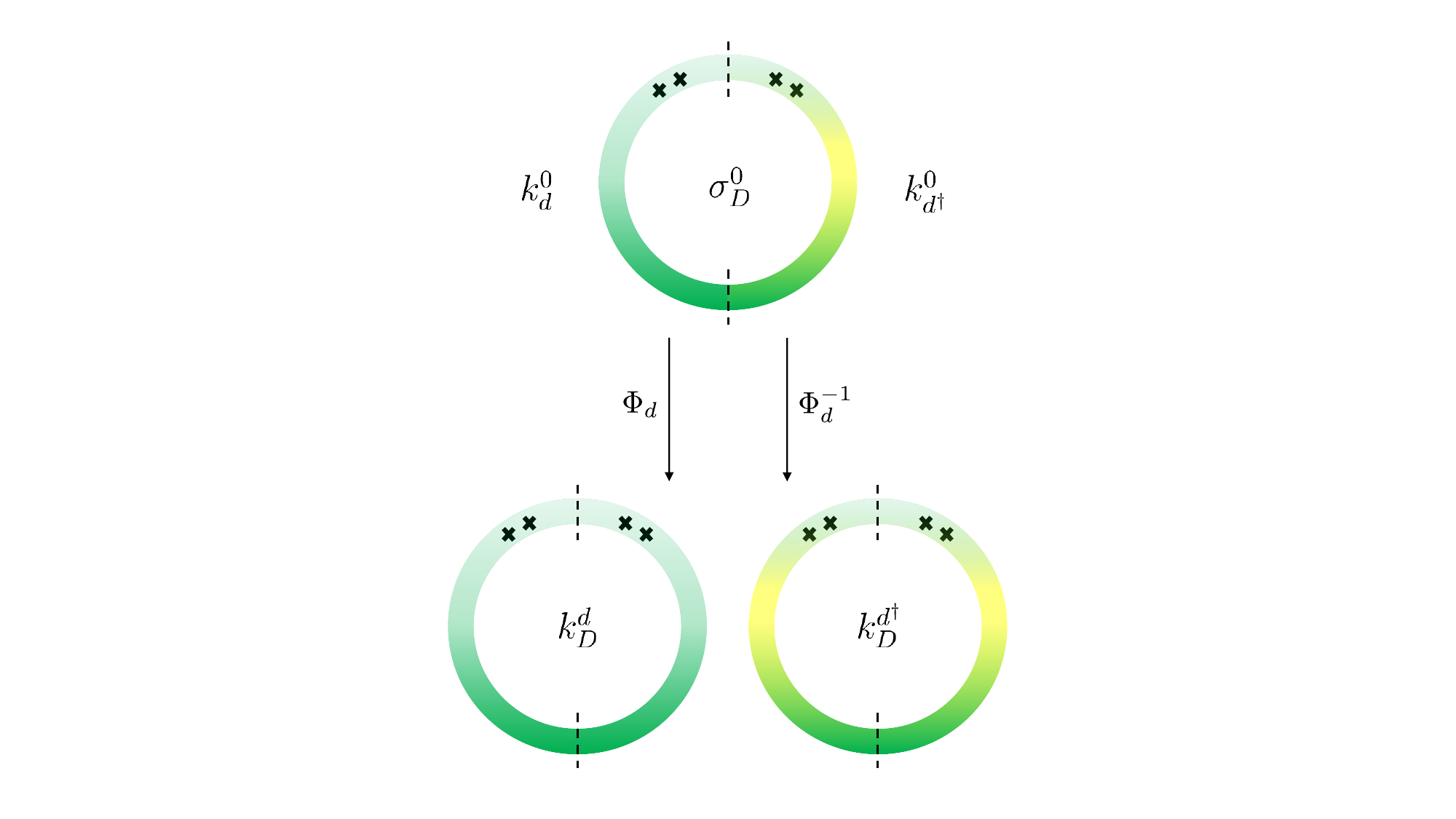}
\caption{For $D=dd^\dagger$, and when considering the path integral that computes $\Tr D$, a saddle $\sigma_D^0$ with no reflection symmetry can be used to construct two reflection symmetric saddles.  The top figure shows such a situation where the background fields (whose features are indicated by the $\times$ symbols) are left-right symmetric but the saddle (described by the colorations) is not.  The background fields for the top figure define $\Tr D$, and the background fields on the left and right halves define respectively $d$, $d^\dagger$. To construct the new saddles, one simply cuts $\sigma_D^0$ into pieces $k^0_d, k^0_{d^\dagger}$, operates on each with the reflection maps $\Phi_d$ or $\Phi_d^{-1}$, and forms new configurations by sewing together each such reflection with the corresponding $k^0_d, k^0_{d^\dagger}$.  Here the vertical arrows indicate the actions of $\Phi_d$ and $\Phi_d^{-1}$, each of which acts as a left/right reflection on its argument (which is then glued to another copy of the corresponding $k^0_d, k^0_{d^\dagger}$).  The resulting reflection-symmetric configurations are called $k_D^d, k_D^{d^\dagger}$.  Equations \eqref{eq:add1}-\eqref{eq:add3} then require one of $k_D^d, k_D^{d^\dagger}$ to have action no greater than that of the original saddle $\sigma^0_D$. }
\label{fig:reflect}
\end{figure}
If the saddle $\sigma^0_D$ breaks the ${\mathbb Z}_2$ symmetry of the background fields, then $k^0_d$ and $k^0_{d^\dagger}$ will not be related by $\Phi_d$.  In this case we can use $k^0_d$ and $k^0_{d^\dagger}$ to build new configurations for the path integral over $\tilde M_D$.  In particular, as illustrated in figure \ref{fig:reflect} (above), the action of $\Phi_d$ on  $k^0_d$ defines a new configuration $k^R_d  \eqqcolon \Phi_d(k^0_d)$  on $M_{d^\dagger}$, and gluing this to  $k^0_d$ defines a new configuration
$k_D^d$ for the path integral over $\tilde M_D$.  We can also define a corresponding $k_D^{d^\dagger}$ by gluing $k^0_{d^\dagger}$ to its image under the inverse of $\Phi_d$.  Note that  $k^d_D$,  $k_D^{d^\dagger}$ will not generally solve the equations of motion at the surface where $M_d$ meets $M_{d^\dagger}$, and in fact that derivatives of fields in $k^d_D$,  $k_D^{d^\dagger}$ will not generally be continuous at these surfaces.

By construction, both $k^d_D$  and $k_D^{d^\dagger}$ are invariant under $\Phi_d$.  The key observation needed to prove our Lemma is then that the action $S$ is additive, in the sense that
\begin{eqnarray}
\label{eq:add1}
I(\sigma^0_D) &=& I(k^0_d) +I(k^0_{d^\dagger}), \ {\rm while}\\
I(k_D^d) &=& I(k^0_d) +  I(\Phi_d[k^0_d]) = 2 I(k^0_d) \ {\rm and} \\
I(k_D^{d^\dagger}) &=& I(k^0_{d^\dagger}) +  I(\Phi_d^{-1}[k^0_{d^\dagger}]) = 2 I(k^0_{d^\dagger}).
\label{eq:add3}
\end{eqnarray}
This additivity follows from the fact that $S = \int L$, together with the requirement that $L$ depend only on fields and their first derivatives.  The point is that $\sigma^0_D$ must be smooth since it solves the Euclidean equations of motion  with smooth boundary conditions. (We assume these equations to be elliptic.) Furthermore, by construction, the values of fields at the boundaries of $k^0_d$ will agree with those at the boundaries of $\Phi_d[k^0_d]$, and similarly for $k^0_{d^\dagger}$ and $\Phi^{-1}_d[k^0_{d^\dagger}]$.  This means that the fields defined by either  $k_D^d$ or $k_D^{d^\dagger}$ are continuous.  And while the first derivatives may not be continuous at the boundaries of $M_d$ and $M_{d^\dagger}$, they have well-defined limits from each side; i.e., the first-derivatives have at worst step-function discontinuities.  This means that $L$ is bounded, and in particular has no delta-function contributions at the boundaries between the $M_d$ and $M_{d^\dagger}$ regions of $M_D$.

It follows that the action does indeed satisfy \eqref{eq:add1}-\eqref{eq:add3}.  Comparing these equations shows that the smaller of
$I(k_D^d)$ and $I(k_D^{d^\dagger})$ must be less than or equal to $I(\sigma^0_D)$, and that it is strictly less if $I(k^0_d) \neq I(k^0_{d^\dagger})$.  Furthermore, if the final $k_D^d$, $k_D^{d^\dagger}$ are not saddles then they cannot minimize the action and the action of the dominant saddle must be even smaller; i.e.,
\begin{equation}
\label{eq:Isigmaswref}
I(\sigma^{\mathrm{dom}}_D) \le \min \{ I(k_D^d), I(k_D^{d^\dagger})\} \le I(\sigma^0_D).
\end{equation}

As a result, if $\sigma^0_D$ is a dominant saddle, then $k_D^d$, $k_D^{d^\dagger}$ are also saddles with $I(k_D^d) = I(k_D^{d^\dagger}) =  I(\sigma^{0}_D) $.  Noting that $k_D^d$, $k_D^{d^\dagger}$ are invariant under the ${\mathbb Z}_2$ symmetry then establishes Lemma \ref{lemma:sym}.

We are now in a position to prove our main result \eqref{eq:trineq3} at leading semi-classical order.  As stated above,
at this order, comparing path integrals over $\tilde M_B$, $\tilde M_C$, and $\tilde M_{BC}$ is equivalent to comparing the dominant saddles $\sigma_B$, $\sigma_C$, and $\sigma_{BC}$ on these source manifolds.  In particular, at this order we have
\begin{equation}
\label{eq:leadingQM}
\Tr B = e^{-I(\sigma_B)}, \ \  \Tr C = e^{-I(\sigma_C)}, \ \ {\rm and}  \ \ \Tr BC = e^{-I(\sigma_{BC})}.
\end{equation}

We are interested in the case $\tilde M_B = \tilde M_{b^\dagger b}$, $\tilde M_C=\tilde M_{c^\dagger c}$, so that we may also write $\tilde M_{BC} = \tilde M_{d^\dagger d}$ using $d=bc^\dagger$ and the fact that $\tilde M_{b^\dagger b c^\dagger c} = \tilde M_{c b^\dagger b c^\dagger}$ (which in turn follows from the fact that our gluing operation is invariant under cyclic permutations).   Applying Lemma \ref{lemma:sym} to $\tilde M_{BC} = \tilde M_{d^\dagger d}$, we may take $\sigma_{BC}$ to have a $\mathbb Z_2$ reflection symmetry that exchanges $d=cb^\dagger$ and $d^\dagger=bc^\dagger$.  We may then cut the saddle $\sigma_{BC}$  into the 4 pieces $k_b, k_{b^\dagger}, k_c, k_{c^\dagger}$ associated with the $M_b, M_{b^\dagger}, M_c, M_{c^\dagger}$ regions of $\tilde M_{BC}$.

\begin{figure}[t]
	\centering
\includegraphics[width=\linewidth]{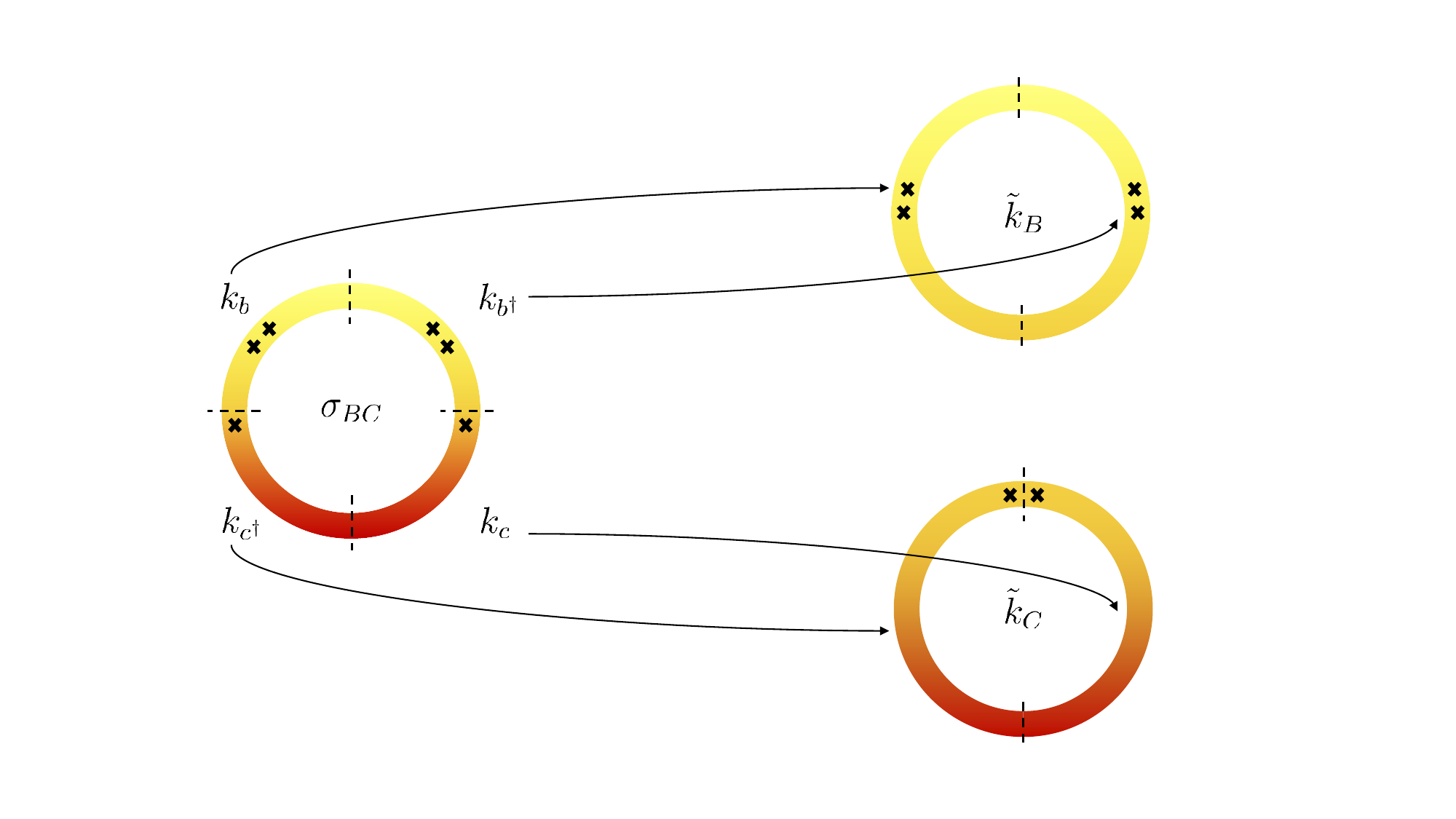}
\caption{For $B=bb^\dagger, C=cc^\dagger$, a ${\mathbb Z}_2$-symmetric saddle $\sigma_{BC}$ for $\Tr BC$ (shown at left) can be cut into four pieces $k_c, k_{c^\dagger}, k_{b}, k_{b^\dagger}$.  The pieces can then be recombined to make a pair of $\mathbb Z_2$-symmetric configurations $\tilde k_B$ and $\tilde k_C$ (shown at right) that contribute to the path integrals for, respectively, $\Tr B$ and $\Tr C$.}
\label{fig:BandCtoBC}
\end{figure}

We may now glue the resulting $k_b$ and $k_{b^\dagger}$ together to define a ${\mathbb Z}_2$-symmetric configuration $\tilde k_B =k_{bb^\dagger}$ for the path integral over $\tilde M_B$; see figure \ref{fig:BandCtoBC}.
 Note that ${\mathbb Z}_2$ symmetry requires the fields on $k_B$ to be continuous at the boundaries between $k_b$ and $k_{b^\dagger}$.  Thus the action $I(k_B)$ is well-defined.  We may also define the analogous configuration
$k_C = k_{cc^\dagger}$ for the path integral over $\tilde M_C$, whose action $I(k_C)$ is again well-defined.

Furthermore, as in the proof of Lemma \ref{lemma:sym} we have $I(k_B) + I(k_C) = I(\sigma_{BC})$.  And since the dominant saddles $\sigma_B, \sigma_C$ must have actions no larger than $I(k_B)$, $I(k_C)$, using the leading semiclassical approximation \eqref{eq:leadingQM}  we find
\begin{equation}
\label{eq:trin1}
\Tr(BC) \le \Tr(B) \Tr(C)
\end{equation}
as desired.

However, so far we have
required each operator $b,c$ to be given by a single path integral.  We would also like to discuss operators given by a linear combination of path integrals.  I.e., we wish to allow $b = \sum_i b_i$, and $c = \sum_i c_i$,
where each of the $b_i, c_j$ are single path-integrals as above.
This generalization is straightforward when there is a single dominant saddle $\sigma_{BC}$ for $\Tr (BC)$, which we as usual assume to be the case.

To proceed, we first write $B = \sum_{ij} b_i b^\dagger_j$,
 $C = \sum_{ij} c_i c^\dagger_j$, and $BC = \sum_{i,j,k,l}  b_i b^\dagger_j c_k c^\dagger_l$.
We then note that a dominant saddle $\sigma_{BC}$ for $\Tr BC$ will be associated with some particular term $b_i b^\dagger_j c_k c^\dagger_l$ (and also with its equally-dominant adjoint if this term is not real).
As in the proof of Lemma \ref{lemma:sym} we may then cut this saddle into two pieces corresponding to $M_{b^\dagger_j c_k}$ and $M_{ c^\dagger_l b_i}$.  Gluing each of these to its reflection\footnote{\label{foot:adjoints} The result of such a reflection gives a corresponding piece of the analogous saddle for the adjoint term.  Thus we can also think of our construction as pasting of from the $ijkl$ term together with pieces of the $jilk$ term.} then defines ${\mathbb Z}_2$-symmetric
configurations for the diagonal terms given by path integrals over $\tilde{M}_{b_j b^\dagger_j c_k c^\dagger_k}$
and for  $\tilde{M}_{b_i b^\dagger_i c_l c^\dagger_l}$.  Since the original saddle $\sigma_{BC}$ was dominant (with some action $I(\sigma_{BC})$) and our saddles are real,
 additivity again requires that the pieces corresponding to $M_{b^\dagger_j c_k}$ and $M_{ c^\dagger_l b_i}$ both have actions $\frac{1}{2}I(\sigma_{BC})$, and
 that the new ${\mathbb Z}_2$-symmetric configurations are saddles with actions equal to $I(\sigma_{BC})$.  (This follows from the analogue of \eqref{eq:Isigmaswref} when $\sigma^0_D$ is dominant so that the left and right hand sides are equal.) As a result, the new saddles may be used as dominant saddles.
  Using either saddle in this way then reduces us to consideration of a single $\mathbb Z_2$-symmetric saddle, whence the rest of the argument follows as above. 
It now remains to study higher-derivative corrections about a saddle $\sigma_{BC}$.  We will use $I_0$ to denote the original action without such corrections.  We will treat corrections to $I_0$ perturbatively, which means that
at each order the saddles are found by solving a 2nd derivative equation of motion with sources determined by the lower-order parts of the solution.

At the off-shell level needed for our argument, at each order $n$ in perturbation theory we may treat the action as being a 2nd order polynomial functional $I_n$ of the fields.  The coefficients of the quadratic terms in this functional are given by the second variations of $I_0$ about the lower order saddle. The action is thus positive semi-definite for perturbations about a dominant saddle.
The coefficients in the linear term are given by varying the higher derivative corrections at linear order.  We shall assume that any zero-modes of the linearized theory about $\sigma_{BC}$ are associated with symmetries of $I_0$ that are preserved by the higher-derivative terms, and thus in particular that they are preserved by the linear term in $I_n$.  It then follows that $I_n$ is bounded from below, and that it is in fact minimized at the desired saddle.

Furthermore, in a perturbative treatment
there can be no danger of violating \eqref{eq:trin1} unless it is saturated by the classical 2-derivative theory.  As a result, if we again suppose that the dominant 2-derivative saddle for each path integral is unique\footnote{In particular, it is unclear how to control the possibility that two a priori unrelated saddles might have precisely the same action at the two-derivative level, but might then have  this degeneracy lifted by higher derivative corrections in a manner unfavorable to our argument.  We leave consideration of this interesting-but-finely-tuned possibility open for future study.}, then we need only consider perturbations around saddles $\tilde \sigma_B, \tilde \sigma_C, \sigma_{BC}$, where $\tilde \sigma_B, \tilde \sigma_C$ are constructed from $\sigma_{BC}$ by using the above cut-and-paste procedure.  In particular, at any point $p_B$ on the $B$ side of $\tilde M_{BC}$, the sources for the first correction will precisely match those at a corresponding point $p_B$ on $\tilde M_B$, and similarly on any point $p_C$ on the $C$ side.  It follows that the setting for computing the first-order corrections is of precisely the same form as the zero-order problem defined above, where the sources for this problem on $\tilde M_{BC}$ may again be reproduced by gluing $\tilde M_B$ to $\tilde M_C$.  We may thus argue in exactly the same way that \eqref{eq:trin1} also holds at first order in higher derivative corrections, and in fact iteratively at every higher order as well.

\subsection{The trace inequality for semiclassical gravity}
\label{sec:bulk}
\label{subsec:main}

We now turn our attention to bulk gravity theories.  For convenience of notation we continue to suppose that the bulk theory is dual to a hypothetical non-gravitational theory ${\cal D}$, or to an ensemble of such theories, though in the end our arguments will be entirely in the bulk.  In particular, the arguments apply even to bulk theories for which dual non-gravitational boundary theories are not known to exist.

On the ${\cal D}$ side of the duality, the path integrals for $\langle \Tr_{\scriptscriptstyle {\cal D}}(B), \Tr_{\scriptscriptstyle {\cal D}}(C)\rangle$, and $\langle \Tr_{\scriptscriptstyle {\cal D}}(BC)\rangle$ may be formulated  as path integrals over source manifolds $\tilde M_B, \tilde M_C$, and $\tilde M_{BC}$ just as in section \ref{sec:nongrav}, and in particular with $\tilde M_B = \tilde M_{b^\dagger b}$, $\tilde M_C=\tilde M_{c^\dagger c}$ and $\tilde M_{BC} = \tilde M_{d^\dagger d}$ for $d=cb^\dagger$.  We again confine the discussion to the case where the boundary conditions defined by any such source manifold are real.  By this we mean that, if the formalism allows complex bulk configurations $k$ to be considered, then if $k_B$ satisfies the boundary conditions defined by $\tilde M_B$, so does the complex conjugate $k^*_B$.  We also again require each of the associated source-manifolds-with-boundary $M_b,M_c$ to have ${\cal C}_\epsilon$ rims for some $\epsilon >0$ as described in section \ref{sec:nongrav}.

The AdS/CFT dictionary of \cite{Witten:1998qj} (or its extrapolation to ensembles) then states that $\langle \Tr_{\scriptscriptstyle {\cal D}}(B) \Tr_{\scriptscriptstyle {\cal D}}(C) \rangle$, and $\langle \Tr_{\scriptscriptstyle {\cal D}}(BC) \rangle$ may equivalently be computed as bulk path integrals that sum over all bulk spacetimes with boundary conditions determined by the above source manifolds $\tilde M_B, \tilde M_C$, and $\tilde M_{BC}$.  As stated in the introduction, we take this bulk path integral to be normalized by dividing by the no-boundary state or, equivalently, we take the bulk path integral with boundary conditions set by some $\tilde M$ to sum {\it only} over bulk spacetimes in which every point can be connected to the boundary at $\tilde M$.  Disconnected closed universes are {\it not} included in our sum.

At leading semiclassical order the basic structure of our arguments will closely follow those of section \ref{sec:nongrav}.  In particular, we will again restrict to situations far from phase transitions by requiring the bulk path integral for
$\langle \Tr_{\scriptscriptstyle {\cal D}}(B) \Tr_{\scriptscriptstyle {\cal D}}(C) \rangle$ to have a single dominant saddle.  However, we will need to deal with two new inter-related further complications.  The first is that gravitational actions are generally not bounded below on the space of real Euclidean fields.  The second is that, as a result of the issue just described, the so-called ``Euclidean'' gravitational path integral cannot actually be taken to be defined as the integral over real Euclidean fields.

To allow the casual reader to focus on the big picture, the present section presents an overview of the argument for \eqref{eq:trineq3} and deals with the above issues by simply making assumptions about the gravitational path integral as needed.  We then return to address those assumptions in section \ref{sec:asstatus}.

We begin by discussing the leading-order result, in which we take each path integral to be dominated by a smooth bulk saddle.  Higher order corrections will be discussed later, at the end of this section.

 We are free to call the dominant bulk saddles for each path integral $\sigma_B, \sigma_C$, and $\sigma_{BC}$ in direct analogy with section \ref{sec:nongrav}.  We thus have
\begin{equation}
\label{eq:leadinggrav}
\Tr_{\scriptscriptstyle {\cal D}} B = e^{-I(\sigma_B)}, \ \  \Tr_{\scriptscriptstyle {\cal D}} C = e^{-I(\sigma_C)}, \ \ {\rm and}  \ \ \Tr_{\scriptscriptstyle {\cal D}} BC = e^{-I(\sigma_{BC})}.
\end{equation}
In particular, we suppose that the semiclassical approximation to our path integral satisfies the following assumption:
\begin{assumption}
\label{as:min}
For a bulk path integral specified by boundary conditions defined by a (compact) closed source manifold $\tilde M$ with real sources,
we assume that there is a class of configurations ${\cal K}_{\tilde M}$ such that i) the bulk fields described by any $k\in {\cal K}_{\tilde M}$ are continuous, ii) the bulk action $I$ is a real-valued functional on ${\cal K}_{\tilde M}$ and iii) in the semiclassical limit, the path integral is dominated by a real saddle $\sigma \in {\cal K}_{\tilde M}$ that minimizes the action $I$ over ${\cal K}_{\tilde M}$.  In particular, we have $I(\sigma) = \min_{k\in {\cal K}_{\tilde M}} I(k)$.  Furthermore, if ${\cal K}_{\tilde M}$ includes a complex configuration $k$, then the complex conjugate $k^*$ also lies in the same ${\cal K}_{\tilde M}$.    We similarly assume that the class
${\cal K}_{\tilde M}$ is invariant under a corresponding action of any symmetry of $\tilde M$.
\end{assumption}
As described  in section \ref{sec:nongrav}, this assumption is naturally satisfied in  contexts where the Euclidean path integral over real fields converges.  In that case, ${\cal K}_{\tilde M}$ is just the class of real field configurations.  But this is not generally the case in gravitational theories.  We thus emphasize that
Assumption \ref{as:min} does not require ${\cal K}_{\tilde M}$ to contain {\it all} real configurations, and in fact does not generally require configurations $k \in {\cal K}_{\tilde M}$ to be real at all (except for the dominant saddle in the semiclassical limit).  Instead, it requires only that  $I(k)$ be real-valued on ${\cal K}_{\tilde M}$.  This flexibility will be useful in later sections where we discuss several different possible choices of  ${\cal K}_{\tilde M}$ associated with different approaches to defining the path integral.

Since the present section addresses a general theory of gravity, we will make no attempt to write down an explicit action.  However, we do require the action to satisfy the following additivity property which can be checked in any particular theory (and which will be discussed for familiar examples in section \ref{subsec:FALPI}):
\begin{assumption}
\label{as:add}
Consider two boundary source manifolds $\tilde M_{bc}$, $\tilde M_{c^\dagger d}$, where $\tilde M_{bc}$ is given by cyclicly gluing together the input and output boundaries of some $M_{b}, M_{c}$, and where $\tilde M_{c^\dagger d}$ is similarly constructed from $M_{c^\dagger }, M_{d}$. Given any real bulk saddles $\sigma_{bc} \in {\cal K}_{\tilde M_{bc}}$,
$\sigma{}_{c^\dagger d} \in {\cal K}_{\tilde M_{c^\dagger d}}$, we assume there is a prescription for slicing $\sigma_{bc}$ into two pieces $k_b, k_c$,
and of similarly slicing $\sigma{}_{c^\dagger d}$ into two pieces $k_{c^\dagger}, k_d$ which satisfy
\begin{equation}
\label{eq:bulkadd1}
I(\sigma_{bc}) = I(k_b) + I(k_c),  \ \ \ {\rm and} \ \ \ I(\sigma_{c^\dagger d}) = I(k_{c^\dagger}) + I(k_d).
\end{equation}
We emphasize that this condition need only be satisfied by real saddles and not by general configurations in $ {\cal K}_{\tilde M_{bc}}$ and ${\cal K}_{\tilde M_{c^\dagger d}}$.
\begin{figure}[t]
	\centering
\includegraphics[width=\linewidth]{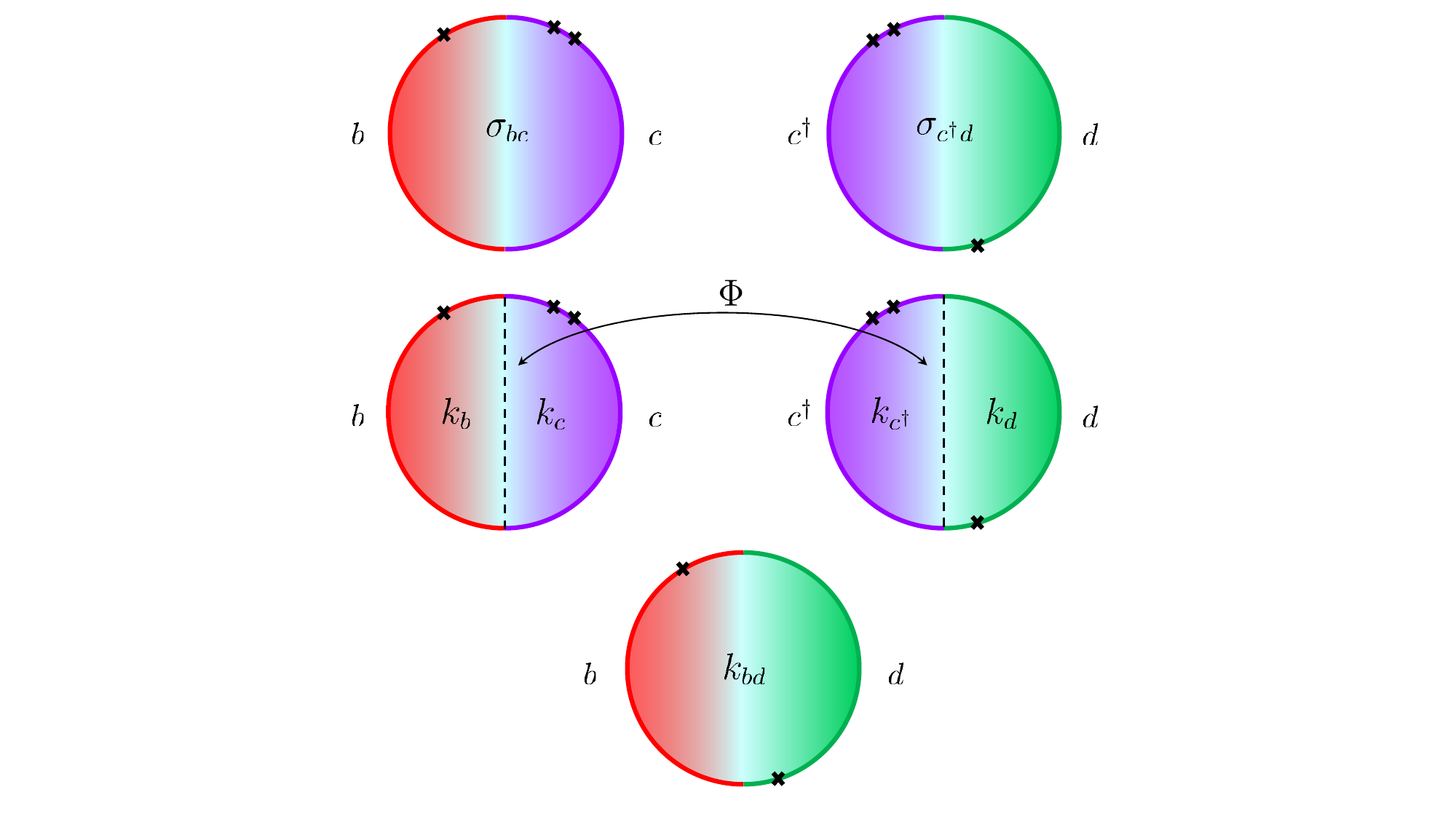}
\caption{Real bulk saddles $\sigma_{bc}$ and $\sigma_{c^\dagger d}$ for $\Tr_{\scriptscriptstyle {\cal D}} bc$ and $\Tr_{\scriptscriptstyle {\cal D}} c^\dagger d$ can be cut into pieces. Note that the cutting step generally creates new boundaries (dashed lines) not restricted by the asymptotically AdS boundary conditions.  When the data on the two new boundaries are related by a diffeomorphism $\Phi$, the pieces can be pasted together to make
a bulk configuration $k_{bd}$ for  $\Tr_{\scriptscriptstyle {\cal D}} bd$. }
\label{fig:bulkcutandpaste}
\end{figure}
We also assume that the slicing prescription preserves any symmetries of the bulk saddle $\sigma_{bc}$.

As shown in figure \ref{fig:bulkcutandpaste}, the cutting of $\sigma_{bc}$ into $k_b, k_{c}$ generally creates new boundaries not restricted by properties of $\tilde M_{bc}$.  As a result, we require the action $I$ to be defined on such bulk configurations.  This may require the specification of appropriate boundary terms at the new boundaries.  It may also require corner terms where the new boundaries intersect $\tilde M_{bc}$; see related discussions in  \cite{Hayward:1993my,Hawking:1996ww,Brown:2000dz}.

Furthermore, suppose that there is a diffeomorphism $\Phi$  from the new boundaries of $k_c$ to the new boundaries of $k_{c^\dagger}$ that preserves the values of all bulk fields (though which need not preserve normal derivatives of bulk fields). Then we can glue the new boundaries of $k_b$ to those of $k_d$ to create a new configuration $k_{bd} \in {\cal K}_{\tilde M_{bd}}$ whose actionwe assume to be
\begin{equation}
\label{eq:bulkadd}
I(k_{bd}) = I(k_b) + I(k_d);
\end{equation}
see again figure \ref{fig:bulkcutandpaste}.
\end{assumption}

As in section \ref{sec:nongrav}, we first consider the case where each object in \eqref{eq:trineq3} is computed by a single path integral, returning later to cases that involve linear combinations of path integrals. We will need the analogue of Lemma \ref{lemma:sym} for the gravitational context:

\begin{lemma}
\label{lemma:symG}
Consider an operator $D = d^\dagger d$ in ${\cal D}$, where $d$ is computed in ${\cal D}$ by a Euclidean path integral over a source manifold $M_d$ with real sources.  The source manifold $\tilde M_{D} = \tilde M_{d^\dagger d}$ then clearly enjoys a ${\mathbb Z}_2$ symmetry as discussed above. This symmetry is in fact preserved by any bulk saddle $\sigma_D$ that dominates the {\it bulk} path integral for  $\Tr_{\scriptscriptstyle {\cal D}} \tilde M_D$.  In cases where several allowed bulk saddles share this minimum value of the action,  the symmetry is preserved by at least one such $\sigma_D$.
\end{lemma}

\begin{figure}[t]
	\centering
\includegraphics[width=\linewidth]{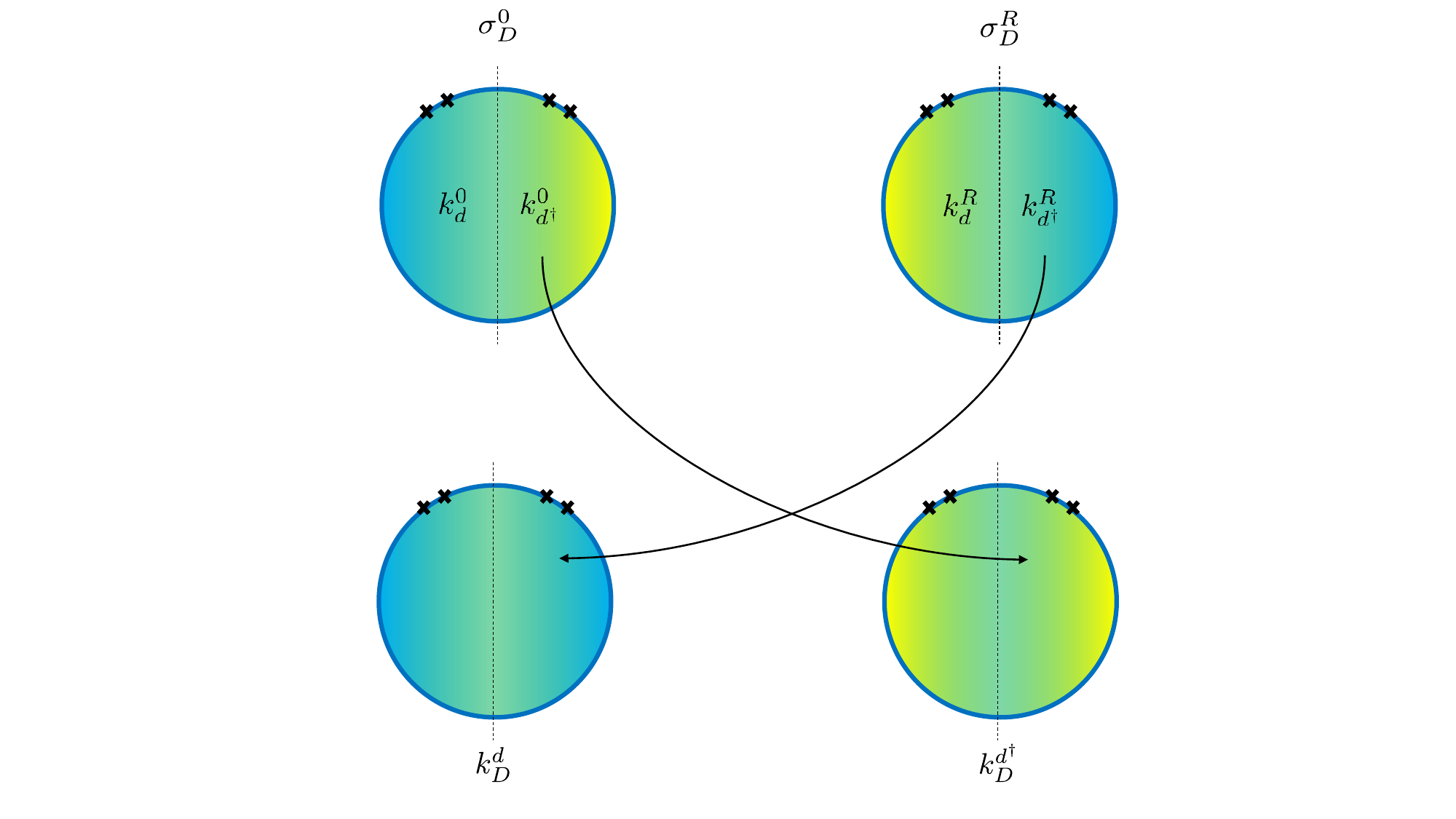}
\caption{When $D = d^\dagger d$, an arbitrary bulk saddle $\sigma^0_D$ for $\Tr_{\scriptscriptstyle {\cal D}} D$ can be reflected to give another saddle $\sigma^R_D$ for the same bulk path integral.  We can cut $\sigma^0_D$ into parts $k^0_d, k^0_{d^\dagger}$, and we can similarly cut $\sigma^R_D$ into $k^R_d, k^R_{d^\dagger}$. Gluing $k^0_d$ to $k^R_{d^\dagger}$ defines a $\mathbb{Z}_2$-symmetric configuration $k_D^d$ for  $\Tr_{\scriptscriptstyle {\cal D}} D$, and similarly gluing $k^R_d$ to $k^0_{d^\dagger}$ defines a $\mathbb{Z}_2$-symmetric   $k_D^{d^\dagger}$.  One of these must have action less than or equal to that of the original saddle $\sigma^0_D$.  If $\sigma^0_D$ was dominant, Assumption \ref{as:min} would imply that we have now constructed two $\mathbb{Z}_2$-symmetric saddles both having precisely the same action as $\sigma^0_D$.}
\label{fig:bulksymmsaddle}
\end{figure}

Using assumptions \ref{as:min} and \ref{as:add}, we can give a proof of Lemma \ref{lemma:symG} that directly parallels the proof of Lemma \ref{lemma:sym} in section \ref{sec:nongrav}.  The argument is depicted in figure \ref{fig:bulksymmsaddle}.
We first consider an arbitrary saddle $\sigma^0_D$ that lies in ${\cal K}_{{\tilde M}_D}$ and use assumption \ref{as:add} to divide it into $k^0_{d^\dagger}$, $k^0_d$.  Note that the values of the bulk fields on the new boundaries of $k^0_{d^\dagger}$ agree with those on the new boundaries of $k^0_d$ by continuity  on $\Sigma^0_D$; see again Assumption \ref{as:min}.  But we can use the reflection symmetry of $\tilde M_D$ to construct a reflected saddle $\sigma^R_D$ that again lies in
${\cal K}_{{\tilde M}_D}$, and which we then divide into
$k^R_{d^\dagger}$, $k^R_d$.  Because $k^0_d$ and $k^R_{d^\dagger}$ are related by the reflection symmetry, the field values on their new boundaries agree.  We may thus paste these pieces together to form a new configuration $k_D^d \in {\cal K}_{{\tilde M}_D}$ with an explicit bulk reflection symmetry, and we may also construct the analogous $k_D^{d^\dagger} \in {\cal K}_{{\tilde M}_D}$ from $k^R_d$ and $k^0_{d^\dagger}$.  As in section \ref{sec:nongrav}, the additivity properties \eqref{eq:bulkadd1}, \eqref{eq:bulkadd} applied to the current pieces then imply that either $I(k_D^d) \le I(\sigma^0_D)$ or  $I(k_D^{d^\dagger}) \le I(\sigma^0_D)$.  As a result,  if $\sigma^0_D$ is a dominant saddle, then either $k_D^d$ or $k_D^{d^\dagger}$ must be an equally-dominant saddle that preserves the desired symmetry.

Lemma \ref{lemma:symG} will soon allows us to prove the trace inequality \eqref{eq:trineq3} at leading semi-classical order.  As stated above,
at this order we have
\begin{equation}
\label{eq:leadingQMD}
\Tr_{\scriptscriptstyle {\cal D}} B = e^{-I(\sigma_B)}, \ \  \Tr_{\scriptscriptstyle {\cal D}} C = e^{-I(\sigma_C)}, \ \ {\rm and}  \ \ \Tr_{\scriptscriptstyle {\cal D}} BC = e^{-I(\sigma_{BC})}.
\end{equation}

We are interested in the case $\tilde M_B = \tilde M_{b^\dagger b}$, $\tilde M_C=\tilde M_{c^\dagger c}$, so that we may also write $\tilde M_{BC} = \tilde M_{d^\dagger d}$ using $d=bc^\dagger$ and the fact that $\tilde M_{b^\dagger b c^\dagger c} = \tilde M_{c b^\dagger b c^\dagger}$. (This follows from the fact that our gluing operation is invariant under cyclic permutation of the parts to be glued).   Applying Lemma \ref{lemma:symG} to $\tilde M_{BC} = \tilde M_{d^\dagger d}$, we may take $\sigma_{BC}$ to have a $\mathbb Z_2$ symmetry that exchanges $d=bc^\dagger$ and $d^\dagger=cb^\dagger$.  We may then cut the saddle $\sigma_{BC}$  into the two pieces $k_B, k_C$ associated with the $M_B, M_C$ source manifolds.    Furthermore, since Assumption \ref{as:add} required the slicing prescription to preserve symmetries of the original bulk saddle, the boundaries $\partial k_B$, $\partial k_C$ will be invariant under corresponding reflection symmetries.  This will in particular be true for the new boundaries created by slicing $\sigma_{BC}$ into parts.

We now make a final monotonicity assumption regarding our action.
\begin{assumption}
\label{as:mono}
Consider again the setting of assumption \ref{as:add} and the pieces $k_b, k_c$ described there. Let $\partial_{new} k_b$ denote the new boundaries of $k_b$ created by slicing $\sigma_{bc}$ in two; i.e., these are the boundaries of $k_b$ that were not boundaries in $\sigma_{bc}.$  We assume that when $k_b$ is invariant under a ${\mathbb Z}_2$ symmetry, we may use this symmetry to glue any point of $\partial_{new} k_b$ to its image to define a configuration $\tilde k_b \in {\cal K}_{{\tilde M}_b}$ associated with the bulk path integral for $\Tr_{\scriptscriptstyle {\cal D}}(b)$; see figure \ref{fig:bulkfold} below.  We further assume that this gluing operation does not increase the action.  In other words, we assume
\begin{equation}
\label{eq:mono}
I(\tilde k_b)  \le I(k_b).
\end{equation}
\end{assumption}
\begin{figure}[h!]
	\centering
\includegraphics[width=.8\linewidth]{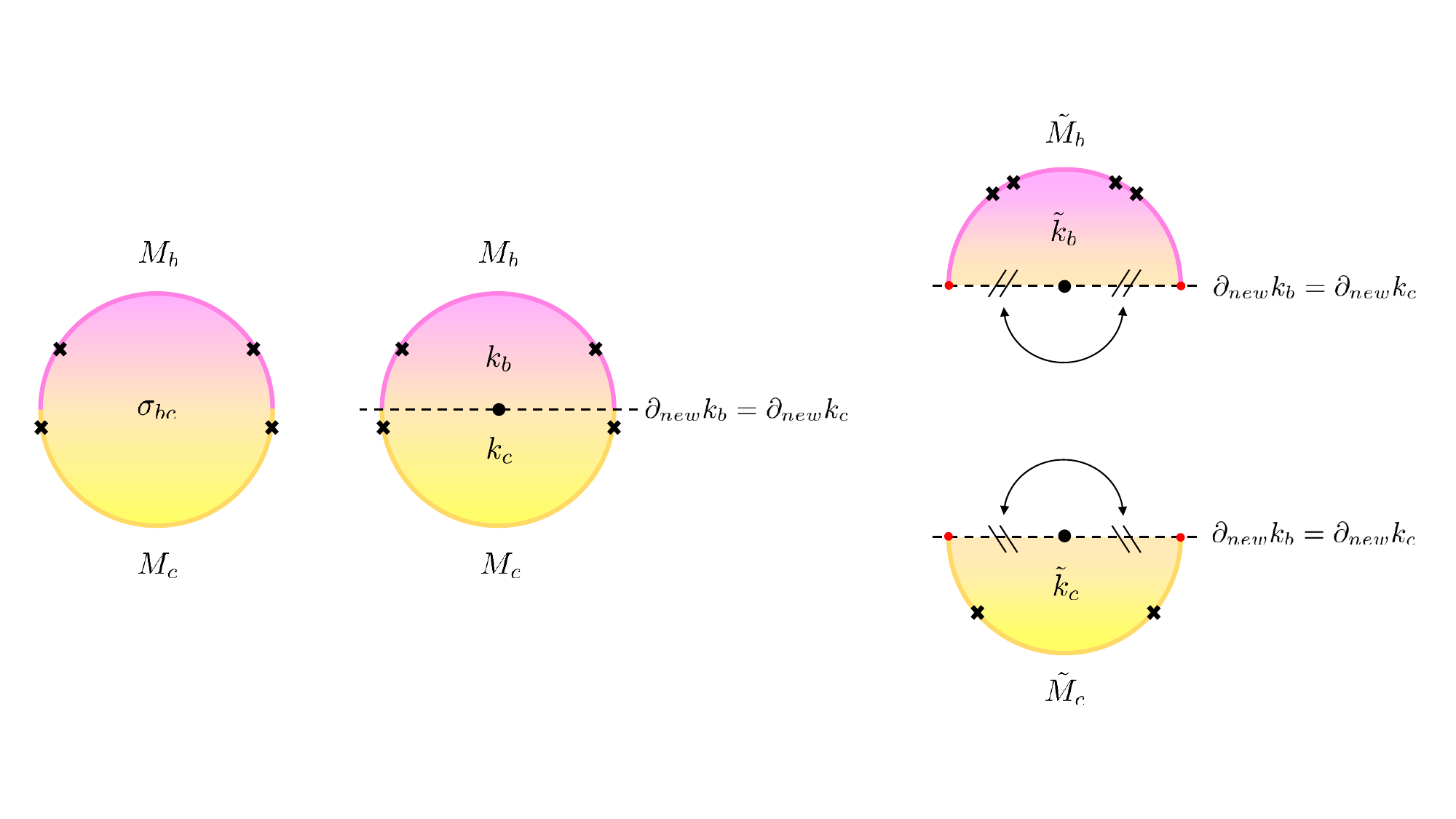}
\caption{ We consider a bulk saddle $\sigma_{bc}$ with a ${\mathbb Z}_2$ reflection symmetry that preserves both $M_b$ and $M_c$.  Such a saddle can be cut into two reflection-symmetric pieces $k_b$ and $k_c$. This creates new boundaries but, in each piece, the reflection symmetry allows one to remove the new boundaries by making identifications.  The results define configurations $\tilde k_b \in {\cal K}_{{\tilde M}_b}, \tilde k_c \in {\cal K}_{{\tilde M}_c}$ that satisfy boundary conditions appropriate to computing $\Tr_{\scriptscriptstyle {\cal D}}(b)$, $\Tr_{\scriptscriptstyle {\cal D}}(c)$.  If the reflections have fixed points on the new boundaries of $k_b,k_c$, then $\tilde k_b, \tilde k_c$ will have conical singularities with deficit angle $\pi$.}
\label{fig:bulkfold}
\end{figure}

Before using this assumption, it is important to explain why the relation \eqref{eq:mono} is natural, and in particular why it is not generally an equality.  In the nongravitational discussion of section \ref{sec:nongrav}, the
topology of any saddle was always that of the corresponding source manifold $\tilde M$ that defined the relevant path integral.  As a result, the equivalent of $\partial_{new} k_b$ always separated cleanly into input and output boundaries.  In particular, in the non-gravitational case the reflection symmetry that acted on $\partial_{new} k_b$ had no fixed points.  Thus the equivalent of $\tilde k_b$ was always smooth.

In the gravitational context, we may indeed expect that $I(\tilde k_b) = I(k_b)$ when $\tilde k_b$ is smooth.  However, the dimensionality of the bulk saddle is typically greater than that of the source manifold.  In particular, the topology of source manifold no longer dictates the topology of the bulk.
As a result, the construction of $\tilde k_b$ from $k_b$ will  introduce a conical deficit of $\pi$ at any fixed points of the reflection symmetry that lie on the new boundary $\partial_{new} k_b$ of $k_b$; see again figure \ref{fig:bulkfold}.  In such cases, the monotonicity assumption \eqref{eq:mono} amounts to the condition that conical deficits give a non-positive contribution to the Euclidean gravitational action.  This is consistent with the standard sign choices for the Euclidean Einstein-Hilbert and Jackiw-Teitelboim actions (see e.g. \cite{Gibbons:1978ac}).  In fact, for later purposes it is useful to add a further assumption which essentially states that the contribution of conical deficits is strictly negative:
\begin{assumption}
\label{as:strictmono}
Consider again the context of Assumption \ref{as:mono}.  If the reflection symmetry of $k_b$ has fixed points on $\partial_{new} k_b$, then we in fact have
\begin{equation}
\label{eq:strictmono}
I(\tilde k_b)  < I(k_b).
\end{equation}
\end{assumption}
Assumption \ref{as:strictmono} will be of use when we consider perturbative corrections, though we will set it aside for now.

\begin{figure}[h!]
	\centering
\includegraphics[width=\linewidth]{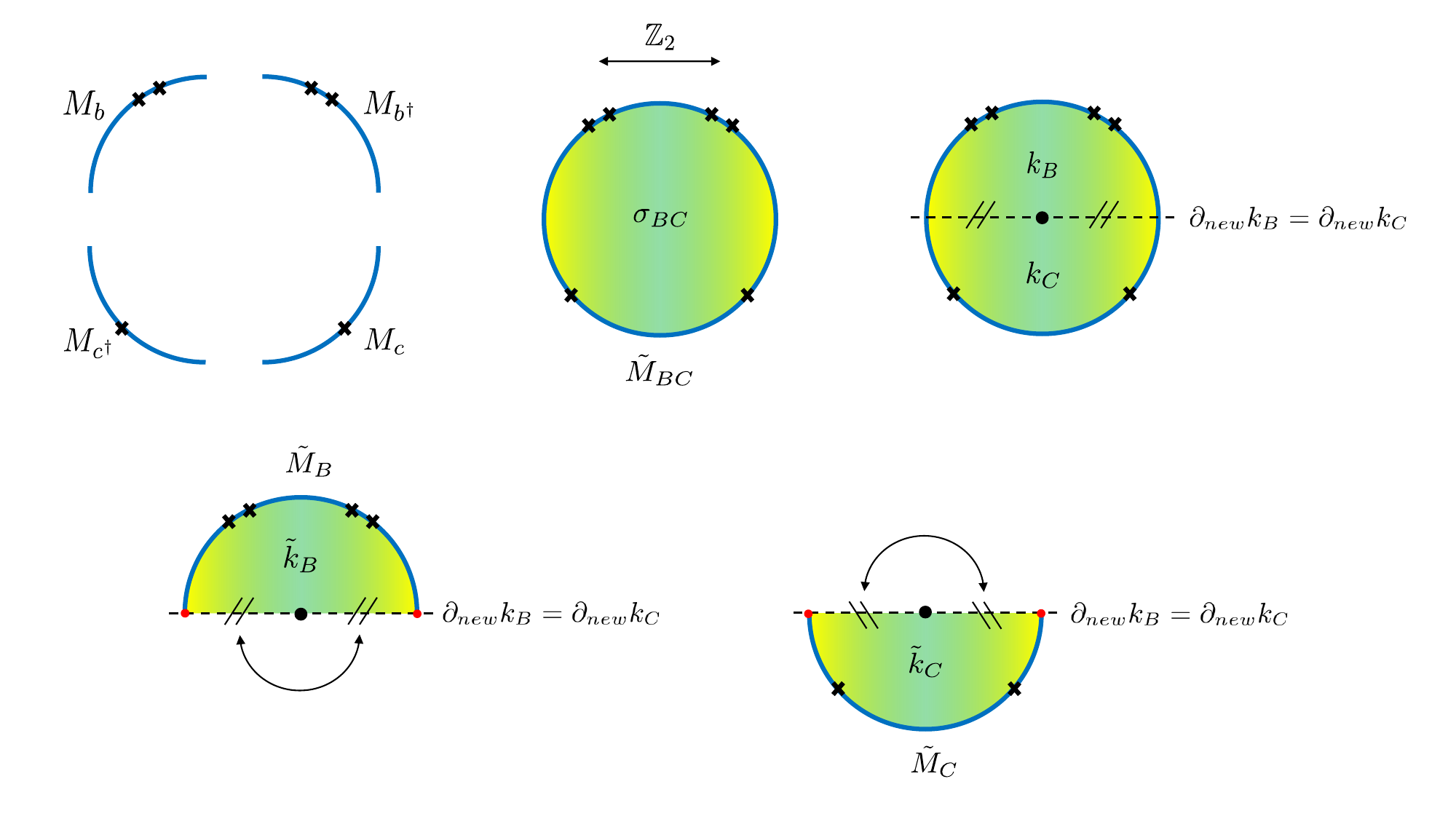}
\caption{This diagram depicts a bulk version of the non-gravitational argument that was illustrated in figure \ref{fig:BandCtoBC}. {\bf Upper left panel:} The boundary source manifolds $M_b$, $M_{b^\dagger}$, $M_c$ and $M_{c^\dagger}$ which are glued together to form $\tilde M_{BC}$. {\bf Upper middle panel:} By Lemma \ref{lemma:symG} we can consider a dominant bulk saddle $\sigma_{BC}$ for the bulk path integral with boundary conditions $\tilde M_{BC}$ such that $\sigma_{BC}$ has a $\mathbb Z_2$ reflection symmetry. {\bf Upper right panel:} The saddle $\sigma_{BC}$ can be cut into two pieces $k_B, k_C$.  These pieces are {\it not} generally related by any symmetry. Instead, by Assumption \ref{as:add}, the new boundaries $\partial_{new} k_B= \partial_{new} k_C$ created by the cut are each separately invariant under the $\mathbb Z_2$ reflection symmetry and $I(\sigma_{BC})=I(k_B) + I(k_C)$. {\bf Lower panel:} By gluing every point of $\partial_{new} k_B$ to its image under the $\mathbb Z_2$ symmetry we define a configuration $\tilde k_B$ for the bulk path integral with boundary conditions $\tilde M_{B}$.  In particular, this operation identifies pairs of red dots to construct $\tilde M_B$.  The configuration $\tilde k_C$ is also constructed in the same way from $k_C$. Assumptions \ref{as:mono} and \ref{as:min} then imply$ I(\sigma_{BC})=I(k_B) + I(k_C)\ge I(\tilde k_B) + I(\tilde k_C) \ge I(\sigma_B) + I(\sigma_C)$, where $\sigma_B, \sigma_C$ are the dominant saddles for
$ \Tr_{\scriptscriptstyle {\cal D}} B, \Tr_{\scriptscriptstyle {\cal D}} C$.}
\label{fig:TI_proof}
\end{figure}
Returning to the main argument, we may use the above procedure to construct configurations $\tilde k_B, \tilde k_C$ for $\Tr_{\scriptscriptstyle {\cal D}}(B)$, $\Tr_{\scriptscriptstyle {\cal D}}(C)$  from the pieces $k_B,k_C$ that were cut from $\sigma_{BC}$.
We then apply Assumption \ref{as:mono}, replacing $b$ in \eqref{eq:mono}  by either  $B$ or $C$.  Finally, we apply the minimization assumption (Assumption \ref{as:min}) to find that the dominant saddles $\sigma_B, \sigma_C$ for  $\Tr_{\scriptscriptstyle {\cal D}}(B)$, $\Tr_{\scriptscriptstyle {\cal D}}(C)$ satisfy
\begin{equation}
\label{eq:subaddD}
I(\sigma_{BC})=I(k_B) + I(k_C) \ge I(\tilde k_B) + I(\tilde k_C) \ge I(\sigma_B) + I(\sigma_C).
\end{equation}
By \eqref{eq:leadingQMD}, this is then equivalent to  the desired trace inequality \eqref{eq:trineq3} at leading order in the semiclassical expansion.   The important steps of the above argument are illustrated in figure \ref{fig:TI_proof}.

The above reasoning suffices for the case where  $B,C$ each represent a single boundary condition.  The remaining case where
they are linear combinations of boundary conditions then follows just as at the end of section \ref{sec:nongrav}.  Starting with a general saddle for any term in the sum associated with $\Tr_{\scriptscriptstyle {\cal D}} (BC)$, Assumptions \ref{as:min} and \ref{as:add} imply that there is another saddle with equal or lesser action that is associated with one of the diagonal terms in the sum.  And this diagonal term can then be used as above to construct saddles $\sigma_B, \sigma_C$ for $\Tr_{\scriptscriptstyle {\cal D}} B, \Tr_{\scriptscriptstyle {\cal D}} C$ that satisfy \eqref{eq:subaddD}.  So, again, the desired result holds.

We may also address perturbative quantum corrections to \eqref{eq:trineq3}.   This turns out to be straightforward since we take the path integral for $\langle \Tr_{\scriptscriptstyle {\cal D}} (BC) \rangle$ to be dominated by a {\it single} saddle.
A key point is that perturbative quantum corrections are explicitly given by quantum field theory in the curved spacetime backgrounds defined by our saddles.  In particular, they are computed by {\it non-gravitational} path integrals, or by path integrals  that include only perturbative gravitons, of the general form described in section \ref{sec:nongrav}, but where the leading-order bulk saddles $\sigma_{BC}, \sigma_B, \sigma_C$ now play the role of $\tilde M_{BC}, \tilde M_B, \tilde M_C$ from section \ref{sec:nongrav}.

A second key point is that,  in any strictly perturbative framework,  quantum corrections can lead to violations of \eqref{eq:trineq3} only if this inequality is saturated at leading semi-classical order.  Since we assume unique saddles $\sigma_{BC}, \sigma_B, \sigma_C$ for, respectively, $\Tr_{\scriptscriptstyle {\cal D}}(BC)$, $\Tr_{\scriptscriptstyle {\cal D}}(B)$, and $\Tr_{\scriptscriptstyle {\cal D}}(C)$, our arguments above require that $\sigma_B$, $\sigma_C$ can be obtained by slicing $\sigma_{BC}$ into two pieces, each of which is separately invariant under a reflection symmetry.  The saddle $\sigma_B$ is then obtained by using the reflection symmetry of the $B$ piece to glue together any new boundaries created by the slicing operation.  The saddle $\sigma_C$ is also constructed in the analogous fashion.

Furthermore, the above argument also shows that  strict saturation of \eqref{eq:trineq3} at leading semiclassical order requires one to be able to reconstruct $\sigma_{BC}$ from $\sigma_B, \sigma_C$ by a procedure directly analogous to that building $\tilde M_{BC}$ from $\tilde M_B$, $\tilde M_C$ (shown previously in figure \ref{fig:BandCtoBC}); i.e., $\sigma_B = k)B$ and $\sigma_C = k_C$.  Moreover, the objects $k_B, k_C$ used to construct $\tilde k_B = \sigma_B, \tilde k_C = \sigma_C$ now play the roles of $M_B, M_C$ from section \ref{sec:nongrav}.  Thus, for example, the quantum correction to $\Tr_{\scriptscriptstyle {\cal D}} B$ is precisely $\Tr_{pert \ bulk} B_{pert \ bulk}$ where this is a trace over the perturbative bulk Hilbert space and where $B_{pert \ bulk}$ is an operator on that Hilbert space.  Furthermore, the reflection symmetry of $\sigma_B$ implies the operator $B_{pert \ bulk}$ to be positive.  Indeed, since the analogous statements hold for $C$ and $BC$, the simple quantum-mechanical argument given by \eqref{eq:Trinder} can be used to write
\begin{equation}
\Tr_{pert \ bulk} (B_{pert \ bulk}C_{pert \ bulk}) \le (\Tr_{pert \ bulk} B_{pert \ bulk})(\Tr_{pert \ bulk} C_{pert \ bulk}).
\end{equation}
Thus we see that,  at each order in the semiclassical expansion, quantum corrections cannot induce violations of \eqref{eq:trineq3}.

Since we have not specified the gravitational theory, it is not natural at this stage to separate out discussions of higher derivative terms.  We will instead address related issues in section \ref{sec:asstatus} when we discuss the status of our assumptions in various classes of theories.

\section{The status of our assumptions in general gravitational theories}
\label{sec:asstatus}
\label{subsec:EH}

We now turn to a discussion of assumptions \ref{as:min}-\ref{as:strictmono} from section \ref{subsec:main} for general theories of gravity. These assumptions require the semiclassical approximation to Euclidean quantum gravity to be determined by minimizing an action functional over appropriate classes ${\cal K}_{\tilde M}$ of spacetimes satisfying boundary conditions given by some $\tilde M$. At first glance, this idea may appear to be famously false in Euclidean Einstein-Hilbert gravity due to the conformal factor problem \cite{Gibbons:1978ac}.  In particular, one can find smooth Euclidean bulk spacetimes satisfying arbitrary boundary conditions that make the Euclidean Einstein-Hilbert action arbitrarily negative, so that minimal action configurations do not exist.

Any attempt to establish the assumptions used in section \ref{subsec:main} must thus begin with some viewpoint on how the conformal factor problem is to be addressed.  We have already discussed the Saad-Shenker-Stanford paradigm for JT gravity (with dilaton-free matter couplings) in section \ref{subsec:SSS}, where we showed that it leads to the desired trace inequality in the semiclassical limit.  While we see no simple argument that such a paradigm satisfies our assumption \ref{as:mono} or assumption \ref{as:strictmono}, we argue below that our assumptions {\it are} in fact satisfied within two other (perhaps overlapping) paradigms for dealing with the conformal factor issue.   The first, which we call the Gibbons-Hawking-Perry paradigm,  is a hypothetical non-linear generalization of the contour rotation prescription described in \cite{Gibbons:1978ac} for linearized fluctuations about Euclidean Schwarzschild.  The second follows \cite{Marolf:2022ybi} in taking the Lorentzian path integral to be fundamental, evaluating the Lorentzian path integral with ``fixed-area boundary conditions," and arguing that the result reduces to an integral over Euclidean spacetimes that are on-shell up to the presence of conical singularities.

We discuss each of these paradigms in turn below.  The first discussion (section \ref{subsec:GHP}) is necessarily brief and schematic due to the hypothetical nature of the supposed extension of known results.  More details will be provided when considering the second  paradigm in section \ref{subsec:FALPI}.
This will allow key elements of the assumptions either to be proven or to be reformulated as precise conjectures concerning the classical action which should be amenable to future mathematical and numerical studies.

\subsection{The Gibbons-Hawking-Perry Contour Rotation Paradigm}
\label{subsec:GHP}

As shown long ago by Gibbons, Hawking, and Perry \cite{Gibbons:1978ac},
at the linearized level for familiar cases one can obtain physically reasonable results (see also \cite{Allen:1984bp,Prestidge:1999uq,Kol:2006ga,Headrick:2006ti,Monteiro:2008wr,Monteiro:2009tc,Monteiro:2009ke,Marolf:2021kjc,Cotler:2021cqa,Marolf:2022ntb,Marolf:2022jra}) by `rotating the contour of integration.'  This in fact means that one defines the path integral to integrate over some non-trivial contour $\Gamma$ in the space of complex metrics.  In the linearized cases mentioned above, the action is real and bounded-below on the $\Gamma$ chosen in \cite{Gibbons:1978ac}.   This last point contrasts with the Saad-Shenker-Stanford paradigm which also uses a non-real contour, but on which the action is manifestly complex.  In the case considered by Gibbons, Hawking and Perry, the action also diverges to $+\infty$ in all asymptotic regions of $\Gamma$.  As a result, the action on $\Gamma$ is necessarily minimized at some finite smooth saddle-point that dominates the path integral in the semi-classical limit.  If this same structure persists in the non-linear theory, then Assumption \ref{as:min} is clearly satisfied if we simply redefine configurations on $\Gamma$ to be `real' for the purposes of that assumption. See also the discussions of contour rotation for the full theory in \cite{Dasgupta:2001ue,Ambjorn:2002gr}.

Now, Assumptions \ref{as:mono} and \ref{as:strictmono} require the full space ${\cal K}_{\tilde M}$ to admit configurations with conical singularities.
If the saddles are known to be smooth, and if the construction of $\Gamma$ respects symmetries of the boundary conditions, then we can take ${\cal K}_{\tilde M}$ to be given by those spacetimes lying on $\Gamma$ which can be formed from smooth spacetimes by applying a single cut-and-paste operation of the type described in Assumption \ref{as:add}.
This choice allows us to restrict attention to spacetimes that are `not too wild' and on which we can hope to have some control over the action as a function on ${\cal K}_{\tilde M}$.
Furthermore, if the specification of the desired contour $\Gamma$ is sufficiently local in spacetime, then cutting spacetimes $\gamma_1, \gamma_2 \in \Gamma$ into pieces and pasting them together to build a new configuration $\gamma$ will also naturally yield $\gamma \in \Gamma$.  As a result, the above definition of ${\cal K}_{\tilde M}$ would then be manifestly invariant under such operations.  So long as the spacetimes satisfy appropriate boundary conditions, Assumption \ref{as:add} will then be satisfied if our action includes appropriate boundary terms.  Explicit discussions of such boundary conditions and boundary terms for JT and Einstein-Hilbert gravity will appear in section \ref{subsec:FALPI} below.

It thus remains only to discuss Assumptions \ref{as:mono} and \ref{as:strictmono}.  As described between \eqref{eq:mono} and \eqref{eq:strictmono}, for Euclidean geometries in Einstein-Hilbert gravity the two sides of \eqref{eq:mono} differ only by contributions from the conical singularities.  A standard calculation shows that this gives an extra factor of $e^{-A/4G}$ on the left hand side, where $A$ is the area of the conical singularity.  Similarly, in JT gravity (normalized as in \eqref{eq:JTaction}), the difference is a factor of $e^{-4\pi \phi}$ evaluated at the singularity.   As a result, in either of these theories, so long as $A$ (or $\phi$) is positive on the contour $\Gamma$, these assumptions will be satisfied as well.

As a final comment on this paradigm, let us address the question of perturbative higher derivative corrections to either JT gravity or Einstein-Hilbert gravity.  Rather than attempt to discuss Assumptions \ref{as:min}-\ref{as:strictmono} for the full path integral with higher-derivative corrections, we will instead take perturbative treatment of such terms to mean that their corrections to the two-derivative theory are computed by first finding saddles that would dominate the semiclassical computation in the two-derivative theory, and then using the higher derivative terms to compute perturbative corrections to the relevant actions.
So long as we suppose that the dominant 2-derivative saddle for each path integral is unique, we may then argue that the trace inequality \eqref{eq:trineq3} is preserved by higher derivative corrections in direct analogy with the non-gravitational discussion at the end of section \ref{sec:nongrav}.  The only comment needed to promote that argument to the gravitational context is to again note that assumption \ref{as:mono} (applied at the level of the two-derivative theory) means that we may indeed confine discussion of higher derivative corrections to perturbation theory about saddle points for $\Tr_{\scriptscriptstyle {\cal D}} (B)$ , $\Tr_{\scriptscriptstyle {\cal D}} (C)$  in the two-derivative theory that are constructed from the two-derivative saddle for
$\Tr_{\scriptscriptstyle {\cal D}} (BC)$ using the cut-and-paste procedure above.
As in the non-gravitational discussion at the end of section \ref{sec:nongrav}, we leave open for future study the more general but finely-tuned case where the saddles fail to be unique.

\subsection{Euclidean path integrals from fixed-area Lorentzian path integrals}
\label{subsec:FALPI}

While the discussion of the Gibbons-Hawking-Perry paradigm in section \ref{subsec:GHP} was straightforward, it also relied on the conjectured existence of a hypothetical contour $\Gamma$ with certain properties.  Furthermore, since the form of the presumed $\Gamma$ is not known, it is difficult to perform further checks within that approach.  In contrast, we shall see that the paradigm described in \cite{Marolf:2022ybi} allows a more detailed discussion of assumptions \ref{as:min}-\ref{as:strictmono} and also presents more well-defined opportunities for further consistency checks. For lack of a better name, we will refer to the approach of \cite{Marolf:2022ybi} as the Lorentzian fixed-area paradigm.

\subsubsection{The Lorentzian fixed-area paradigm}

The treatment of \cite{Marolf:2022ybi} considered the special case of computing partition functions $Z(\beta) =\Tr(e^{-\beta H})$ for time-independent gravitational systems.  However, it did so by taking the Lorentz-signature path integral to be fundamental, and to be defined as an integral over spacetimes that were both real and Lorentz-signature up to the presence of certain codimension-2 singularities that one may call ``Lorentzian conical singularities" following \cite{Colin-Ellerin:2020mva,Colin-Ellerin:2021jev};  see also \cite{Hartle:2020glw,Schleich:1987fm,Mazur:1989by,Giddings:1989ny,Giddings:1990yj,Marolf:1996gb}, as well as \cite{Dasgupta:2001ue,Ambjorn:2002gr} and \cite{Feldbrugge:2017kzv,Feldbrugge:2017fcc,Feldbrugge:2017mbc} for earlier arguments that treating the Lorentzian formalization as fundamental is essential to resolving the Euclidean conformal factor problem.  As a result, much as in section \ref{subsec:pureJT}, $Z(\beta)$ was first written as an integral transform of distributional quantities that one may call $\Tr(e^{it H})$.  Due to their distributional nature, the quantities $\Tr(e^{it H})$ are generally not well-defined for any fixed $t$, though integrating over $t$ gives a well-defined result.

The suggestion of \cite{Marolf:2022ybi} was to first integrate over the real Lorenz-signature metrics while holding fixed the areas of the codimension-2 conical singularities.  In practice, this was done using the stationary phase approximation.  It is an interesting point that the Jackiw-Teitelboim and Einstein-Hilbert actions define good variational principles with such fixed-area boundary conditions \cite{Dong:2019piw}, and that the associated saddles may have arbitrary conical singularities at the fixed-area surface (as suggested in \cite{Akers:2018fow,Dong:2018seb}); similar statements also hold in the presence of perturbative higher derivative corrections \cite{Dong:2019piw}.   Evaluating the above-mentioned integral transform then led to a result that could be expressed as a final integral over {\it Euclidean}-signature metrics that satisfied the Euclidean equations of motion everywhere away from the fixed-area codimension-2 conical singularities, and which were thus known as Euclidean fixed-area saddles.  Since the saddles were parameterized by the here-to-fore-fixed areas of the conical singularities, the final integral was simply an integral over the associated areas.  For simple gravitational partition functions, this process was shown in \cite{Marolf:2022ybi} to yield the standard results.

Let us therefore imagine that, in the semiclassical limit, a similar paradigm can be applied to any Euclidean path integral.  In particular, given any operator $B$ in the dual theory ${\cal D}$, we imagine that $\Tr_{\scriptscriptstyle {\cal D}}B$ can be computed semiclassically as
\begin{equation}
\label{eq:intoverA}
\Tr_{\scriptscriptstyle {\cal D}}B \approx \int_{A \in {\mathbb{R}^+}} dA\, e^{-\tilde I_A(s_A)},
\end{equation}
where $A\ge 0$ parameterizes the possible codimension-2 areas of a set of conical singularities, $\tilde I_A$ is an action that gives a good variational principle when the area $A$ is fixed, and the argument $s_A$ denotes the real Euclidean saddle of $\tilde I_A$ having the lowest action $\tilde I_A(s_A)$ for the given value of $A$ that is consistent with satisfying the boundary condition at infinity.  This paradigm can also be applied to JT gravity with matter (where a codimension-2 surface is a discrete set of points) by replacing the area $A$  by the value of the dilaton $\phi$ summed over conical singularies.  Here we assume $\phi_0+\phi \ge 0$ at each singularity.

In writing \eqref{eq:intoverA}, it is assumed that the integral on the right-hand-side converges and that no further contour rotations are required.  This is not at all obvious from a cursory study of the gravitational action.  However, as argued in \cite{Dong:2018seb} (see also \cite{Akers:2018fow}), the quantities $e^{-\tilde I_A(s_A)}/\Tr_{\scriptscriptstyle {\cal D}}B$ are expected to represent the probabilities $p(A)$ of finding an extremal surface with area $A$ in a quantum gravity state with boundary conditions determined by the operator $B$.  Since probabilities sum to unity, this would then require the right-hand-side of \eqref{eq:intoverA} to converge as desired.  This idea has by now been investigated in a variety of contexts which appear to support this conclusion; see e.g.
\cite{Penington:2019kki,Marolf:2020vsi,Dong:2021clv,Dong:2022ilf,Marolf:2022ybi,Chandrasekaran:2022eqq,Kudler-Flam:2022jwd,Blommaert:2023vbz}.

For clarity, we formalize this assumption as follows:
\begin{assumption}
\label{as:fixed}
We assume that, in the UV-completion of either JT gravity or Einstein-Hilbert gravity with minimally-coupled matter, the integral over fixed-area-saddles on the right-hand-side of \eqref{eq:intoverA} converges and gives a good approximation to the left-hand-side in the semiclassical limit.
\end{assumption}

Assumption \ref{as:fixed} is now {\it almost} sufficient to allow us derive assumptions \ref{as:min}-\ref{as:strictmono} for both JT and Einstein-Hilbert gravity.
 However, recall that -- just as in the non-gravitational setting of section \ref{sec:nongrav} -- the cut-and-paste operations of section \ref{subsec:main} can produce surfaces on which certain equations of motion do not hold, and in particular at which first derivatives of fields fail to be continuous (though such derivatives admit well-defined limits when approaching the surface from either side).

As a result, we will need to further strengthen Assumption \ref{as:fixed} by adding yet another assumption.  We motivate this addition using an idea similar to the motivation for Assumption \ref{as:fixed} itself.  In particular, let us note that there is a set of diffeomorphism-invariant observables defined by the conformal geometry of a minimal surface anchored to particular cuts of the asymptotically AdS boundary.  Furthermore, the same is true for the minimal surface that is anchored to both the fixed-area surface {\it and} to particular
cuts of the asymptotically AdS boundary, and it is again true when the surface is minimal only within any given homotopy class. Similarly, as will be discussed further in the next paragraph,  one would expect states to be orthogonal when they have distinct such conformal geometries. One therefore expects that one can assign a probability to each possible conformal geometry in this context, and that the full path integral is given by integrating over such conformal geometries in analogy with \eqref{eq:intoverA}.  (As we will see below, it will be convenient to take the slicing prescription of Assumption \ref{as:add} to be defined by minimal surfaces.)

Here we have restricted discussion of the metric on the minimal surface to conformal equivalence classes of geometries since requiring the surface to be minimal is a form of gauge-fixing.  After such gauge-fixing, the full induced metric will not form a set of commuting observables.  Instead, one component of the induced metric becomes a function of the other coordinates and momenta by solving the Hamiltonian constraint.  Since it is canonically conjugate to the trace of the extrinsic curvature (which has been fixed to zero), one solves this constraint for the conformal factor of the induced metric.  The remaining conformal geometry on the minimal surface continues to define a set of commuting observables.

The above comments then motivate the following assumption, which is a generalization of Assumption \ref{as:min}:

\begin{assumption}
\label{as:disc}
Let ${\cal K}_{\tilde M, A}$ be the class of spacetimes that i) satisfy asymptotic boundary conditions specified by $\tilde {M}$, ii) satisfy fixed-area boundary conditions at $A$,  iii) have fields that are continuous everywhere, and iv) satisfy the conditions to be a fixed-area saddle everywhere except on a single codimension-1 minimal surface $\Sigma$ anchored both to the fixed-area surface and to some cut of the boundary.  As usual, we restrict to the case where the sources on $\tilde M$ are real.  In this case we assume that the fixed-area action on ${\cal K}_{\tilde M, A}$ is minimized by real saddles; i.e., every $k \in {\cal K}_{\tilde M, A}$ has action equal to or greater than that of some real saddle $k_s \in {\cal K}_{\tilde M, A}$ of the fixed-area action $\tilde I_A$.
\end{assumption}
Note that minimal surfaces in Euclidean signature are often well-defined even when the spacetime is not smooth.  We assume that this is the case in the above context though, if needed, we could further elaborate on this definition by requiring the spacetime to be built from smooth spacetimes using cut-and-paste along minimal surfaces.

Now, as discussed recently in both \cite{Stanford:2022fdt,Penington:2023dql} in the context of JT gravty, there are various subtleties and possible choices involved in using minimal surfaces to construct observables (or, equivalently in the language of those references, to fully fix a gauge in the Euclidean path integral).  Such subtleties may in the end require further refinements to assumption \ref{as:disc}.  But it is also plausible that such issues are not important at the level of our current discussion.  We have thus formulated assumption \ref{as:disc} without taking such issues into account.  Similarly, while minimal surfaces are smooth when the bulk spacetime dimension satisfies $D\le 7$ \cite{Fdrer1970TheSS}, they can be singular for $D \ge 8$ \cite{Simons1968MinimalVI}.  This is another reason why a useful conjecture for $D\ge 8$ could require further modification and/or additional work to describe a useful notion of the Einstein-Hilbert action on ${\cal K}_{\tilde M, A}$.  However, at least for $10 \ge D\ge8$ it turns out that such singularities are non-generic \cite{Li2020GenericRO,Chodosh2023GenericRF}.

\subsubsection{Aside:  A new positive action conjecture}

We now make a small digression in order to gather further supporting evidence for Assumption \ref{as:disc}.  One may note that a particular consequence of Assumption  \ref{as:disc} is a new positive action conjecture. We use the term conjecture in order to indicate both that it is not to be added to the existing list of assumptions needed to derive our trace inequality, and also that it may be more amenable to study in the near future than are the assumptions listed above.
\begin{conjecture}
\label{as:posS}
Let ${\cal K}_{\tilde M, A}$ be as in Assumption \ref{as:disc}.  Then the fixed-area action $\tilde I_A$ is bounded below on ${\cal K}_{\tilde M, A}$.
\end{conjecture}
As will be explained in section \ref{sec:posI}, Conjecture \ref{as:posS} is equivalent to what may naturally be called a positive-action conjecture for gravitational wavefunctions.

While Conjecture \ref{as:posS} remains to be proven for general theories, we show in appendix \ref{app:JT} that the analogous holds for JT gravity with dilaton-free couplings to matter.  In JT gravity, the value of the dilaton at a point plays the same role as the area of a codimension-2 surface in higher dimensional gravitational theories; see e.g \cite{Maldacena:2016upp,Harlow:2018tqv}.  Furthermore, in the same way that we might fix the area of a disconnected codimension-2 surface in higher dimensions, in JT we should allow the specification of a fixed-dilaton (fixed-$\phi$) set, in the sense that we fix the sum $\phi_{total} = \sum_i \phi_i$ where $\phi_i$ are the dilaton values at each of the singular points.  In discussing JT gravity we thus write ${\cal K}_{\tilde {M}, \phi_{total}}$ instead of ${\cal K}_{\tilde M, A}$.

The important point in the JT argument is that, since the bulk spacetime dimension is $D=2$, the minimality of a codimension-1 surface $\Sigma$ implies that its extrinsic curvature tensor vanishes.  As a result, for any $k \in {\cal K}_{\tilde M, A}$ the spacetime metric is in fact $C^1$ (except at the fixed-dilaton conical singularity) and the Ricci scalar cannot contain codimension-1 delta-functions localized on $\Sigma$.  Since $R=-2$ on each side of $\Sigma$, we then find $R=-2$ on $\Sigma$ as well (again, except at the fixed-dilaton conical singularity).
Furthermore,  as reviewed in appendix \ref{subsec:JTSchwarz}, if we ignore the conical singularities then the JT action on
${\cal K}_{\tilde {M}, \phi_{total}}$ would be given by the Schwarzian action.  It would thus be bounded below by the analysis of appendix \ref{subsec:JTpos}.

But it is also easy to include the contribution from the conical singularities.  This  is just $I_{sing} = -2 \pi \sum_i \phi_i \delta_i$, where $\delta_i$ is the conical deficit at each singularity.  Conical excesses are also allowed, but those are just deficits with $\delta_i < 0$.  Since each deficit must satisfy $\delta_i \le 2\pi$ we have the bound\
\begin{equation}
I_{sing} \ge -4 \pi \sum_i \phi_i  = - 4 \pi \phi_{total}.
\end{equation}
Combining this with the bound on the Schwarzian action and the positivity of the matter action then establishes Conjecture \ref{as:posS} in this context.

\subsubsection{Assumptions 5 \& 6 are sufficient}

Having motivated Assumptions  \ref{as:fixed} and \ref{as:disc}, we now turn to the issue of showing that -- together with known results for JT and Einstein-Hilbert gravity -- they imply Assumptions \ref{as:min}-\ref{as:strictmono}.  As a result, they also imply the desired bulk dual \eqref{eq:trineq3} of the trace inequality.  Let us first note that Assumptions \ref{as:fixed} and \ref{as:disc} immediately imply Assumption \ref{as:min}, as they were designed to do.  Furthermore, assumptions \ref{as:mono} and \ref{as:strictmono} were already shown to be true for JT and Einstein-Hilbert gravity by the discussion of section \ref{subsec:main} (between \eqref{eq:mono} and \eqref{eq:strictmono}).

This then leaves only Assumption \ref{as:add}.    There are two parts to showing that this assumption holds.  The first is to specify a cutting rule for Assumption \ref{as:add} so that pasting the various pieces together yields a spacetime that satisfies the desired asymptotic boundary conditions and, in particular, for which the new spacetime has a sufficiently smooth asymptotic boundary for which the action is finite and for which the action also defines a good variational principle. As noted above,
for either JT or Einstein-Hilbert gravity, it will be convenient to take the cuts to be minimal surfaces.  For JT gravity, appendix \ref{subsec:JTBC} then shows that the associated cut-and-paste construction preserves the boundary conditions of the variational principle (which were also shown to imply finiteness in appendix \ref{subsec:JTSchwarz}).  For Einstein-Hilbert gravity, this property is even easier to verify and is established in appendix \ref{app:cd1minsurface}.

The second task is then to establish the additivity property \eqref{eq:bulkadd}.   This is again straightforward for both JT and Einstein-Hilbert gravity.  The main point is that, since the boundaries we sew together are now taken to have $K=0$, there can be no delta-function contribution to the Ricci scalar at the seam where the sewing occurs\footnote{It is interesting to note that,  as described in appendix \ref{subsec:JTadd}, in JT gravity it turns out that even when the slicing surfaces are {\it not} minimal the delta-function in the Ricci scalar is such that it does not spoil \eqref{eq:bulkadd}.}.

Furthermore, the matter action is additive for the reasons explained in section \ref{sec:nongrav}.  It thus remains only to show additivity for the boundary terms at asymptotic boundaries.  One such term is always the Gibbons-Hawking term, while the rest are boundary counter-terms.  Due to the requirement that each operator have an appropriate ``rim'', the boundary metric and other analogous boundary conditions are manifestly smooth.  Thus the boundary counter terms are integrals of smooth functions and their additivity is also manifest.

\begin{figure}[h!]
	\centering
\includegraphics[width=0.7\linewidth]{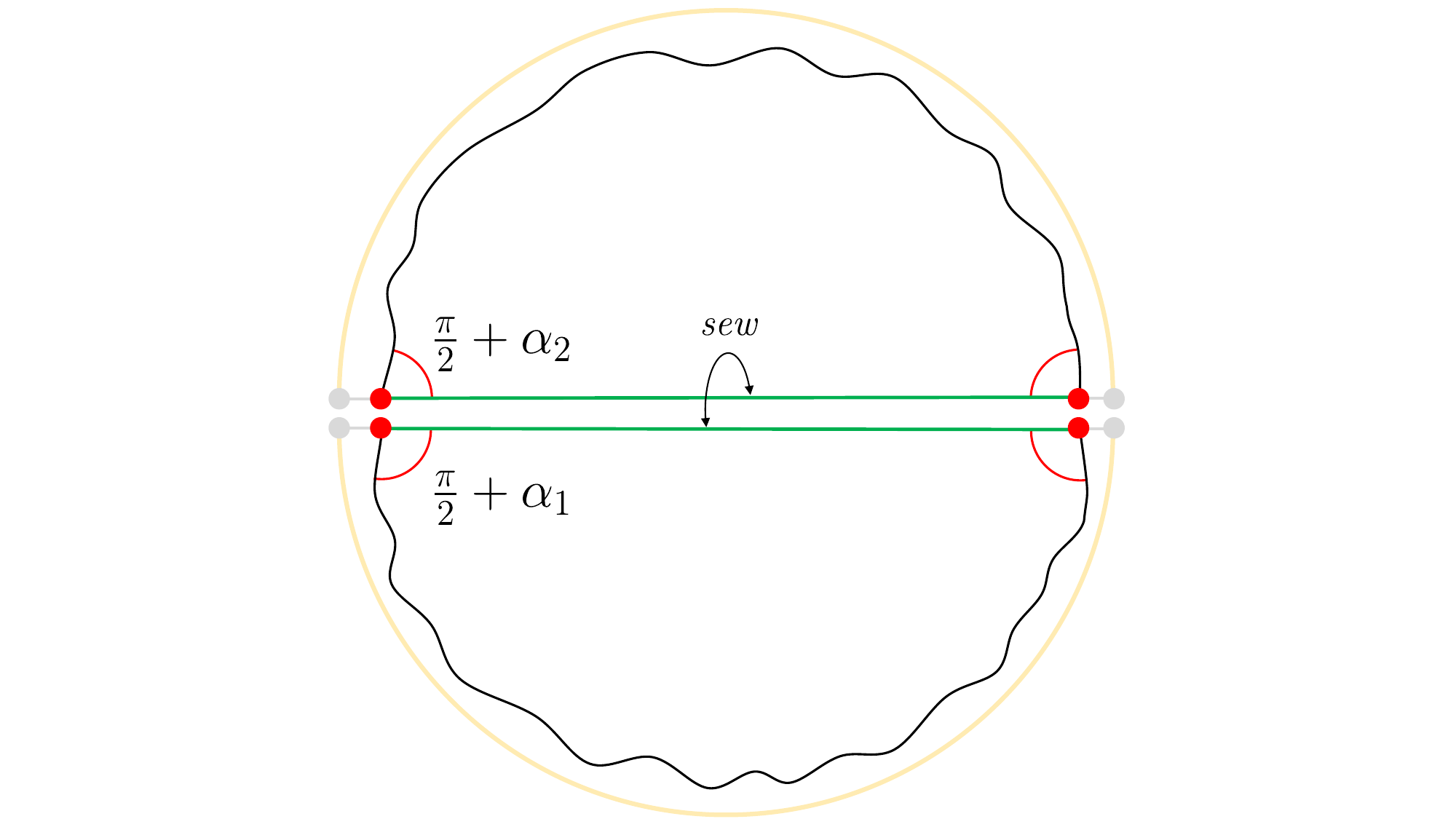}
\caption{Our cut-and-paste construction can join pieces of the asymptotic boundary together in a way that is not smooth.  We illustrate this here for a two-dimensional example (e.g., as appropriate to JT gravity).  In particular, when two pieces are sewn together,  two corners  (with, say, associated interior intersection angles $\pi/2+\alpha_1, \pi/2+\alpha_2$) of the individual pieces can merge. When this occurs, the extrinsic curvature density $\sqrt{h} K$ of the resulting $\partial {\cal M}$ will contain a delta-function of strength $\alpha_1 + \alpha_2.$ }
\label{fig:Kdelta}
\end{figure}

The final term to consider is then the Gibbons-Hawking term.  For a regulated version of the spacetime where the boundary has been moved inward to a finite value $\epsilon$ of the appropriate Fefferman-Graham coordinate (or of the defining function of the conformal frame in the language used for JT gravity in appendix \ref{app:JT}), the extrinsic curvature of the regulated boundary generally has a delta-function at the seam; see figure \ref{fig:Kdelta}.  But since the strength of this delta-function is always determined by the angles at which the asymptotic boundary meets the seam, we can render this part of the action additive by simply adding an appropriate `corner term' to the definition of the action for the cut space.

This procedure is discussed in great detail for JT gravity in appendix \ref{subsec:JTadd}.  The Einstein-Hilbert case is then discussed in appendix \ref{app:cd1minsurface}.  Appendix  \ref{app:cd1minsurface} in fact shows that, with the usual boundary conditions and with the choices we have made, in the $AlAdS_d$ Einstein-Hilbert case (with $d\ge 3$) the above delta-function turns out to make no contribution to the action in the limit $\epsilon \rightarrow 0$.  Thus the corner terms also vanish in the $\epsilon \rightarrow 0$ limit and are not strictly needed in the Einstein-Hilbert case.  This concludes the argument for both JT and Einstein-Hilbert gravity that Assumptions \ref{as:fixed} and \ref{as:disc} imply Assumptions \ref{as:min}-\ref{as:strictmono}, and thus also the desired result \eqref{eq:trineq3}.

\section{On a positive-action conjecture for quantum gravity wavefunctions}
\label{sec:posI}

As a slight aside from the main discussion, this section elaborates further on the status of Conjecture \ref{as:posS} in Einstein-Hilbert and other theories of gravity.  We will first describe how it is equivalent to what is naturally called a positive-action conjecture for quantum gravity wavefunctions.  We then point out that, at least when the the surface $\Sigma$ is a slice of a foliation of the bulk spacetime that is smooth away from $\Sigma$ and which also smoothly foliates a compact AlAdS boundary, the
conjecture is implied by the requirement that the gravitational Hamiltonian $H$ is bounded-below.  While the known asymptotically AdS positive energy theorems \cite{Abbott:1981ff,Gibbons:1983aq,Horowitz:1998ha} are not sufficient to prove positivity of $H$ at this level, the connection nevertheless provides additional physical reasons to believe that Conjecture \ref{as:posS} will hold.

\begin{figure}[h!]
	\centering
\includegraphics[width=0.7\linewidth]{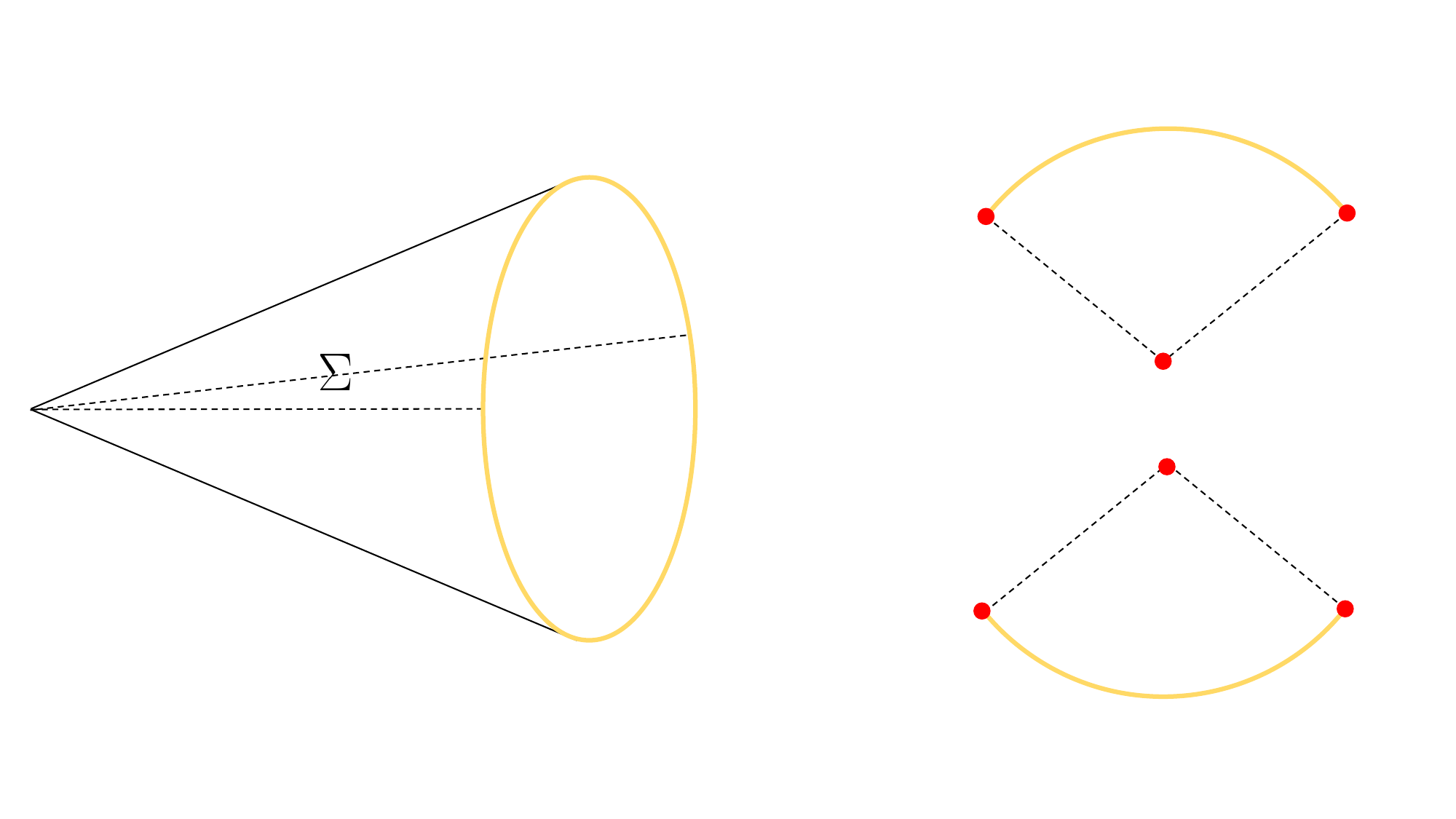}
\caption{A cone is shown along with a minimal surface $\Sigma$ anchored both to the conical defect and to the asymptotic boundary.  Slicing the cone open along $\Sigma$ gives a spacetime without conical defects, but where the new boundary may contain `corners' (represented a red dots) at which the extrinsic curvature contains delta-functions.}
\label{fig:nocone}
\end{figure}

Recall that Conjecture \ref{as:posS} referred to bulk spacetimes ${\cal M}$ which have only asymptotic boundaries, but which may contain a surface $\Sigma$ on which derivatives of fields are not continuous.  Furthemore, away from $\Sigma$ the Euclidean equations of motion are satisfied.   As a result, if we cut the spacetime along $\Sigma$ then each piece gives a smooth extremum of the standard Euclidean action so long as an appropriate boundary term is included at the cutting surface $\Sigma$ and corresponding boundary conditions are imposed at $\Sigma$.  In particular, even though ${\cal M}$ may have a conical singularity, the resulting pieces do not (though the boundaries of these pieces at $\Sigma$ may contain `corners' at which the extrinsic curvature of $\Sigma$ contains delta-functions; see figure \ref{fig:nocone}).

 For Einstein-Hilbert gravity, it is convenient to take this boundary term to be a Gibbons-Hawking term at $\Sigma$, and thus to think of the resulting action as defining a variational principle for the space of configurations defined by fixing the induced metric on $\Sigma$ to its value in ${\cal M}$.  Note that such pieces are precisely the spacetimes that appear as saddles in Euclidean path integral computations of gravitational wavefunctions in the induced-metric representation, where for a $(d+1)$-dimensional bulk, one thinks of the state as a functional $\Psi(h^{(d)})$ of the $d$-dimensional (Riemannian-signature) metric induced on a Cauchy surface.  Furthermore, the induced metric $h^{(d)}$ on $\Sigma$ that minimizes this action will correspond to the peak of that wavefunction, which one expects to be finite in the semiclassical limit.  As a result, one could also motivate Assumption \ref{as:posS} from the belief that semiclassical Euclidean path integral calculations of such wavefunctions should give sensible answers.

Now, much as in the discussion of section \ref{subsec:main}, if $\Sigma$ divides ${\cal M}$ into two pieces ${\cal M}_1$ and ${\cal M}_2$, we can also glue ${\cal M}_1$ to a reflected manifold ${\cal M}_1^R$ with the same action to define a new member ${\cal M}_1^{double}$ of an appropriate space ${\cal K}_{\tilde M, A}$.
Moreover, as argued in section \ref{sec:asstatus}, Einstein-Hilbert gravity satisfies the additivity condition \eqref{eq:bulkadd}, so that
\begin{equation}
I({\cal M}_1^{double}) =  2I({\cal M}_1).
\end{equation}
In particular, having a lower bound for  $I({\cal M}_1^{double})$ is equivalent to having a lower bound for the action $I({\cal M}_1) $ of the piece ${\cal M}_1$.  Furthermore, using
\begin{equation}
I({\cal M}) =  I({\cal M}_1)+ I({\cal M}_2),
\end{equation}
 having a uniform lower bound for all such pieces would imply a lower bound for the action on ${\cal K}_{\tilde M, A}$.

In fact, we can also drop the requirement that $\Sigma$ be minimal.  The reason for this is explained in detail for JT gravity in appendix \ref{subsec:JTSchwarz}, though it holds equally well for general theories of gravity.
As described there, if one thinks of each piece ${\cal M}_1, {\cal M}_2$  as being part of a larger saddle that extends beyond $\Sigma$, then the Hamiltonian constraint requires the on-shell action to be invariant under continuous deformations of $\Sigma$ that do not move the anchor set $\partial \Sigma$ on the asymptotic boundary.  More specifically, the previous statement is true so long as the action includes sufficient `corner terms' so that it defines a good variational principle under boundary conditions for $\Sigma$ consistent with the desired deformation.

It thus follows that Conjecture \eqref{as:posS} is in fact equivalent to the following conjecture which, due to the above-mentioned connection with $\Psi(h^{(d)})$, we call the positive-action conjecture for quantum gravity wavefunctions:

\begin{conjecture}
\label{conj:posS}
Consider the space of smooth Euclidean spacetimes having both an Asymptotically locally AdS (AlAdS) boundary (associated with some cosmological constant $\Lambda <0$) and an additional finite-distance boundary at some surface $\Sigma$.  In the usual way, we use a Fefferman-Graham expansion to fix a `boundary metric' at the AlAdS boundary. We also impose some class of boundary conditions at the `corners' where $\Sigma$ meets the AlAdS boundary.  We require that
the boundary conditions allow $\Sigma$ to be deformed to a minimal surface.  Note, however, that we impose no boundary conditions on $\Sigma$.

Let us now further restrict to such spacetimes that solve the vacuum Euclidean Einstein equations with cosmological constant $\Lambda$.  On such solutions we consider the Euclidean Einstein-Hilbert action with cosmological constant $\Lambda$, together with the standard Gibbons-Hawking term on all smooth parts of the boundary, the standard boundary counter-terms on the AlAdS boundary, and
`corner terms' appropriate to the above-chosen boundary conditions at the corners. For fixed such boundary conditions we require the above action functional is bounded below.
\end{conjecture}
Here the qualification that the corner boundary conditions must allow $\Sigma$ to be deformed to a minimal surface is needed in order to preserve equivalence with Conjecture \ref{as:posS}, but is not obviously critical for the existence of a lower bound.  We also emphasize that we have {\it not} fixed a particular induced metric on $\Sigma$, so that our lower bound is required to be independent of that induced metric.  Furthermore, if Conjecture \ref{conj:posS} holds then one can clearly also couple the system to positive-action matter with similar results.

This conjecture generalizes Hawking's original positive action conjecture \cite{Gibbons:1978ac} from asymptotically flat to AlAdS spacetimes, and also by introducing the finite boundary $\Sigma$ (appropriate to thinking of the spacetime as a Euclidean saddle for $\Psi(h^{(d)})$ instead of a partition function).  Such generalizations are natural in the spirit of the original conjecture.  The above conjecture is also weaker than that of \cite{Gibbons:1978ac} in the sense that we require only that each set of boundary conditions lead to a lower bound, but we allow this lower bound to depend on the choice of boundary conditions and, in particular, we allow the possibility that for some boundary conditions the greatest lower bound is less than zero.

The positive action conjecture for asymptotically Euclidean spacetimes was proven by realizing that the Euclidean action in that context is equal to the Hamiltonian for a higher-dimensional Lorentzian-signature theory of gravity evaluated on a Riemannian-signature Cauchy surface \cite{cmp/1103905050}.  This trick fails in the asymptotically AdS context, so a new proof strategy is needed.

While it is unclear to us how to give a complete proof in general, there is a simple context in which the conjecture follows from having a lower bound for the gravitational Hamiltonian.  To see the connection, consider a bulk spacetime ${\cal M}$ subject to boundary conditions as stated in the conjecture, and suppose that ${\cal M}$ admits a smooth foliation such that  $\Sigma = \Sigma_1 \cup \Sigma_2$ with $\Sigma_1$ diffeomorphic to $\Sigma_2$ and with both $\Sigma_1, \Sigma_2$ being limiting cases of the slices in the foliation.  We will refer to $\Sigma_1$ as the time $t_1$ and to $\Sigma_2$ as the time $t_2$ with slices in the foliation labeled by $t \in (t_1,t_2)$.  Consider then the action $I_{[t_1,t]}$ of for the region defined by slices with $t \in [t_1,t]$. Clearly the zero-volume region $[t_1,t_1]$ has  $I_{[t_1,t_1]}=0$.  Furthermore,  the usual Hamilton-Jacobi argument gives $\partial_{t_2} I_{[t_1,t_2]} =  H(t_2)$, where $H(t_2)$ is the standard (time-dependent) gravitational Hamiltonian defined by the boundary conditions on the asymptotic boundary and evaluated on the initial data defined by the surface $\Sigma_{t_2}$.  As a result, if the boundary Hamiltonian $H(t)$ has a $t$-independent lower bound $E_0$, then the action will satisfy
\begin{equation}
I_{[t_1,t_2]} \ge \left(|t_2-t_1|\right) \  E_0.
\end{equation}
For any given $t$ it is natural to believe the corresponding $H(t)$ to be bounded below so, since we consider a case in which the range of $t$ is compact, it is also natural to expect this lower bound to be uniform\footnote{However, it should be noted that  due to issues related to footnote \ref{foot:Trev} in section \ref{sec:nongrav}, the boundary conditions on certain slices of a Euclidean solution may not be simply related to boundary conditions for any Lorentz-signature gravitational Hamiltonian.  Stability of the Lorentz-signature theory is thus insufficient to motivate the above belief.  Furthermore, the known AlAdS positive-energy theorems address only a limited set of AlAdS boundary conditions.  Indeed, techniques that follow \cite{Witten:1981mf} are likely limited to contexts that allow supersymmetry.}.

However, there are many situations which are not of the above form.   Consider, for example, Euclidean AdS$_3$ in the conformal frame where the boundary metric is that of $S^2$.  Slicing the $S^2$ along surfaces of constant polar angle $\theta$, the boundary anchor sets $\partial \Sigma_t$ are then circles of time-dependent size that pinch off at the poles.   Furthermore, due to the Casimir energy of AdS$_3$ \cite{Balasubramanian:1999re, deHaro:2000vlm}, the lower bound on the Hamiltonian diverges as the size of the circle shrinks to zero.

While spherical AdS$_3$ gives an example where $H(t)$ has no uniform bound, it is nevertheless a context where the total action is finite and, moreover, where one very much expects the given spacetime to minimize the action.  We are therefore hopeful that further study of this example may suggest how the above sketch of a proof might be improved to deal with more general contexts.  We may also hope to learn to deal with the loci where the bulk topology forces the above foliations break down.  However, we leave such investigations for future work.

\section{Discussion}

\label{sec:disc}
The above work discussed the trace inequality
\begin{equation}
\label{eq:disctrin}
 \Tr_{\scriptscriptstyle {\cal D}}(BC) \le
 \Tr_{\scriptscriptstyle {\cal D}}(B) \Tr_{\scriptscriptstyle {\cal D}}(C),
\end{equation}
which applies to positive operators $B,C$ on any Hilbert space ${\cal H}$.   The symbol ${\cal D}$ denotes the non-gravitational CFT dual of a bulk theory, and we write $\Tr_{\scriptscriptstyle {\cal D}}$ to emphasize that the trace is the standard trace on the ${\cal D}$ side of the duality.  In particular,
$\Tr_{\scriptscriptstyle {\cal D}}$ denotes the familiar operation computed by introducing any orthonormal basis $|i\rangle$ on the ${\cal D}$ Hilbert space and performing the sum \eqref{eq:trdef}. Averaging over an ensemble gives
\begin{equation}
\label{eq:disctrin2}
\langle \Tr_{\scriptscriptstyle {\cal D}}(BC)\rangle \le
 \langle \Tr_{\scriptscriptstyle {\cal D}}(B) \Tr_{\scriptscriptstyle {\cal D}}(C) \rangle.
\end{equation}

Our goal was to understand the status of the above inequality on the bulk side of the AdS/CFT duality.  In particular, we studied the conjectured inequality
\begin{equation}
\label{eq:disctrineq3}
\zeta\left( \tilde M_{bc^\dagger c b^\dagger} \right) \le  \zeta\left( \tilde M_{b^\dagger b} \sqcup \tilde M_{c^\dagger c} \right),
\end{equation}
where $\zeta(\tilde M)$ denotes the gravitational path integral with boundary conditions given by the closed boundary source-manifold $\tilde M$ and  $\sqcup$ denotes disjoint union.  Here $\tilde M_{bc^\dagger c b^\dagger}$ is a smooth closed manifold specifying boundary conditions for our bulk theory on a Euclidean Asymptotically locally Anti-de Sitter boundary that can be broken into four pieces $M_b, M_{b^\dagger}, M_c, M_{c^\dagger}$.  Furthermore, we require that connecting $M_b$ and  $M_{b^\dagger}$ gives a new smooth closed manifold  $M_{bb^\dagger}$ that is invariant under a reflection-symmetry that exchanges the $b$ and $b^\dagger$ pieces (and which complex-conjugates any complex boundary conditions) and similarly for $M_{cc^\dagger}$.

At the level of the semiclassical expansion, and when the operators $B,C$ define Eucldean bulk path integrals with connected boundaries, we argued that the natural bulk dual of \eqref{eq:disctrineq3} was in fact satisfied to all orders in the semiclassical expansion in two important contexts.  The first is the case of JT gravity with a dilaton-free coupling to two-derivative matter, with the possible further addition of perturbative higher derivative terms.  The second is given by Einstein-Hilbert gravity minimally coupled to two-derivative matter, where again higher derivative terms can also be included perturbatively.

In all cases we assumed the bulk path integral defined by $\tilde M_{bc^\dagger c b^\dagger}$ (dual to $\langle \Tr_{\scriptscriptstyle {\cal D}}(BC)\rangle$) to be dominated by a single bulk saddle.  When several bulk saddles are equally dominant,  formally non-perturbative effects associated with additional saddles and/or mixing between saddles can be more important than perturbative corrections and are subtle to analyze; see e.g. recent discussions in \cite{Vidmar:2017pak,Murthy} for condensed matter analogues and in \cite{Penington:2019kki,Marolf:2020vsi,Dong:2020iod,Akers:2016ugt}.  We thus save further consideration of this case for future study.

For pure JT gravity, much can be said using explicit calculations based on standard Euclidean saddles.  In addition, the non-perturbative definition of the Euclidean path integral described by Saad, Shenker, and Stanford \cite{Saad:2019lba} can be used to give a general derivation of \eqref{eq:disctrineq3}.

For more general cases the Euclidean  path integral is sufficiently poorly understood that we cannot use the term ``proofs'' to refer to  our arguments.  Instead, we proceeded by stating various assumptions that we argued were plausibly true in regimes where a Euclidean gravitational path integral emerges from a more UV-complete theory.  In particular, we considered three rather distinct paradigms for such path integrals. One of these was the possible extension of the Saad-Shenker-Stanford approach mentioned above \cite{Saad:2019lba} to (some UV-completion of) JT gravity coupled to positive action matter.   Another paradigm involved a nonlinear generalization of the Gibbons-Hawking-Perry contour rotation prescription \cite{Gibbons:1978ac}. The third was the paradigm described in \cite{Marolf:2022ybi} that took a real-time formulation as fundamental but then described a procedure for transforming semiclassical computations into what were often sums over Euclidean saddles.   We argued that all three paradigms lead to a bulk version of \eqref{eq:disctrin} in the semiclassical context described above, though our arguments were based on a set of assumptions about the semiclassical limit of a supposed UV-complete theory of quantum gravity.

A significant restriction in our arguments was that we considered only boundary conditions which are real in Euclidean signature, and which thus cannot include Lorentzian components.  There is a sense in which extending our analysis of general gravitational theories to complex boundary conditions would be trivial, since we need only to suitably extend the various assumptions we made along the way.  This, however, would miss an important point that arises for both the non-gravitational path integrals studied in section \ref{sec:nongrav} and the JT case analyzed semiclassically in section \ref{subsec:pureJT}.  In particular, once the sources are complex, the relevant saddles will generally not lie on the original contour of integration.  In that context, the most direct analogue of the argument given here would attempt to show that cutting and pasting a valid complex saddle (through which the integration can be deformed to pass) for the left-hand-side of \eqref{eq:disctrineq3} yields a configuration $k$ that lies on the steepest descent curve  $\Gamma_{ds}$ through the dominant saddle for the right-hand-side (so that the dominant saddle then has lower action).  Since it is not at all clear to us why that $k$ should lie on the relevant $\Gamma_{sd}$, we have not attempted to formulate a gravitational argument in this language.  Instead, we leave further consideration of complex sources for future work.

While it may be difficult to verify our assumptions about the Euclidean path integral, such assumptions imply other properties of the {\it classical} Euclidean action that are more amenable to study in the near future, and which in particular might be investigated numerically.  The most tangible prediction is the positive action conjecture for gravitational wavefunctions described in section \ref{sec:posI}.  This conjecture generalizes Hawking's original positive action conjecture \cite{Gibbons:1978ac} to the AdS context, and also generalizes it further by allowing spacetimes with an extra finite-distance boundary.  However, we require only that the action be bounded below for each set of AlAdS bondary conditions, and not that the action be strictly non-negative.  As some evidence in support of this conjecture, we were able to prove that the corresponding result holds in the simpler case of JT gravity (with general dilaton-free couplings to positive-action matter).

Let us now return to the discussion of the bulk analogue \eqref{eq:disctrineq3} of the trace inequality \eqref{eq:disctrin} in a dual field theory.
The main physical lesson from our investigations appears to be that this inequality is closely associated with positivity of entropy in the sense of having a positive density of states.  To be more precise, we saw in various ways that the right-hand-side of \eqref{eq:disctrin} tends to be {\it much} larger than the left when the spectral densities of $B$ and $C$ are large.  This is manifest from the CFT-side argument surrounding equation \eqref{eq:Trinder}, as well as from the thermodynamic discussion in section \ref{subsec:pureJT}.  However, a corresponding feature also appeared in our gravitational arguments where in the Einstein-Hilbert case we found the left-hand-side to be suppressed relative to the right by a factor of $e^{-A/4G}$ associated with the area of an extremal surface.  Similarly, in the JT context with the action \eqref{eq:JTaction} we found a similar suppression by $e^{-4\pi \phi_0}$.  These expressions are readily recognized as being associated with the RT/HRT entropy \cite{Ryu:2006bv, Ryu:2006ef,Hubeny:2007xt} of a boundary region.  We thus see that, even if the trace inequality were violated in contexts where such entropies are small, the violations would disappear in the limit where these entropies are large.  In other words, we see that the large entropy regime is a relatively poor probe of whether a gravitational theory can admit a non-gravitational dual.

While the trace inequality \eqref{eq:disctrin} is a very weak constraint in the context familiar quantum mechanical operators, we noted in the introduction that it can have fundamental implications.  For example, it is deeply associated with the fact that the algebra of bounded operators on a Hilbert space is a type I von Neumann algebra.  We will return both to this connection and to the fundamental status of \eqref{eq:disctrin} in bulk gravitational theories in a forthcoming work \cite{VN}.

\section*{Acknowledgements}

DM thanks Clifford Johnson for discussion of JT gravity, Jorma Louko for conversations related to positive action conjecture for wavefunctions, and Gary Horowitz for discussions of existing positive action theorems. He also thanks the Perimeter Institute for its hospitality during the final stages of this work.   EC thanks Alexey Milekhin for conversations related to the trace inequality and positivity of entropy.
The work of DM and ZW was supported by NSF grant PHY-2107939, and by funds from the University of California.
ECs participation in this project was made possible by a DeBenectis Postdoctoral Fellowship and through the support of the ID\# 62312 grant from the John Templeton Foundation, as part of the \href{https://www.templeton.org/grant/the-quantuminformation-structure-ofspacetime-qiss-second-phase}{‘The Quantum Information Structure of Spacetime’ Project (QISS)}. The opinions expressed in this project/publication are those of the authors and do not necessarily reflect the views of the John Templeton Foundation.

\appendix
\section{Properties of the JT action}
\label{app:JT}

This appendix discusses a number of details regarding asymptotically AdS$_2$ Jackiw-Teitelboim gravity.  After defining the theory  by stating the action and boundary conditions in section \ref{subsec:JTBC}, additivity of the JT action in the sense of \eqref{eq:bulkadd} is shown in section \ref{subsec:JTadd}.  The relation to the Schwarzian action is then reviewed in section \ref{subsec:JTSchwarz}, and the Schwarzian form is shown to be bounded below in section \ref{subsec:JTpos}.

\subsection{Action and Boundary conditions for JT gravity}
\label{subsec:JTBC}

Much of the later analysis in this appendix will involve study of the boundary conditions for asymptotically-AdS$_2$ JT gravity.  The purpose of this section is to describe such boundary conditions in detail.  We consider here the pure JT gravity theory consisting of only a dilaton
$\phi$ and a metric $g$ on a 2d spacetime ${\cal M}$, without additional matter fields.    While our boundary conditions are just those of  e.g. \cite{Saad:2019lba} (which are the Euclidean versions of those of \cite{Maldacena:2016upp}), we take this opportunity to rewrite them in a form more similar to that commonly used to describe asymptotically locally Anti-de Sitter spacetimes in higher dimensional theories of gravity.  For use with our cut-and-paste constructions,  we also allow a slight extension of the usual boundary conditions to in which certain fields can be non-smooth on a codimension-1 surface. Below, we set the AdS$_2$ scale $\ell$ to $1$.

\begin{figure}[h!]
	\centering
\includegraphics[width=\linewidth]{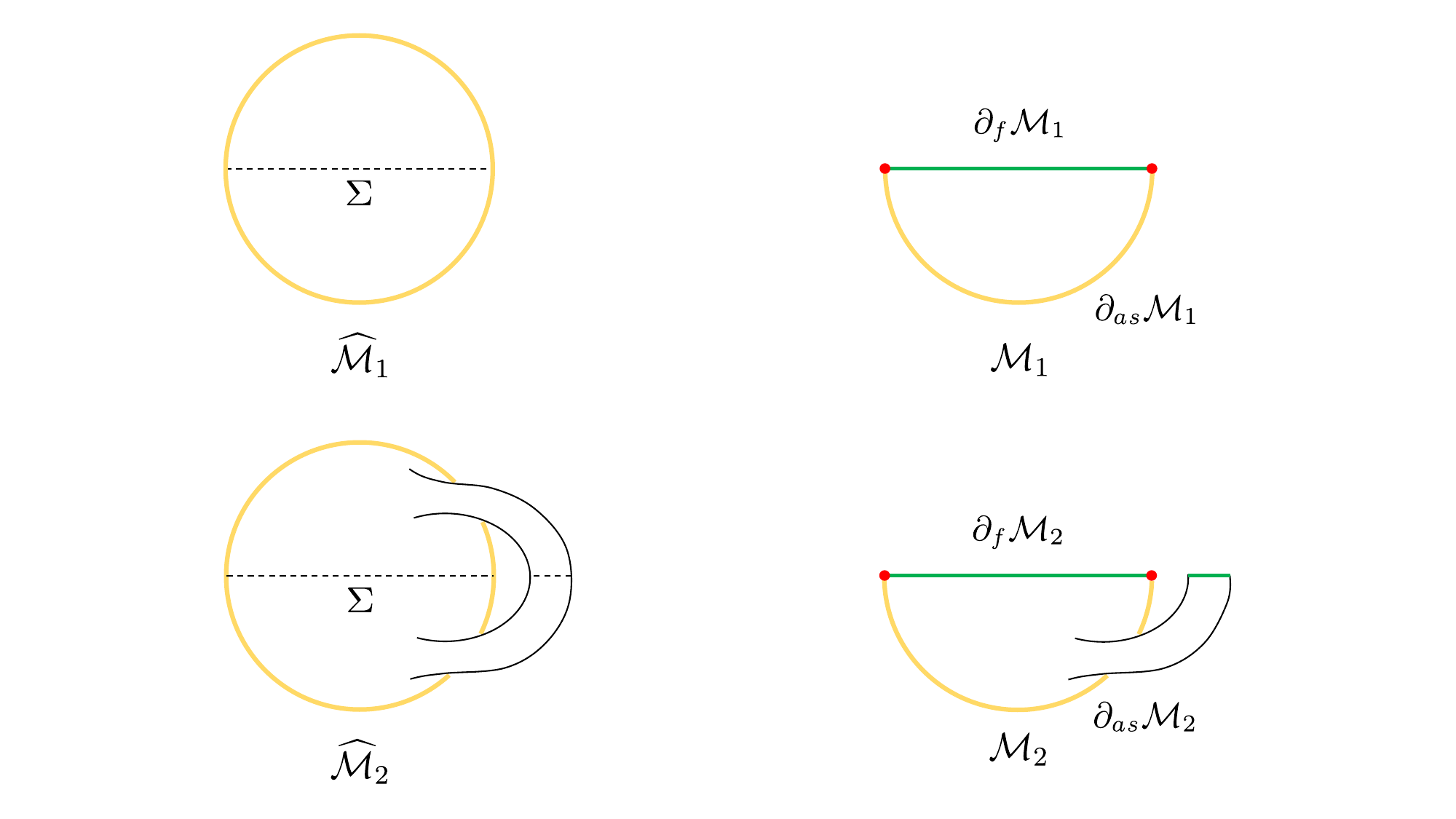}
\caption{Two examples of parent spaces $\widehat {\cal M}$ are shown at left.  These $\widehat {\cal M}$ have only asymptotic boundaries (yellow).  Cutting away the part beyond a surface $\Sigma$ gives a manifold with boundaries and corners that we call ${\cal M}$.  Such ${\cal M}$ generally still have asymptotic boundaries $\partial_{as}{\cal M}$ (yellow) inherited from $\widehat {\cal M}$, but also have finite boundaries $\partial_f {\cal M}$ (green) created by the cut. The two types of boundaries will generally meet at corners (red).}
\label{fig:surfaces}
\end{figure}

The first step in defining configurations of our theory is to consider 2-dimensional compact manifolds $\widehat {\cal M}$ with boundaries.  These are perhaps most simply defined as the spaces obtained from $S^2$ by i) removing $n$ open disks, which creates $n$ circular boundaries that we label $i=1,\dots,n$, ii) choosing $g< n/2$ and then for $i\le g$ identifying the $i$th circular boundary with the $(i+g)$th circular boundary, iii) perhaps filling in one of the remaining circular boundaries with a cross-cap in order to obtain the non-orientable cases.  Examples are shown in the left panel of figure \ref{fig:surfaces}.

However, in order to accommodate the cut-and-paste constructions described in the main text, we also consider certain 2d manifolds with corners at their boundaries.  In practice, it will be sufficient to define these by starting with one of the above 2d manifolds $\widehat {\cal M}$ with boundary (and without corners), choosing any smooth $1d$ surface $\Sigma$ in that manifold that divides $\widehat {\cal M}$ into two parts, and removing one of the parts.   We call what is left a manifold ${\cal M}$ with boundaries and corners; see the right panel of figure \ref{fig:surfaces}.

The boundary of the final ${\cal M}$ now consists of two types of segments.   The first type, whose union we we call the asymptotic boundary $\partial_{as} {\cal M}$, consists of those segments in $\partial {\cal M}$ which also lie on the boundary of the parent space $\widehat {\cal M}$ from which ${\cal M}$ was cut:  $\partial_{as} {\cal M} = \partial {\cal M} \cap \partial \widehat {\cal M}$.  The second type, whose union we call the finite boundary $\partial_f {\cal M}$, consists of the remainder which we see must in fact form the slicing surface $\Sigma$ (so that $\partial_f {\cal M} = \Sigma$).  In general, $\partial_{as} {\cal M}$ and $\partial_f {\cal M}$ will intersect at a finite number of points.

We wish to consider metrics $g$ on ${\cal M}$ which are the restriction of metrics $\hat g$ on the parent space $\widehat {\cal M}$ and which in some region near $\partial_{as} \widehat {\cal M}$ can be written in the form
\begin{equation}
\label{eq:asymptmetric}
{ds}^2 = \frac{dz^2 + h(\theta,z) d\theta^2}{z^2},
\end{equation}
where both the function $h$ and the coordinates $z, \theta$ are smooth on $\widehat {\cal M}$, where $z$ has a first-order zero at $\partial_{as}\widehat {\cal M}$, and where both $\partial_z h$ and $\partial_z^2 h$ vanish at $z=0$.  Similarly, we consider dilaton fields of the form
\begin{equation}
\label{eq:asymptdil}
\phi = \frac{f(\theta,z)}{z},
\end{equation}
with $f$ smooth on $\widehat {\cal M}$.  Note that, for a given metric and dilaton, the form of \eqref{eq:asymptmetric} and \eqref{eq:asymptdil} are preserved by appropriate smooth coordinate transformations which satisfy
\begin{eqnarray}
\label{eq:changez}
z &\rightarrow & a(\theta)z + O(z^3), \\
\theta &\rightarrow & b(\theta) + O(z^2),
\end{eqnarray}
and under which we find that the associated $h$ on the boundary transforms as $h|_{z=0} \rightarrow (h [\frac{1}{a}\frac{db}{d\theta}]^2)|_{z=0}.$
As a result, the coordinates $(z,\theta)$ that give the form \eqref{eq:asymptmetric} and \eqref{eq:asymptdil} are far from unique.

We also wish to impose further boundary conditions. To do so, we endow each of our manifolds with a special preferred scalar function $\Omega$ which will be used to translate between the physical metric and dilaton (which diverge at $\partial_{as} {\cal M}$) and an unphysical rescaled metric and dilaton that will be used to specify the additional boundary conditions. It is convenient to define $\Omega$ to be a function on the parent space $\widehat {\cal M}$ such that $\Omega$ vanishes on $\partial \widehat {\cal M}=\partial_{as} \widehat {\cal M}$ but has $d\Omega$ nowhere vanishing on $\partial \widehat {\cal M}= \partial_{as} \widehat {\cal M}$. We require $\Omega$ to be smooth on most of the spacetime, though -- for use with our cut-and-paste constructions -- we also allow the existence of a finite number of smooth codimension-1 surfaces $\Sigma$ on which $\Omega$ is continuous and limits of first derivatives from either side are well-defined, but where $d\Omega$ can have discontinuities across the surface.  The various such surfaces $\Sigma$ are allowed to intersect at a finite number of points.
We will refer to $\Omega$ as the defining function of the conformal frame, or simply as the conformal factor.  In terms of any given set of coordinates $(z, \theta)$ satisfying the conditions above, we may write
 \begin{equation}
 \Omega = z \omega( z, \theta)
 \end{equation}
 for some smooth positive function $\omega$ that does not vanish anywhere on $\partial \widehat {\cal M}=\partial_{as} \widehat {\cal M}$.
We then use $\Omega$ to define rescaled (unphysical) fields
 \begin{equation}
 \label{eq:unphysfields}
 d\tilde s^2 := \Omega^{2} {ds}^2,  \ \ \ {\rm and} \ \ \ \tilde \phi := \Omega \phi.
 \end{equation}
We may thus introduce a coordinate $u$ on each connected component of $\partial_{as} {\cal M}$ such that $u$ measures the unphysical proper distance defined by $d\tilde s^2$, and we may then require $\tilde \phi$ on $\partial_{as} {\cal M}$ to be some fixed function $\phi_b(u)$; i.e., we impose
\begin{equation}
\label{eq:UBCs}
\tilde ds^2\big|_{\partial_{as} {\cal M}} = du^2,  \ \ \  \tilde \phi|_{\partial_{as} {\cal M}} = \phi_b(u).
\end{equation}
In doing so, we must also fix the range of $u$ and thus the total (unphysical) length of the boundary. Indeed, the second statement in \eqref{eq:UBCs} is meaningful only if we also label each boundary point once-and-for-all with a value of $u$ as part of our boundary conditions.

We note for future reference that \eqref{eq:UBCs} implies
\begin{equation}
\label{eq:omega}
\omega^{-1}|_{z=0} = \sqrt{h|_{z=0}} \frac{d\theta}{du}.
\end{equation}
Since \eqref{eq:UBCs} are coordinate invariant, they are in particular preserved by coordinate transformations of the form \eqref{eq:changez}.

Finally, as a further boundary condition, on every piece of $\partial_f {\cal M}$ we require $\theta(z)$ to have an expansion of the form
\begin{equation}
\label{eq:partialf}
\theta = \theta_0 + \theta_2 z^2 + ...
\end{equation}
near $\partial_{as} {\cal M}$ in terms of coordinates in which the metric takes the form \eqref{eq:asymptmetric}. Since the $O(z)$ term in \eqref{eq:partialf} vanishes, this in particular requires all intersections between the finite and asymptotic boundaries to be orthogonal as defined by the unphysical metric $\tilde g$. Note that this condition is again preserved by coordinate transformations of the form \eqref{eq:changez}.

This completes our discussion of (asymptotic) boundary conditions for $JT$ gravity.  Note that, since we allowed discontinuities in $d\Omega$, these boundary conditions are manifestly invariant under the cut-and-paste construction associated with Assumption \ref{as:add} of section \ref{subsec:main}.

  %Along each component of $\partial_{as} {\cal M}$ we then introduce a normalized coordinate $\theta = 2\pi u/\beta$ and use it to build a Gaussian normal coordinate system by firing geodesics orthogonally inward from $\partial_{as} {\cal M}$.  If we call the associated Gaussian normal coordinate $\tilde z$ and choose $\partial_{as} {\cal M}$ to be the surface $\tilde z=0$, then in some neighborhood of $\partial_{as} {\cal M}$ we have with $f$ smooth and $f=1$ at $\tilde z=0$.  Since the intersections of $\partial_{as} {\cal M}$ and $\partial_f {\cal M}$ are orthogonal, at $\partial_{as} {\cal M}$ every component of $\partial_f {\cal M}$ becomes tangent to some constant $\theta$ curve.

%We will take the parent space $\widehat {\cal M}$ (and thus also ${\cal M}$) to be equipped with a preferred smooth positive scalar function $\Omega$, which we call the defining function of the conformal frame, or simply the conformal factor for short. We require $\Omega$ to have a first order zero at $\partial_{as} \widehat {\cal M}$, so that $\Omega$ vanishes there but $d\Omega$ does not.  As a result, in terms of the coordinates $\tilde z, \theta$ we must have

\subsection{Additivity and the JT action}
\label{subsec:JTadd}

Having stated the dilaton and metric boundary conditions above, we can now proceed to discuss the JT action.  Here we focus on the additivity property \eqref{eq:bulkadd}.  We now include possible (dilaton-free) couplings to matter, meaning that the dilaton should not appear in the matter action.  However, we will not spell out the details of the matter action or boundary conditions.  We will simply (and implicitly) assume that the matter action for a given metric is of the form required for non-gravitational systems in section \ref{sec:nongrav} (so that the matter action is separately additive), and that the matter fields fall-off sufficiently quickly at infinity that they do not affect the leading behavior of the dilaton given in \eqref{eq:UBCs} even when the equations of motion are satisfied.

\begin{figure}[h!]
	\centering
\includegraphics[width=0.7\linewidth]{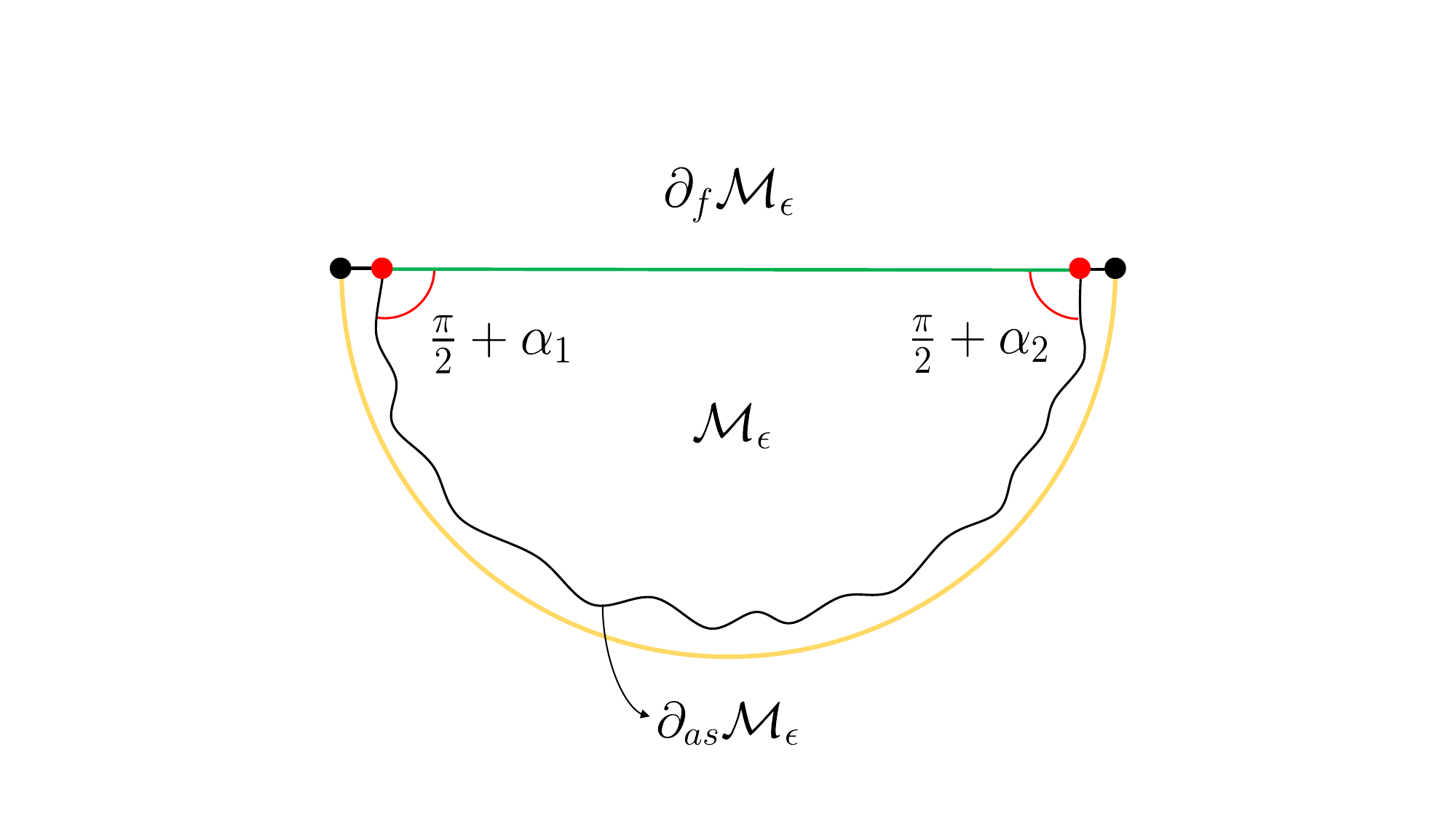}
\caption{A regulated manifold ${\cal M}_\epsilon$ is shown with its finite boundary $\partial_f {\cal M}_\epsilon$ and asymptotic boundary $\partial_{as}{\cal M}_\epsilon$ intersecting at angles $\pi/2+\alpha_1$ and $\pi/2+\alpha_2$.}
\label{fig:Mepsilon}
\end{figure}

Since the metric and dilaton diverge at the asymptotic boundaries $z=0$, the JT action will be defined as the $\epsilon \rightarrow 0$ limit of actions for regulated manifolds ${\cal M}_\epsilon$, each given by the region of ${\cal M}$ with $\Omega \ge \epsilon$ for some $\epsilon > 0$.  The boundary $\partial
{\cal M}_\epsilon$ of the regulated spacetime can again be decomposed into two parts, $\partial_f {\cal M}_\epsilon$ and $\partial_{as} {\cal M}_\epsilon$, the first of which is just the part of
$\partial_f {\cal M}$ that remains in ${\cal M}_\epsilon$, and the second is the closure of the remainder:
\begin{equation}
 \partial_f {\cal M}_\epsilon := {\cal M}_\epsilon \cap \partial_f {\cal M}, \ \ \ {\rm and} \ \ \
 \partial_{as} {\cal M}_\epsilon : = \overline{\partial {\cal M}_\epsilon \setminus \partial_f {\cal M}_\epsilon }.
\end{equation}
Again, the two parts generally intersect in a finite number of points, though the intersections are no longer strictly orthogonal at finite $\epsilon$.  For a given intersection point $i$, we thus let $\pi/2+ \alpha_i>0$ denote the (interior) angle at which the finite and asymptotic boundaries  meet; see figure \ref{fig:Mepsilon}.

We now take the action to be $I: = \lim_{\epsilon \rightarrow 0} I_\epsilon$ with
\begin{eqnarray}
\label{eq:JTe}
I_\epsilon &=& - \phi_0 \left[ \int_{{\cal M}_{\epsilon}} \sqrt{g} R + 2 \int_{\partial_{as} {\cal M}_{\epsilon}} \sqrt{h} K + 2 \int_{\partial_{f} {\cal M}_{\epsilon}} \sqrt{h} K - 2\sum_i \alpha_i\right] \cr
&-& \left[ \int_{{\cal M}_{\epsilon}} \sqrt{g} \phi (R+2)   + 2 \int_{\partial_{as} {\cal M}_{\epsilon}} \sqrt{h} \phi (K-1)
+ 2 \int_{\partial_{f} {\cal M}_{\epsilon}} \sqrt{h} \phi K - 2\sum_i \alpha_i \phi_i \right] \cr &+&I_{matter}.\ \ \ \ \ \
\end{eqnarray}
Here $\phi_0$ is a constant, $h$ is the induced metric on a boundary, $K$ is the extrinsic curvature (a scalar, since the boundary is one-dimensional) defined by the outward-pointing normal,  $i$ ranges over all points where $\partial_{f} {\cal M}_{\epsilon}$ meets $\partial_{as} {\cal M}_{\epsilon}$, and $\phi_i$ are the values of $\phi$ at such meeting points.
Note that, in the first line, we have included separate terms at $\partial_{as} {\cal M}_{\epsilon}$ and $\partial_{f} {\cal M}_{\epsilon}$, neither of which will include effects from corners where they intersect.  The natural delta-function contributions from $K$ at corners have instead been written explicitly in terms of the $\alpha_i$ (up a $\pi/2$ offset for each corner that we now discuss).

The usual calculation then shows \eqref{eq:JTe} to be a good variational principle  when the induced metric and dilaton are fixed on the finite boundaries $\partial_f {\cal M}$ and the boundary conditions of \eqref{subsec:JTBC} are imposed at the asymptotic boundary, and of course when boundary conditions appropriate to $I_{matter}$ are imposed on matter fields.  In particular, while the fact that we allowed $d\Omega$ to be discontinuous across a surface introduces delta-functions in the extrinsic curvature of ${\cal M}_{\epsilon}$ at some values of $\theta$, these delta-functions give finite results when integrated over $\theta$. The discontinuities in $d\Omega$ then have no further impact on the computation. In particular, they do not change any powers of $\epsilon$.

Since we require the matter action to be dilaton-free, the equation of motion obtained by varying $\phi$ in \eqref{eq:JTe} is just $R+2=0$ for any allowed matter.  In particular, the matter field can thus have no effect on the asymptotics of the metric.  This also means that the only positivity property of the matter that we will need is that  $I_{matter}$ be bounded below (say, by zero) for any asymptotically AdS$_2$ constant curvature $R=-2$ Euclidean metric $g$.

We now make a number of comments about the action \eqref{eq:JTe}, in particular regarding its additivity properties.  It is convenient to begin by discussing the first line in \eqref{eq:JTe}, which turns out to purely topological.  Let us denote these terms by $I_{top}$. Since the interior angles at each intersection point $i$ are $\pi/2 +\alpha_i$, the Gauss-Bonnet theorem requires
\begin{equation}
\label{eq:top}
I_{top} = - 4\pi \phi_0 \chi + \pi \phi_0\sum_i(1) = -\pi \phi_0 (4\chi - n_{int} ),
\end{equation}
where $\chi$ is the Euler character of ${\cal M}_{\epsilon}$ and $n_{int}$ is the number of points where $\partial_{f} {\cal M}_{\epsilon}$ and $\partial_{as} {\cal M}_{\epsilon}$ intersect.  The  $\epsilon$-dependence of $I_{top}$ is manifestly trivial, so we will drop $\epsilon$ labels when discussing it below.

Furthermore, given disjoint configurations ${\cal M}_1$ and ${\cal M}_2$,  we see that $I_{top}$ satisfies
\begin{equation}
\label{eq:Itopdu}
I_{top} ({\cal M}_1 \sqcup {\cal M}_2) = I_{top} ({\cal M}_1) + I_{top} ({\cal M}_2).
\end{equation}
Here we use ${\cal M}$ to denote both the underlying manifold with boundaries and corners and the fields carried by that manifold.  We will continue this abuse of notation below.  The symbol $\sqcup$ denotes disjoint union.

\begin{figure}[h!]
	\centering
\includegraphics[width=\linewidth]{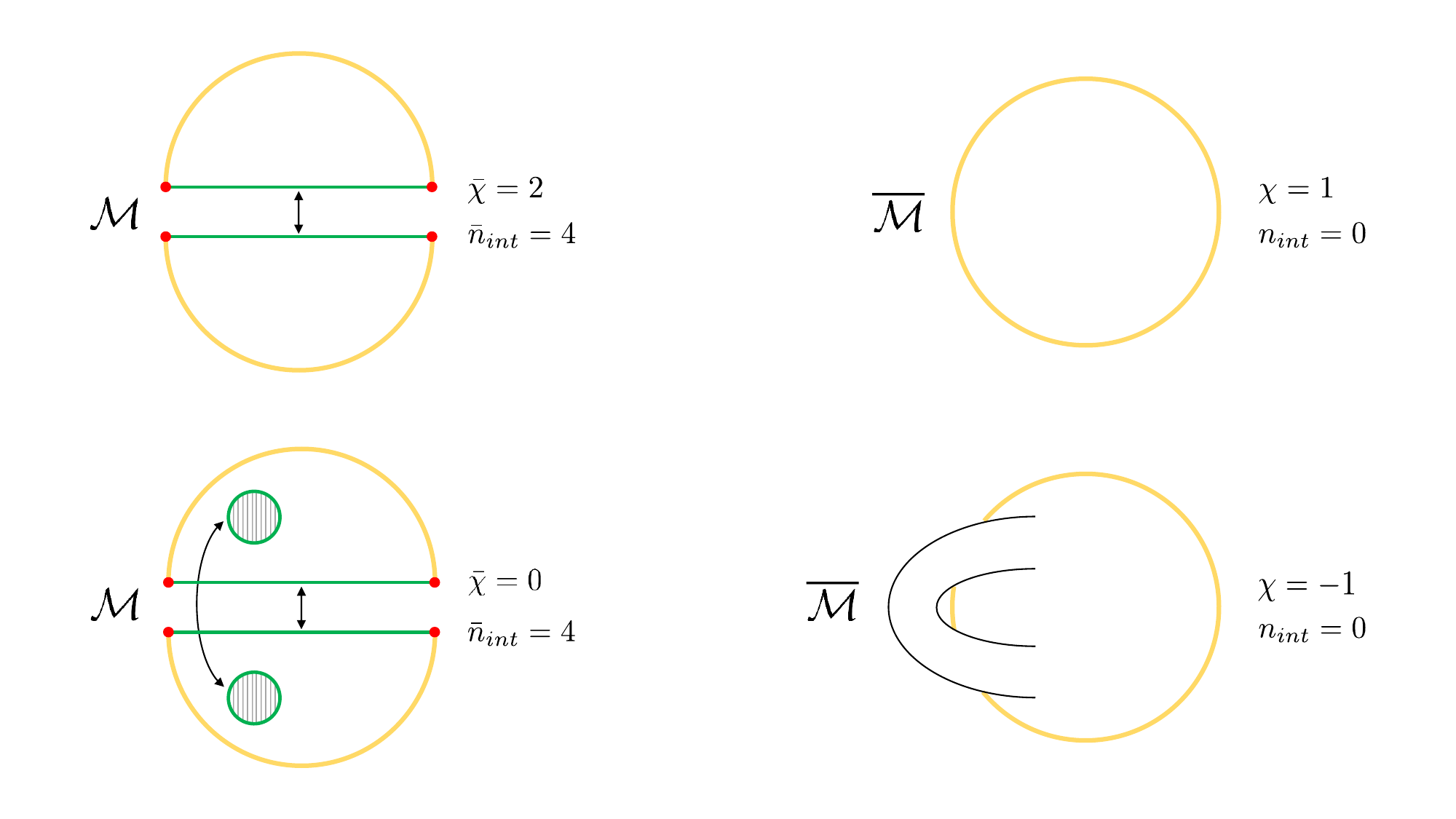}
\caption{ Examples of the gluing operation that constructs a new configuration $\overline {\cal M}$ from ${\cal M}$.  While the gluing changes both $\chi$ and $n_{int}$, it does not change $4 \bar \chi - \bar n_{int}$.  Indeed,  we find  $4 \bar \chi - \bar n_{int} = 4 \chi - n_{int}=4$ on the top line and  $4 \bar \chi - \bar n_{int} = 4 \chi - n_{int}=-4$ on the bottom. The figure shows simple examples in which all finite boundaries disappear from $\overline {\cal M}$, though this is not generic.}
\label{fig:Mbar}
\end{figure}

The term $I_{top}$ also satisfies a second more interesting identity.  To describe this identity, consider a configuration ${\cal M}$ for which $\partial_f {\cal M}$ has $n_f$ connected components $\partial_{f,j} {\cal M}$ for $j = 1, \dots n_f.$  We wish to form a new configuration $\overline {\cal M}$ by identifying pairs of components $\partial_{f,j} {\cal M}$.  In particular, for some $m_f < n_f/2$, suppose we have (surjective) diffeomorphisms $\eta_j: \partial_{f,j} {\cal M} \rightarrow \partial_{f,j+m_f} {\cal M}$ for $j =1, \dots m_f$ that preserve both the induced metric\footnote{Since the conformal factor is also preserved, it does not matter whether we state this definition in terms of $g$ or $\tilde g$.  We also note that we have excluded the possibility of identifying some component of $\partial_{f,j} {\cal M} $ with itself.  When the component is an $S^1$, nontrivial such identifications do exist that yield smooth results, and our analysis below could be generalized to include them, but we will have no need of them.} and the conformal factor $\Omega$.
We define $\overline {\cal M}$ by using each $\eta_j$ to identify its domain with its range, so that
\begin{equation}
n_f(\overline {\cal M}) = n_f({\cal M}) - 2m_f.
\end{equation}
The configuration $\overline {\cal M}$ is not generally smooth, but we can nevertheless evaluate $I_{top}(\overline {\cal M})$ as in \eqref{eq:top} to find
\begin{equation}
\label{eq:toptilde}
I_{top}(\overline {\cal M}) = -\pi \phi_0(4\bar \chi - \bar n_{int} ),
\end{equation}
where $\bar \chi$ and $\bar n_{int}$ are respectively the Euler character of $\overline {\cal M}$ and the number of intersections of $\partial_{as}\overline {\cal M}$ with $\partial_{f}\overline {\cal M}$.  Now, since the components  $\partial_{f,j} {\cal M}$ of  $\partial_{f} {\cal M}$ are one-dimensional, each component is either a line segment or a circle.  We may thus divide the $m_f$ identifications into $m_s$ that identify pairs of line segments and $m_c$ that identify circles (with $m_f = m_c + m_s$).  Identifying circles does not change the number of points where $\partial_{f} {\cal M}_{\epsilon}$ and $\partial_{as} {\cal M}_{\epsilon}$ intersect, but identifying two line segments removes 4 intersections (two on each segment); see figure \ref{fig:Mbar}.  Thus
\begin{equation}
\label{eq:shiftInt}
\bar n_{int} = n_{int} - 4 m_s.
\end{equation}
Similarly, identifying pairs of circles is well-known to leave the Euler character unchanged.  However, identifying a pair of line segments lowers $\chi$ by $1$; see again figure \ref{fig:Mbar}.

Thus we have
\begin{equation}
\label{eq:shiftInt}
\bar \chi  = \chi  - m_s,
\end{equation}
and
\begin{equation}
4 \bar \chi - \bar n_{int} = 4 \chi - n_{int},
\end{equation}
which yield
\begin{equation}
\label{eq:Itopid}
I_{top} (\overline {\cal M}) = I_{top}({\cal M}).
\end{equation}
Because it satisfies \eqref{eq:Itopdu} and
\eqref{eq:Itopid}, we say that $I_{top}$ is sewing-additive.  This in particular means that it satisfies \eqref{eq:bulkadd}.  We will use the same term below for any other functional satisfying analogous identities.  (Note that this as yet says nothing about the operations associated with the yet-to-be-discussed monotonicity relations \eqref{eq:mono} and \eqref{eq:strictmono}.)

In fact, the entire regulated Euclidean action $I_\epsilon$ defined by \eqref{eq:JTe} is sewing-additive.  It is already manifest that $I_\epsilon$ satisfies
\begin{equation}
\label{eq:Idu}
I_\epsilon ({\cal M}_{1} \sqcup {\cal M}_{2}) = I_\epsilon ({\cal M}_{1}) + I_\epsilon({\cal M}_{2}),
\end{equation}
so in establishing sewing-additivity we may focus on the analog of condition \eqref{eq:Itopid}.   Below, we also focus on the remaining terms $I-I_{top}$, since we have already established \eqref{eq:Itopid} for $I_{top}$.

There are now two effects to consider in order to show $I_\epsilon(\overline {\cal M})=I_\epsilon({\cal M})$.  The first is that, as in the discussion of $I_{top}$ above, intersections between the finite and asymptotic boundaries can disappear in pairs, so that there are contributions $\alpha_i \phi_i$ to $I_\epsilon({\cal M})$ that do not appear in $I_\epsilon(\overline {\cal M}).$  However, such a disappearance is associated with the joining of two asymptotic boundary segments as shown in figure \ref{fig:Kdelta} from section \ref{subsec:FALPI}.
The result is generally not smooth, so that the extrinsic curvature density $\sqrt{h} K$ of $\partial_{as}\overline {\cal M}$ contains a new delta-function of a strength defined by the angles $\alpha_i$ associated with the disappearing pair of intersections.  In fact, for disappearing interior intersection angles $\alpha_1, \alpha_2$, the delta-function is of strength $\alpha_1 + \alpha_2.$ This relation can be derived from the Gauss-Bonnet theorem.  In particular, $\sqrt{h}K$ remains smooth when both of the disappearing intersections are orthogonal ($\alpha_1 = \alpha_2=0$).  See again figure \ref{fig:Kdelta} in section \ref{subsec:FALPI}.

The second effect is that, when two boundaries are sewn together, the seam is smooth only when the extrinsic curvatures $K$ match appropriately on the two surfaces.  More generally, the sewing leads to a singularity on the seam which gives a delta-function in $\sqrt{g} R$ related to the discontinuity in extrinsic curvatures. This phenomenon is well-known from the Israel junction conditions; see e.g. \cite{Poisson:2009pwt}.  The particular relation again follows by applying the Gauss-Bonnet theorem to a disk of infinitesimal size (and perhaps strong curvature) bounded by the two surfaces to be sewn together and then taking the limit where the surfaces coincide.  In this way one sees that the contribution to $I_\epsilon(\overline {\cal M})$ from the delta-function in $R$ on $\overline {\cal M}$ precisely compensates for the fact that $I_\epsilon(\overline {\cal M})$ no longer includes explicit contributions from the boundaries of ${\cal M}$ that have been sewn together.

Since the remaining contributions to \eqref{eq:JTe} take identical forms in $\overline {\cal M}$ and ${\cal M}$, this establishes the desired relation
\begin{equation}
I_\epsilon(\overline {\cal M}) = I_\epsilon({\cal M}).
\end{equation}
In particular, we see that the regulated action $I_\epsilon$ is already sewing-additive at finite $\epsilon$.  Taking the limit $\epsilon \rightarrow 0$ then shows that corresponding property again holds for the unregulated action $I$.

\subsection{Relation to the Schwarzian Action}
\label{subsec:JTSchwarz}
%\label{subsec:FALPI}

As shown in \cite{Maldacena:2016upp} (see \cite{Saad:2019lba} for a Euclidean-signature treatment), the on-shell action for pure JT gravity takes a so-called Schwarzian form, which has proved to be extremely useful. We very briefly review this below, though most of the present section is a slight aside that generalizes the above result to our class of manifolds-with-boundaries-and-corners ${\cal M}$.  The extension is not of critical use in the main text, but may be enlightening to some readers.

We also comment briefly on the off-shell extension of this result.  The only equation of motion used to derive the Schwarzian action below is $R+2=0$.  Since the first line of \eqref{eq:JTe} is topological, deviations from the on-shell result are controlled by the term involving $\sqrt{g}\phi(R+2)$.  In the usual way (see e.g. \cite{Marolf:2012vvz} for a review) since we took the asymptotically locally AdS$_2$ boundary conditions to require
both $\partial_z h$ and $\partial_z^2 h$ vanish at $z=0$, we find $R+2 = O(z^3)$. Thus $\sqrt{g}\phi (R+2) = O(1)$ and the discrepancy from the on-shell result is finite.

This discussion is often phrased in terms of a coordinate system in which the function $h$ (defined by \eqref{eq:asymptmetric}) satisfies $h|_{z=0}=1$, with $\theta$ taking values in $[0,2\pi]$.  As noted above, this can always be achieved via coordinate transformations of the form \eqref{eq:changez}, which are indeed gauge redundancies of our system.  The essential point is then that the extrinsic curvature scalar $K$ of the family of surfaces $\Omega = \epsilon$ satisfies the asymptotic expansion
\begin{equation}
K = 1 + \epsilon^2  \operatorname{Sch}(\tan \frac{\theta(u)}{2}, u),
\end{equation}
where
\begin{equation}
\operatorname{Sch}(f(u),u) = \frac{d}{du} \left(\frac{f''(u)}{f'(u)} \right) - \frac{1}{2}\left(\frac{f''(u)}{f'(u)} \right)^2,
\end{equation}
and prime $'$ denotes $\frac{d}{du}$
so that
\begin{equation}
\label{eq:Schdef}
Sch(\tan \frac{\theta(u)}{2},u) = \frac{1}{2} (\theta')^2 - \frac{1}{2} \left(\frac{\theta''}{\theta'} \right)^2 + \frac{d}{du}\left( \frac{\theta''}{\theta'}\right).
\end{equation}
As a result, when ${\cal M}$ has only asymptotic boundaries \eqref{eq:JTe} yields
\begin{equation}
\label{eq:Schresult}
I = I_{top} -2\int \phi_b(u) \left(\frac{1}{2} (\theta')^2 - \frac{1}{2} \left(\frac{\theta''}{\theta'} \right)^2 + \frac{d}{du}\left( \frac{\theta''}{\theta'}\right) \right)du.
\end{equation}

We now wish to obtain analogous results for our manifolds ${\cal M}$ with general finite boundaries and corners where the finite and asymptotic boundaries intersect.  This in particular means that we will need to take due care to include all of the terms in \eqref{eq:JTe} that did not appear in \cite{Saad:2019lba}.  These are the terms that involve $\partial_f {\cal M}$, including the $\alpha_i$ terms associated with the intersection  $\partial_f {\cal M} \cap  \partial_{as} {\cal M}$.  Importantly, we will again take ${\cal M}$ to be on-shell, which in particular means that it was cut from an on-shell parent configuration $\widehat {\cal M}$.

The above-mentioned terms at first appear to depend strongly on the choice of $\partial_f {\cal M}$, or equivalently on the choice of the surface $\Sigma$ used to cut the parent space $\widehat {\cal M}$ to form ${\cal M}$.  However, as usual for gravitational systems, this is not in fact the case.  This is most easily seen by writing the action in the standard Hamiltonian form (see e.g. \cite{Carlip:1993sa} and references therein)
\begin{equation}
\label{eq:IHam}
I = \int_{\cal M} (p_i \dot{q}^i - N^\perp \tilde {\cal H}_\perp -  N^{||} \tilde {\cal H}_{||})) - \int_{\partial_{as} {\cal M}}  N^\perp H_\partial + I_{degen},
\end{equation}
where, $p_i, q^i$ are an appropriate set of coordinates and momenta, $N^\perp, N^{||}$ are the usual lapse and shift, and $I_{degen}$ denotes additional potential contributions from any locus where the foliation used to define the Hamiltonian formalism degenerates.

In \eqref{eq:IHam}, we choose the coordinates $q^i$ to be whatever quantities are to be held fixed on $\partial_f{\cal M}$ for \eqref{eq:JTe} to define a good variational principle. Indeed, it is perhaps useful to recall that the form \eqref{eq:IHam} must hold for any action, and in particular that the above condition on $q^i$ requires any additional boundary term on $\partial_{f} {\cal M}$ to be independent of the momenta $p_i$.  Furthermore, any such boundary terms can then be absorbed into the bulk by an appropriate redefinition of the momenta $p_i$, leaving us with an action of the form \eqref{eq:IHam} as claimed.

The usual Hamilton-Jacobi argument then shows that variations of $I$ with respect to infinitesimal changes in the location of the surface $\Sigma$ (used to cut ${\cal M}$ from $\widehat {\cal M}$) must involve two contributions. Here we restrict attention to variations that preserve the boundary conditions stated above for $I$, and which also preserve the points where $\Sigma$ intersects $\partial_{as} {\cal M}$.   The first contribution is given by varying the location of $\Sigma$ while holding each $q^i$ fixed on $\Sigma$, and the second is given by leaving $\Sigma$ fixed within the manifold by varying that values of $q^i$ on $\Sigma$ as dictated by the appropriate evolution under the equations of motion (in accord with the would-be motion of $\Sigma$ through $\widehat {\cal M}$).  So long as the intersection of $\Sigma$ with $\partial_{as} {\cal M}$ does not change, combining the two terms gives a result in which contributions of the $p_i\dot{q}^i$ term cancel completely.  The remainder is simply linear in the constraints ${\cal H}_\perp$ and ${\cal H}_{||}$.  But the constraints vanish since ${\cal M}$ is on-shell, and we see that the desired variation vanishes as well.

In other words, the action of ${\cal M}$ is in fact invariant under smooth deformations of the surface $\Sigma$ used to slice it from an on-shell $\widehat {\cal M}$, so long as the deformations both leave fixed the intersections of $\Sigma$ with the asymptotic boundary $\partial_{as} \widehat{\cal M}$ and respect the boundary conditions associated with the action $I$. In particular, this requires that the $\alpha_i$ remain of whatever order in $\epsilon$ is specified by the boundary conditions.

This result then allows us to evaluate the action $I({\cal M})$ by choosing any convenient surface $\Sigma$ related to $\partial_{f} {\cal M}$ via smooth boundary-condition preserving deformations within ${\cal M}$.  One choice we can always make is to minimize the physical length of $\Sigma$ (after subtracting the appropriate universal divergence from the region near the boundary).  In two Euclidean dimensions, doing so necessarily results in a smooth surface with vanishing extrinsic curvature \cite{Fdrer1970TheSS}; i.e., in a geodesic.

Having thus set $K=0$ on $\Sigma$, inspection of \eqref{eq:JTe} shows that all boundary terms on the interior of $\partial_f {\cal M}$ now vanish.  The only remaining contributions to $I$ from $\Sigma$ are then those associated with the angles $\alpha_i$. These are straightforward to compute using the well-known fact that geodesics in spacetimes of the form given by \eqref{eq:asymptmetric} are asymptotically of the form $\theta = \theta_0 + \theta_2 z^2+ \dots$, while the proper distance $s$ along the geodesic is $s = \ln z + O(z^2)$.
This fact follows from using \eqref{eq:asymptmetric} and expanding the geodesic equation in powers of $z$ (say, by taking $z$ as a non-affine parameter along the geodesic).
Note that the above expansion shows that taking $K=0$ on $\partial_f {\cal M}$ is consistent with \eqref{eq:partialf}.

\begin{figure}[h!]
	\centering
\includegraphics[width=0.7\linewidth]{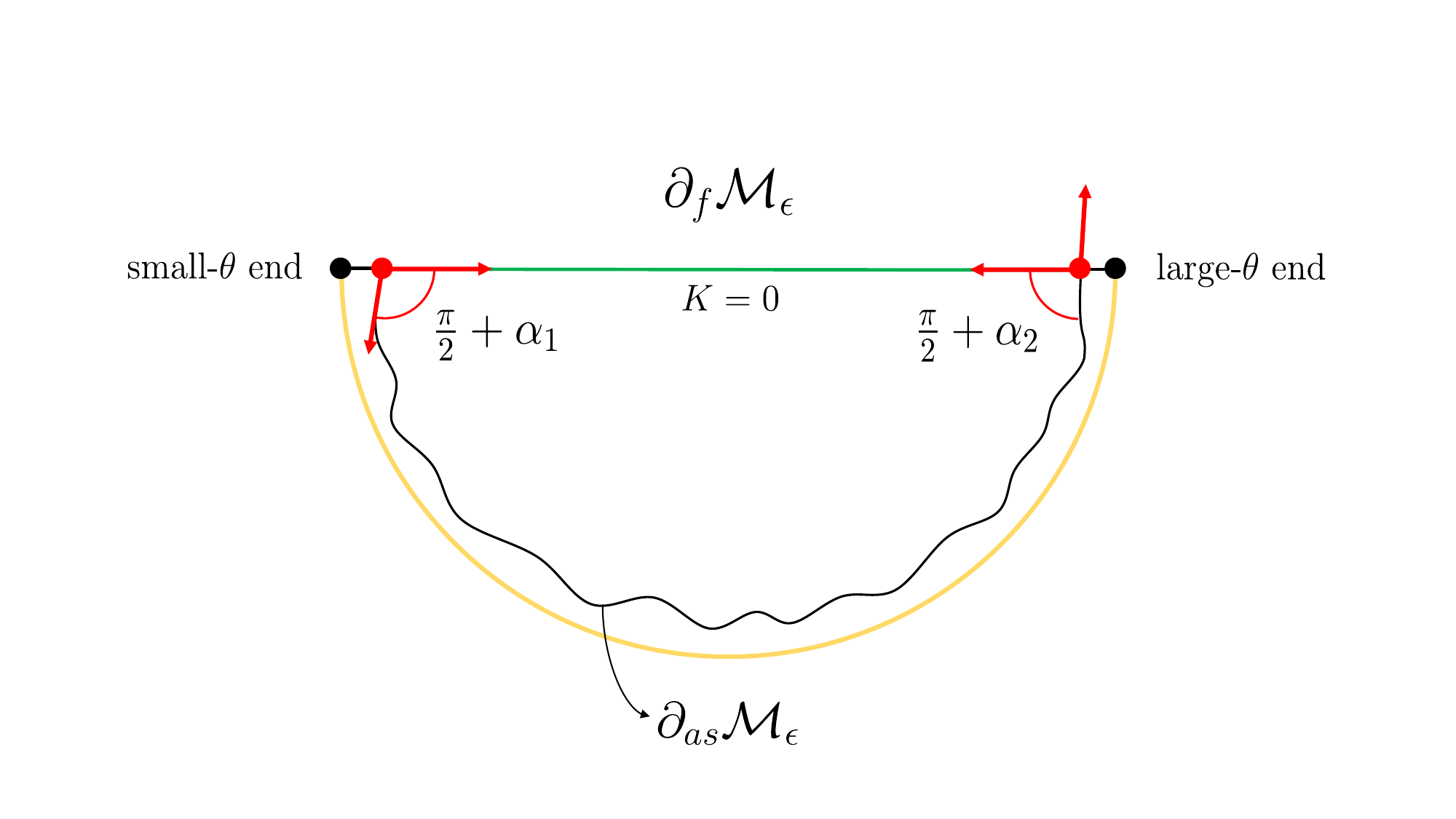}
\caption{The figures shows tangent vectors to $\partial_{as} {\cal M}$ in the direction of increasing $\theta$ and inward-directed tangents to $\partial_f {\cal M}$.  When the finite boundaries are geodesics, the indicated angles $\alpha_i$  satisfy \eqref{eq:alphacos} and \eqref{eq:phialpha} with the signs stated in the text.}
\label{fig:Mbarangles}
\end{figure}

At $\Omega = \epsilon$ the unit-normalized tangent to the inward-directed geodesic is of the form
\begin{equation}
(\frac{d\theta}{ds}, \frac{dz}{ds}) = (\frac{d\theta}{dz}, 1) \frac{dz}{ds} = (2\theta_2 \epsilon/\omega, 1) \frac{\epsilon}{\omega}  (1 + O(\epsilon)).
\end{equation}
On the other hand, defining $\omega_0(\theta) = \omega|_{z=0}$ and noting that $\Omega = z\omega_0 + O(z^2)$, we see that at $\Omega = \epsilon$ the regulated version of the asymptotic boundary $\partial_{as} {\cal M}$ has (in the direction of increasing $\theta$) the unit-normalized tangent
\begin{equation}
(1,\frac{dz}{d\theta}) \frac{d\theta}{ds} = (1,-\epsilon \frac{d}{d\theta} \omega^{-1}) \frac{\epsilon}{\omega}.
\end{equation}
Since we chose the tangent along the asymptotic boundary to be in the direction of increasing $\theta$,
combining these with \eqref{eq:asymptmetric} yields
\begin{eqnarray}
\label{eq:alphacos}
 \cos \left(\alpha + \pi/2 \right) &=& \pm \frac{\epsilon}{\omega} \left(  2 \theta_2  +  \frac{d}{d\theta} \ln \omega^{-1} \right) + O(\epsilon^2)
 \cr
&=& \pm \frac{2\pi \epsilon}{\beta} \left( 2 \theta_2  +  \frac{\theta''}{(\theta')^2}\right) + O(\epsilon^2),
\end{eqnarray}
where the $+$ ($-$) sign corresponds to an intersection point at the large-$\theta$ (small-$\theta$) end of an asymptotic boundary segment; see figure \ref{fig:Mbarangles}.
Applying \eqref{eq:omega} with $h=1$ then yields
\begin{eqnarray}
\alpha  = \mp \epsilon \left( 2 \theta_2 \theta' +  \frac{\theta''}{\theta'}\right) + O(\epsilon^2),
\end{eqnarray}
and
\begin{equation}
\label{eq:phialpha}
\phi \  \alpha = \mp \phi_b(u)  \left( 2 \theta_2 \theta' +  \frac{\theta''}{(\theta')}\right) + O(\epsilon),
\end{equation}
where now the $(-)$ sign is correct for an intersection point at the large-$\theta$ end of an asymptotic boundary segment and the $(+)$ sign holds at the small-$\theta$ end.

As a result, we find

\begin{equation}
I = I_{top} -2\int \phi_b(u) \left(\frac{1}{2} (\theta')^2 - \frac{1}{2} \left(\frac{\theta''}{\theta'} \right)^2 \right)du + 4 \sum_i \phi_b(u_i) \theta_{2,i} \theta'(u_i),
\end{equation}
where $u_i, \theta_{2,i}$ are the $u$-value and $\theta_2$-value of the $i$th intersection point between the finite and asymptotic boundaries.  The natural boundary term in \eqref{eq:Schresult} has been cancelled by the $\phi_i\alpha_i$ contributions, but a new boundary term involving $\theta_2$ remains.  This is related to the fact that, as discussed in the main text, sewing together two boundaries may not result in an asymptotic boundary that is smooth at finite $\epsilon$.  In other words, the sewn-together manifold may not be associated with a smooth conformal factor $\Omega$.  We have seen, however, that the extension to conformal factors that allow $\sqrt{h} K$ to contain delta-functions causes no significant issues.

\subsection{Positivity of the Schwarzian action}
\label{subsec:JTpos}

We now establish the positivity result needed in the main text.  In particular, we consider manifolds ${\cal M}$ satisfying the above boundary conditions and which have {\it only} asymptotic boundaries (i.e., $\partial_f {\cal M}=\emptyset.$).  Positivity of $\alpha, \phi$ and the above-noted fact that we can freely choose any finite boundaries to be geodesics then immediately also imply a similar lower bound for the case of non-empty  $\partial_f {\cal M}$.

As noted previously, we require $I_{matter}$ to be minimally-coupled to the metric $g$.  In particular, the dilaton $\phi$ should not appear in $I_{matter}$.  We also require that $I_{matter}$ be bounded below by zero when the metric $g$ satisfies $R+2=0$ and is locally asymptotically AdS$_2$.  Since the topological term is minimized by taking the spacetime to be a disconnected union of disks, for fixed boundary conditions the full action will be bounded below if we can derive a lower bound for the Schwarzian action \eqref{eq:Schresult} for each disk; i.e., on each circular boundary.  This is certainly to be expected since the Schwarzian action arises \cite{KitaevTalks} (see also \cite{Maldacena:2016hyu}) as an effective description of the low energy limit of (a limit of) the Sachdev-Ye-Kitaev model (first introduced in \cite{PhysRevLett.70.3339}), which is a standard quantum mechanical system.  However, it is reassuring to see a direct argument\footnote{It seems likely that this result is already somewhere in the vast literature concerning the Schwarzian action.  The authors would be happy to receive references to earlier published versions of this result.}.

For general $\phi_b(u)$, we would like to simplify \eqref{eq:Schresult} by defining a new coordinate $\tilde u$ such that
\begin{equation}
\label{eq:uhat}
    d\hat u = \frac{\hat \phi_b}{\phi_b} du.
\end{equation}
We can think of this $\hat u$ as being associated with a different choice of conformal frame defined by $\hat \Omega: = \frac{\hat \phi_b}{\phi_b} \Omega$, in which we see that the new boundary dilaton profile would be given by $\hat \phi_b$.   Since we have required that $\Omega$ be specified as part of the definition of the system, it would not be correct to say that the coordinate transformation \eqref{eq:uhat} actually changes the boundary values of $\phi$, but it nevertheless allows us to rewrite any function of the original $\phi_b$ boundary conditions in terms of another JT system with boundary conditions specified by $\hat \phi_b$.         The Schwarzian action correspondingly becomes
\begin{equation}
\label{eq:JTshift}
\begin{aligned}
I_{Schwarz}&:= -2\int du \, \phi_b(u) \operatorname{Sch}(\tan(\theta/2),u)\\
    &= -2\int d\hat u \,\hat \phi_b(\hat u) \operatorname{Sch}(\tan(\theta/2),\hat u)-\int du \left(  \left( \partial_{\hat u} \frac{\hat \phi_b}{\phi_b} \right)^2 +2\frac{\hat \phi_b}{\phi_b} \partial_{\hat u}^2 \frac{\hat \phi_b}{\phi_b}\right).
\end{aligned}
\end{equation}
The last term becomes manifestly non-negative after integrating by parts.
Thus we only need to show that the first term is bounded below for any convenient $\hat \phi_b(u)$ (and with the period of $\hat u$ dictated by this choice via \eqref{eq:uhat} andwith any convenient the period of $\hat u$). For simplicity, we choose $\hat \phi_b(u)$ to be a constant (which we again call $\bar \phi_b$). We also choose the period of $\hat u$ to be $2\pi$, which then sets the value of $\bar \phi_b$ for a given $\phi_b(u)$.

To clean up the notation,  we will henceforth write $\hat u$ as simply $u$ so that the simplified action (dropping the final term in \eqref{eq:JTshift}) becomes
\begin{eqnarray}
\label{eq:simpleSch}
\bar I_{Schwarz} &=& -2\int d u \,\bar \phi_b(\hat u) \operatorname{Sch}(\tan(\theta/2), u) \cr
&=& - \int d u \,\bar \phi_b(\hat u) \left[   (\theta')^2 -   \left(\frac{\theta''}{\theta'} \right)^2 + 2\frac{d}{du}\left( \frac{\theta''}{\theta'}\right) \right].
\end{eqnarray}
Since we  consider  only circular boundaries, we can ignore the total derivative that gives the final term of \eqref{eq:simpleSch}.   It is also convenient to write $\eta(u)=\theta'(u)$, so that $\bar I_{Schwarz}$ becomes
\begin{equation}
\label{eq:Schwform1}
    \bar I_{Schwarz}=-\bar \phi_b \int du\, \left(  \eta^2 - \left( \frac{\eta'}{\eta} \right)^2\right).
\end{equation}
The expression \eqref{eq:Schwform1} is to be evaluated on functions $\eta$ that satisfy an important constraint, since $\theta$ must increase by $2\pi$ when $u$ increases by $2\pi$.  In other words, we require
\begin{equation}
\label{eq:etaconstraint}
    \int_0^{2\pi} du \, \eta = 2\pi.
\end{equation}
We can take this constraint into account by adding a Lagrange multiplier to the action:
\begin{equation}
\label{eq:Swlambda}
    \bar I_{Schwarz}=-\bar \phi_b \left[\int du\, \left(  \eta^2 - \left( \frac{\eta'}{\eta} \right)^2 \right) - \lambda \left(\int du\,\eta -2\pi \right)\right].
\end{equation}

There is an obvious saddle of this action at $\eta=1, \lambda=2$.  Perturbations around this saddle point may be studied by writing $\eta=1+\Upsilon$.  Expanding \eqref{eq:Swlambda} to quadratic order yields
\begin{equation}
    \Delta \bar I_{Schwarz}= - \int du\, {\bar \phi_b} \left( \Upsilon^2-(\Upsilon')^2\right) = - \int du \, {\bar \phi_b} \left( \Upsilon^2+\Upsilon \Upsilon''\right)+ \text{bdy terms}.
\end{equation}
The eigenfunctions of the operator $\partial_u^2+1$ are
\begin{equation}
    \Upsilon=e^{iku}, \quad k \in \mathbb{Z},
\end{equation}
with eigenvalues $-k^2+1$. The $k=0$ mode thus has negative action, but it is forbidden by the constraint
\begin{equation}
    \int_0^{2\pi} du \, \Upsilon = 0.
\end{equation}
The modes $k=\pm 1$ have  vanishing action, and they turn out to correspond to the SL(2,$\mathbb{R}$) symmetries of the Schwarzian action.  These are in fact gauge symmetries of JT gravity, though they appear as global symmetries of $\bar I_{Schwarz}$ due to the partial gauge-fixing used in deriving $\bar I_{Schwarz}$ \cite{Maldacena:2016upp}. Other modes all have positive quadratic action $\Delta \bar I_{Shcw}$.

The above analysis shows $\eta=1$ to be a local minimum of the action over the space of allowed configurations.  Importantly (and as we will see explicitly below), the same must be true for all of the saddles that can be reached by following the flat directions associated with the SL(2,$\mathbb{R}$) symmetries. These of course share the same minimum value of $\bar I_{Schwarz}$.

However, we wish to show that this value is in fact a {\it global} minimum.  One way to proceed is to note that (since we have fixed the winding number) the space of configurations is connected.  As a result, if any configuration has lower action than $\eta=1$, there is a path through configuration space that connects it to $\eta=1$.  Furthermore, starting at the $\eta=1$ end of the path, the analysis above shows that the action must first increase before it can decrease.  This is the case even if the path starts by following some path in the space ${\cal Z}$ of saddles related to $\eta=1$ by SL(2,$\mathbb{R}$) zero modes, as it must clearly leave the space ${\cal Z}$ at some point and since we have shown that all paths leaving ${\cal Z}$ must first increase the action before the action can decrease.

As a result, along any given such path $p$ there must be a point $P_0$ at which the action reaches a local maximum $\mu(p)$.  Let us now attempt to minimize $\mu(p)$ over all paths $p$.  So long there are no directions in which $\mu(p)$ remains finite at the edge of space of allowed configurations (i.e., when $\eta$ diverges at some $u$), then the minimal $\mu(p)$ must in fact be a saddle $s$ for the action $\bar I_{Schwarz}$ which satisfies $\bar I_{Schwarz}(s) > \bar I_{Schwarz}(\eta=1)$; i.e., it must be a new saddle for the Schwarzian action.

While we will not exclude the runaway possibility with complete rigor, if the divergence in $\eta$ admits any kind of asymptotic expansion at large $\lambda$ it would certainly require the first two terms to cancel against each other at leading order.  But such cancelation requires
\begin{equation}
\label{eq:asympteta}
\eta = \pm \eta'/\eta + \dots,
\end{equation}
where $\dots$ represents lower order terms.  Solving \eqref{eq:asympteta} yields $\pm \eta = \frac{1}{u+\Delta}$ where $\Delta$ is a function whose derivative $\Delta'$ vanishes vanishes as $\eta \rightarrow \infty$.  Without loss of generality we can take the point at which $\eta$ diverges to be $u=0$, in which case $\Delta$ is approximately constant near $u=0$.  As a result, the integral $\int \eta$ must diverge and the constraint \eqref{eq:etaconstraint} cannot be satisfied.

Thus, if the action were to be unbounded below, there must be a new saddle point.  However, we will now seek such new saddles directly and show that they do not exist.  To do so, note that the action can be equivalently written as the action of a particle in a potential by defining $\chi=\ln \eta$ to write
\begin{equation}
\label{eq:chiact}
    \bar I_{Schwarz}={\bar \phi_b} \int du\, \left((\chi')^2 -(e^{2\chi}-\lambda e^\chi +\lambda)\right).
\end{equation}
The solutions to saddle-point equations of motion can of course be labelled by the total energy $E$ which we normalize as
\begin{equation}
\label{eq:chienergy}
    (\chi')^2+e^{2\chi}-\lambda e^\chi+\lambda =E.
\end{equation}
Note further that the potential $V(\chi)=e^{2\chi}-\lambda e^\chi +\lambda$  always approaches $\lambda$ as $u \rightarrow -\infty$, but that it does so in different ways depending on the value of $\lambda$.  For $\lambda \le 0$ the potential increases monotonically and all orbits are unbound.  Here we use the term `orbit' to refer to some $\chi(u)$ that solves the equation of motion obtained from \eqref{eq:chiact} by varying $\chi$, but which does not necessarily satisfy either the periodicity condition $\chi(u) = \chi(u+2\pi)$ or the constraint \eqref{eq:etaconstraint}.  In contrast, for $\lambda > 0$ the potential has a single critical point at $\chi = \ln \lambda/2$ and at which $V = \lambda - \lambda^2/4$.  This critical point is in fact a global minimum of the potential.  Thus the case $\lambda >0$ admits both unbound orbits (with $E \ge \lambda$) and bound orbits (with $E < \lambda$).

The potential remains finite in the asymptotic region of large negative $\chi$, so the velocity in this region is finite.  Since the unbound orbits all run to large negative $\chi$, they can reach $\chi = -\infty$ only at infinite values of $u$.  This means that there is no sense in which they can be periodic, and such orbits are not allowed.

We can therefore focus on the bound orbits.  Let us first consider the special cases that sit at the global minimum for all time.  Such solutions are clearly periodic with any period.  However, they satisfy the constraint \eqref{eq:etaconstraint} only for $\chi=0$, which then requires $\lambda = 2$, $E=1$.  For future reference we note that $E-\lambda = -1$.

Finally, we can investigate the constraints for the other bound states.  The first constraint is that $\chi$ is periodic with period $2\pi$,
\begin{eqnarray}
\label{eq:impose1C}
    2\pi &=& \int_0^{2\pi} du\, \chi' =2\int_{\chi_-}^{\chi_+} \frac{d\chi}{\sqrt{E-\lambda-e^{2\chi}+\lambda e^\chi}}\cr &=& 2\int_{\eta_-}^{\eta_+} \frac{d\eta}{\eta \sqrt{-(\eta-\eta_+)(\eta-\eta_-)}}=2\pi i \frac{1}{\sqrt{E-\lambda}},
\end{eqnarray}
which again requires $E-\lambda=-1$. Here the second step in \eqref{eq:impose1C} used equation \eqref{eq:chienergy} and the quantities $\eta_\pm$ are defined by first defining $\chi_\pm$ to be the two roots of the denominator of the integrand and then setting $\eta_{\pm} = e^{\chi_\pm}=\frac{\lambda \pm \sqrt{\lambda^2+4E-4\lambda}}{2}$. The final answer on the right-hand-side was obtained by noting that the integral over $\eta$ can be expressed as a contour integral around a branch cut from $\eta_-$ to $\eta_+$. Since the contour integral for large $\eta$ vanishes, the final answer is given by the residue of the pole at $\eta=0$.

The second constraint turns out to be trivially satisfied for any $E$ and $\lambda$ since we find
\begin{equation}
    \int du\, \eta =2\int_{\eta_-}^{\eta_+} \frac{\eta}{\eta'}=2\int_{\eta_-}^{\eta_+} \frac{1}{\sqrt{-(\eta-\eta_+)(\eta-\eta_-)}} = 2\pi.
\end{equation}
Again, in the final step the integration is performed using complex contour integration techniques.  Thus we see that all saddles have  $E-\lambda=-1$, but that there is a saddle for each $\lambda \in \mathbb{R}$ (or, equivalently, for each real $E$). The original $\eta=1$ saddle lies in this family with $\lambda=2$, which corresponds to the case where $\eta_+=\eta_-$.

All of these solutions turn out to have the same action, consistent with the earlier statement that there is a family of solutions related by an SL(2,$\mathbb{R}$) symmetry. To see this, note that the Schwarzian action is given by
\begin{equation}
\begin{aligned}
    \bar I_{Schwarz}&={\bar \phi_b} \int_0^{2\pi} du\, \left((\chi')^2 -(e^{2\chi}-\lambda e^\chi +\lambda)\right)\\
    &={\bar \phi_b} \int_0^{2\pi} du\, \left(E- 2(e^{2\chi}-\lambda e^\chi+\lambda)\right)\\
    & = 2\bar \phi_b \int_{\eta_-}^{\eta_+} d\eta\, \frac{E-2\eta^2+2\lambda \eta-2\lambda}{\eta'}\\
    &= 2\bar \phi_b \int_{\eta_-}^{\eta_+} d\eta\, \frac{E-2\eta^2+2\lambda \eta-2\lambda}{\eta \sqrt{E-\eta^2+\lambda \eta-\lambda} },\\
\end{aligned}
\end{equation}
where we still need to set $E-\lambda=-1$. This integral can once again be tackled by the methods of complex analysis. The integrand has poles at $\eta=0, \infty$, so this integral can be evaluated using their residues to find
\begin{equation}
\label{eq:SL2r}
    \bar I_{Schwarz}=\bar \phi_b  2\pi i \sqrt{E-\lambda}=-2\pi \bar \phi_b ,
\end{equation}
which is a constant independent of $\lambda$. Note that this corresponds to the action in section \ref{sec:simple} with $\beta=2\pi$.

Of course, the SL(2,$\mathbb{R}$) symmetry of \cite{Maldacena:2016upp} relates each saddle on ${\cal Z}$ to the $\eta=1$ solution and, in particular, shows that each such saddle is again a local minimum up to the SL(2,$\mathbb{R}$) zero mode.
This observation completes the argument that \eqref{eq:SL2r} is in fact the minimum value of $\bar I_{Schwarz}$, and thus that the Schwarzian action is bounded below.

Allowing a general $\beta$, multiple $S^1$ boundaries labelled by $j$, and restoring the topological term would yield the general bound
\begin{equation}
\label{eq:JTB}
I = I_{top} + I_{Schwarz} \ge - \sum_j \left( 4\pi \phi_0 + 4\pi^2 \bar \phi_b/\beta_j\right).
\end{equation}

\section{Cut-and-paste Asymptotically locally AdS boundary conditions for the Einstein-Hilbert action with boundary counterterms}

\label{app:cd1minsurface}

Standard treatments of Asymptotically locally AdS (AlAdS) boundary conditions, and in particular standard discussions of boundary counterterms, typically assume that all structures should be smooth  (see e.g. \cite{Marolf:2008mf} for a review and references).  However, as described in detail for JT gravity in appendix \ref{app:JT}, our cut-and-paste constructions generally lead to some lack of differentiability.  The purpose of this section is thus to extend the standard boundary conditions to allow this behavior and to establish the associated properties of the action needed for section \ref{sec:asstatus}.

The form of our cut-and-paste construction will make this straightforward. Since we work in Euclidean signature, the relevant conical singularities were already addressed thoroughly in \cite{Dong:2019piw}.  Furthermore, it is readily apparent that they do not affect the asymptotic boundary conditions.  In addition,  as argued in section \ref{sec:asstatus}, the bulk terms in the action remain finite and well-defined under our cut-and-paste operations.  We thus need only consider the effects of these operations on the asymptotic region of the spacetime.

Since  section \ref{sec:asstatus} chooses to slice the smooth spacetimes along minimal surfaces, our study of the asymptotics will be facilitated by understanding the asymptotics of codimension-1 minimal surfaces $\Sigma$ in smooth AlAdS spacetimes.   We are interested in surfaces that are anchored on the asymptotic boundary, in the sense that $\partial \Sigma$ is a smooth codimension-1 submanifold of $\partial {\cal M}$.  These anchor sets are always boundaries of the form $\partial M_a = \partial M_b$ at which two source manifolds-with-boundaries $M_a,M_b$ are sewn together to form some closed manifold $\tilde M_{ab}$. Furthermore, by the rim requirement of section \ref{subsec:main}, such boundaries always lie in cylinders ${\cal C}_\epsilon$ of the form discussed in section \ref{sec:nongrav}.

It is thus convenient to define a Euclidean ``time'' coordinate $t_E$ on the AlAdS boundary $\tilde M$ such that $t_E$ is constant on each cut on which our extremal surface is to be anchored, and for which the (unphysical) AlAdS boundary metric has $g^{(0)}_{t_Et_E}=1$.    We make take any connected component of the anchor set to be of the form $t_E=t_0$.

Consider in particular the part of the extremal surface near this anchor set.   When the bulk AlAdS spacetime has dimension $d+1$, we may introduce $d-1$ coordinates $x^i$ on the $t_E=t_0$ slice and use these (along with $t_E$) to construct a Fefferman-Graham coordinate system near $\tilde M$. The codimension-1 minimal surface can of course be found by minimizing the volume functional

\begin{equation}
\label{eq:V}
V[\Sigma] = \int_\Sigma \sqrt{g_\Sigma} = \int_\Sigma dz dx^{d-1} z^{-d}\sqrt{g^{(0)}_{\partial \Sigma}} \sqrt{1 + (\partial_z t_E)^2 }   (1 + \dots ),
\end{equation}
where $g_\Sigma$ is the determinant of the induced metric on $\Sigma$, $g^{(0)}_\Sigma$ is the determinant of the metric induced on $\partial \Sigma$ by the (unphysical) AlAdS boundary metric $g^{(0)}$, and $\dots$ denotes terms that are subleading as $z\rightarrow 0$.

The Euler-Lagrange equation associated with extremizing \eqref{eq:V} is then proportional to
\begin{equation}
\label{eq:VEL}
0  = \partial_z  \frac{1}{z^{d} \sqrt{1 + (\partial_z t_E)^2 } } \partial_z t_E    (1 + \dots ),
\end{equation}
which for some constant $C$ yields
\begin{equation}
\label{eq:VELsol}
C z^{d} \sqrt{1 + (\partial_z t_E)^2 }   =    \partial_z t_E    (1 + \dots ),
\end{equation}
whence we find
\begin{equation}
\label{eq:VELsol2}
\partial_z t_E  = O( z^d), \ \ \ {\rm and \ thus} \ \ \  t_E =t_0 + O(z^{d+1}).
\end{equation}

As a result, our cut-and-paste construction using minimal surfaces clearly defines spacetimes in which the usual Fefferman-Graham expansion holds up to possible corrections at order $z^{d+1}$ relative to the leading terms.  Since all divergences in the gravitational action are associated with terms that are at most of order $z^{d-1}$, an action defined using the standard boundary counter-terms will remain finite on such spacetimes.  In particular, the traced extrinsic curvature $K$ of a $z=constant$ surface will contain a delta-function $\delta(t-t_0)$, but with a coefficient of order $z^{d+1}$.  Since the volume element on the asymptotic boundary is only $O(z^{-d})$, this means that such a delta-function makes no contribution to the Gibbons-Hawking term at $z=0$.

Furthermore, since the boundary stress tensor $T_{bndy}^{IJ}$ is associated with the term of order $z^d$ in the Fefferman-Graham expansion of the bulk metric, it remains well-defined as well.  Here we use $I, J$ to denote $\{t_E, x^i\}$.  Thus by the usual computation we may write the variation of the action as

\begin{equation}
\delta I  = \int_{\partial {\cal M}} T_{bndy}^{ij} \delta g^{(0)}_{ij} + {\rm EOM \ terms},
\end{equation}
where EOM terms denotes terms proportional to the usual bulk equations of motion.  In particular, from this we see that the standard action continues to give a good variational principle for our cut-and-paste spacetimes.  As a result, we are free to extend the domain of the usual action to include the above non-smooth spacetimes without further modification.

% %~~~~~~~~~~~~~~~~~~~~~~~~~~~~~~~~~~~~~~~~~~~~~~~~~~~~~~~~~~~~~~~~~~~~~
\addcontentsline{toc}{section}{References}
\bibliographystyle{JHEP}
\bibliography{references}

% %~~~~~~~~~~~~~~~~~~~~~~~~~~~~~~~~~~~~~~~~~~~~~~~~~~~~~~~~~~~~~~~~~~~~~
\end{document}